\documentclass[longauth]{aa}

\usepackage{graphicx}
\usepackage{natbib}
\usepackage{scalerel}
\usepackage{ulem}

\usepackage[table]{xcolor}

\usepackage{txfonts}
\usepackage[pdfencoding=auto,psdextra]{hyperref}
\AtBeginDocument{%
}
\hypersetup{
    colorlinks=true,
    linkcolor=blue,
    filecolor=blue,      
    urlcolor=blue,
    citecolor=blue,
}
\makeatletter
\renewcommand*\aa@pageof{, page \thepage{} of \pageref{LastPage}}
\makeatother

\usepackage[utf8]{inputenc}

\usepackage[switch, modulo]{lineno}

\usepackage{euclid}

\usepackage{bm}
\usepackage{tabularx}

\newcommand{\borg}{\texttt{BORG}\xspace}
\newcommand{\fnl}{{f_\mathrm{NL}}}
\newcommand{\bpd}{{b_{\phi , \, \delta}}}
\newcommand{\euclid}{\textit{Euclid}}

\newcommand{\fnll}{f_{\mathrm{NL}}^{\mathrm{local}}}

\newcommand{\Mpch}{\ensuremath{h^{-1}\,\text{Mpc}}}
\newcommand{\hMpc}{\ensuremath{h\;\text{Mpc}^{-1}}}
\newcommand{\mvec}[1]{{\mathbf{#1}}}

\definecolor{darkgreen}{rgb}{0.0, 0.4, 0.26}

\newcommand{\replace}[2]{#2}
\newcommand{\new}[1]{#1}
\begin{document} 
  \title{\Euclid: Field-level inference of primordial non-Gaussianity and cosmic initial conditions\thanks{This paper is published on behalf of the Euclid Consortium.}}
   \titlerunning{\Euclid: Field-level inference of PNG and Cosmic I.Cs.}
\newcommand{\orcid}[1]{}		   
\author{A.~Andrews\thanks{\email{adam.andrews@inaf.it}}\inst{\ref{aff1},\ref{aff2}}
\and J.~Jasche\orcid{0000-0002-4677-5843}\inst{\ref{aff3},\ref{aff4}}
\and G.~Lavaux\orcid{0000-0003-0143-8891}\inst{\ref{aff5}}
\and F.~Leclercq\orcid{0000-0002-9339-1404}\inst{\ref{aff5}}
\and F.~Finelli\orcid{0000-0002-6694-3269}\inst{\ref{aff1},\ref{aff2}}
\and Y.~Akrami\orcid{0000-0002-2407-7956}\inst{\ref{aff6},\ref{aff7}}
\and M.~Ballardini\orcid{0000-0003-4481-3559}\inst{\ref{aff8},\ref{aff1},\ref{aff9}}
\and D.~Karagiannis\orcid{0000-0002-4927-0816}\inst{\ref{aff10},\ref{aff11}}
\and J.~Valiviita\orcid{0000-0001-6225-3693}\inst{\ref{aff12},\ref{aff13}}
\and N.~Bartolo\inst{\ref{aff14},\ref{aff15},\ref{aff16}}
\and G.~Ca\~nas-Herrera\orcid{0000-0003-2796-2149}\inst{\ref{aff17},\ref{aff18}}
\and S.~Casas\orcid{0000-0002-4751-5138}\inst{\ref{aff19},\ref{aff20}}
\and B.~R.~Granett\orcid{0000-0003-2694-9284}\inst{\ref{aff21}}
\and F.~Pace\orcid{0000-0001-8039-0480}\inst{\ref{aff22},\ref{aff23},\ref{aff24}}
\and D.~Paoletti\orcid{0000-0003-4761-6147}\inst{\ref{aff1},\ref{aff2}}
\and N.~Porqueres\inst{\ref{aff25}}
\and Z.~Sakr\orcid{0000-0002-4823-3757}\inst{\ref{aff26},\ref{aff27},\ref{aff28}}
\and D.~Sapone\orcid{0000-0001-7089-4503}\inst{\ref{aff29}}
\and N.~Aghanim\orcid{0000-0002-6688-8992}\inst{\ref{aff30}}
\and A.~Amara\inst{\ref{aff31}}
\and S.~Andreon\orcid{0000-0002-2041-8784}\inst{\ref{aff21}}
\and C.~Baccigalupi\orcid{0000-0002-8211-1630}\inst{\ref{aff32},\ref{aff33},\ref{aff34},\ref{aff35}}
\and M.~Baldi\orcid{0000-0003-4145-1943}\inst{\ref{aff36},\ref{aff1},\ref{aff37}}
\and S.~Bardelli\orcid{0000-0002-8900-0298}\inst{\ref{aff1}}
\and D.~Bonino\orcid{0000-0002-3336-9977}\inst{\ref{aff24}}
\and E.~Branchini\orcid{0000-0002-0808-6908}\inst{\ref{aff38},\ref{aff39},\ref{aff21}}
\and M.~Brescia\orcid{0000-0001-9506-5680}\inst{\ref{aff40},\ref{aff41},\ref{aff42}}
\and J.~Brinchmann\orcid{0000-0003-4359-8797}\inst{\ref{aff43},\ref{aff44}}
\and S.~Camera\orcid{0000-0003-3399-3574}\inst{\ref{aff22},\ref{aff23},\ref{aff24}}
\and V.~Capobianco\orcid{0000-0002-3309-7692}\inst{\ref{aff24}}
\and C.~Carbone\orcid{0000-0003-0125-3563}\inst{\ref{aff45}}
\and J.~Carretero\orcid{0000-0002-3130-0204}\inst{\ref{aff46},\ref{aff47}}
\and M.~Castellano\orcid{0000-0001-9875-8263}\inst{\ref{aff48}}
\and G.~Castignani\orcid{0000-0001-6831-0687}\inst{\ref{aff1}}
\and S.~Cavuoti\orcid{0000-0002-3787-4196}\inst{\ref{aff41},\ref{aff42}}
\and A.~Cimatti\inst{\ref{aff49}}
\and C.~Colodro-Conde\inst{\ref{aff50}}
\and G.~Congedo\orcid{0000-0003-2508-0046}\inst{\ref{aff51}}
\and C.~J.~Conselice\orcid{0000-0003-1949-7638}\inst{\ref{aff52}}
\and L.~Conversi\orcid{0000-0002-6710-8476}\inst{\ref{aff53},\ref{aff54}}
\and Y.~Copin\orcid{0000-0002-5317-7518}\inst{\ref{aff55}}
\and F.~Courbin\orcid{0000-0003-0758-6510}\inst{\ref{aff56},\ref{aff57},\ref{aff58}}
\and H.~M.~Courtois\orcid{0000-0003-0509-1776}\inst{\ref{aff59}}
\and A.~Da~Silva\orcid{0000-0002-6385-1609}\inst{\ref{aff60},\ref{aff61}}
\and H.~Degaudenzi\orcid{0000-0002-5887-6799}\inst{\ref{aff62}}
\and G.~De~Lucia\orcid{0000-0002-6220-9104}\inst{\ref{aff33}}
\and A.~M.~Di~Giorgio\orcid{0000-0002-4767-2360}\inst{\ref{aff63}}
\and J.~Dinis\orcid{0000-0001-5075-1601}\inst{\ref{aff60},\ref{aff61}}
\and F.~Dubath\orcid{0000-0002-6533-2810}\inst{\ref{aff62}}
\and C.~A.~J.~Duncan\inst{\ref{aff52}}
\and X.~Dupac\inst{\ref{aff54}}
\and S.~Dusini\orcid{0000-0002-1128-0664}\inst{\ref{aff15}}
\and M.~Farina\orcid{0000-0002-3089-7846}\inst{\ref{aff63}}
\and S.~Farrens\orcid{0000-0002-9594-9387}\inst{\ref{aff64}}
\and F.~Faustini\orcid{0000-0001-6274-5145}\inst{\ref{aff65},\ref{aff48}}
\and S.~Ferriol\inst{\ref{aff55}}
\and M.~Frailis\orcid{0000-0002-7400-2135}\inst{\ref{aff33}}
\and E.~Franceschi\orcid{0000-0002-0585-6591}\inst{\ref{aff1}}
\and S.~Galeotta\orcid{0000-0002-3748-5115}\inst{\ref{aff33}}
\and B.~Gillis\orcid{0000-0002-4478-1270}\inst{\ref{aff51}}
\and C.~Giocoli\orcid{0000-0002-9590-7961}\inst{\ref{aff1},\ref{aff37}}
\and P.~G\'omez-Alvarez\orcid{0000-0002-8594-5358}\inst{\ref{aff66},\ref{aff54}}
\and A.~Grazian\orcid{0000-0002-5688-0663}\inst{\ref{aff16}}
\and F.~Grupp\inst{\ref{aff67},\ref{aff68}}
\and S.~V.~H.~Haugan\orcid{0000-0001-9648-7260}\inst{\ref{aff69}}
\and W.~Holmes\inst{\ref{aff70}}
\and F.~Hormuth\inst{\ref{aff71}}
\and A.~Hornstrup\orcid{0000-0002-3363-0936}\inst{\ref{aff72},\ref{aff73}}
\and P.~Hudelot\inst{\ref{aff5}}
\and S.~Ili\'c\orcid{0000-0003-4285-9086}\inst{\ref{aff74},\ref{aff27}}
\and K.~Jahnke\orcid{0000-0003-3804-2137}\inst{\ref{aff75}}
\and M.~Jhabvala\inst{\ref{aff76}}
\and B.~Joachimi\orcid{0000-0001-7494-1303}\inst{\ref{aff77}}
\and E.~Keih\"anen\orcid{0000-0003-1804-7715}\inst{\ref{aff78}}
\and S.~Kermiche\orcid{0000-0002-0302-5735}\inst{\ref{aff79}}
\and A.~Kiessling\orcid{0000-0002-2590-1273}\inst{\ref{aff70}}
\and B.~Kubik\orcid{0009-0006-5823-4880}\inst{\ref{aff55}}
\and M.~Kunz\orcid{0000-0002-3052-7394}\inst{\ref{aff80}}
\and H.~Kurki-Suonio\orcid{0000-0002-4618-3063}\inst{\ref{aff12},\ref{aff13}}
\and S.~Ligori\orcid{0000-0003-4172-4606}\inst{\ref{aff24}}
\and P.~B.~Lilje\orcid{0000-0003-4324-7794}\inst{\ref{aff69}}
\and V.~Lindholm\orcid{0000-0003-2317-5471}\inst{\ref{aff12},\ref{aff13}}
\and I.~Lloro\orcid{0000-0001-5966-1434}\inst{\ref{aff81}}
\and E.~Maiorano\orcid{0000-0003-2593-4355}\inst{\ref{aff1}}
\and O.~Mansutti\orcid{0000-0001-5758-4658}\inst{\ref{aff33}}
\and O.~Marggraf\orcid{0000-0001-7242-3852}\inst{\ref{aff82}}
\and K.~Markovic\orcid{0000-0001-6764-073X}\inst{\ref{aff70}}
\and M.~Martinelli\orcid{0000-0002-6943-7732}\inst{\ref{aff48},\ref{aff83}}
\and N.~Martinet\orcid{0000-0003-2786-7790}\inst{\ref{aff84}}
\and F.~Marulli\orcid{0000-0002-8850-0303}\inst{\ref{aff85},\ref{aff1},\ref{aff37}}
\and R.~Massey\orcid{0000-0002-6085-3780}\inst{\ref{aff86}}
\and E.~Medinaceli\orcid{0000-0002-4040-7783}\inst{\ref{aff1}}
\and S.~Mei\orcid{0000-0002-2849-559X}\inst{\ref{aff87}}
\and Y.~Mellier\inst{\ref{aff3},\ref{aff5}}
\and M.~Meneghetti\orcid{0000-0003-1225-7084}\inst{\ref{aff1},\ref{aff37}}
\and E.~Merlin\orcid{0000-0001-6870-8900}\inst{\ref{aff48}}
\and G.~Meylan\inst{\ref{aff56}}
\and M.~Moresco\orcid{0000-0002-7616-7136}\inst{\ref{aff85},\ref{aff1}}
\and L.~Moscardini\orcid{0000-0002-3473-6716}\inst{\ref{aff85},\ref{aff1},\ref{aff37}}
\and C.~Neissner\orcid{0000-0001-8524-4968}\inst{\ref{aff88},\ref{aff47}}
\and S.-M.~Niemi\inst{\ref{aff17}}
\and J.~W.~Nightingale\orcid{0000-0002-8987-7401}\inst{\ref{aff89}}
\and C.~Padilla\orcid{0000-0001-7951-0166}\inst{\ref{aff88}}
\and S.~Paltani\orcid{0000-0002-8108-9179}\inst{\ref{aff62}}
\and F.~Pasian\orcid{0000-0002-4869-3227}\inst{\ref{aff33}}
\and K.~Pedersen\inst{\ref{aff90}}
\and V.~Pettorino\inst{\ref{aff17}}
\and S.~Pires\orcid{0000-0002-0249-2104}\inst{\ref{aff64}}
\and G.~Polenta\orcid{0000-0003-4067-9196}\inst{\ref{aff65}}
\and M.~Poncet\inst{\ref{aff91}}
\and L.~A.~Popa\inst{\ref{aff92}}
\and L.~Pozzetti\orcid{0000-0001-7085-0412}\inst{\ref{aff1}}
\and F.~Raison\orcid{0000-0002-7819-6918}\inst{\ref{aff67}}
\and R.~Rebolo\inst{\ref{aff50},\ref{aff93},\ref{aff94}}
\and A.~Renzi\orcid{0000-0001-9856-1970}\inst{\ref{aff14},\ref{aff15}}
\and J.~Rhodes\orcid{0000-0002-4485-8549}\inst{\ref{aff70}}
\and G.~Riccio\inst{\ref{aff41}}
\and E.~Romelli\orcid{0000-0003-3069-9222}\inst{\ref{aff33}}
\and M.~Roncarelli\orcid{0000-0001-9587-7822}\inst{\ref{aff1}}
\and R.~Saglia\orcid{0000-0003-0378-7032}\inst{\ref{aff68},\ref{aff67}}
\and A.~G.~S\'anchez\orcid{0000-0003-1198-831X}\inst{\ref{aff67}}
\and B.~Sartoris\orcid{0000-0003-1337-5269}\inst{\ref{aff68},\ref{aff33}}
\and M.~Schirmer\orcid{0000-0003-2568-9994}\inst{\ref{aff75}}
\and P.~Schneider\orcid{0000-0001-8561-2679}\inst{\ref{aff82}}
\and T.~Schrabback\orcid{0000-0002-6987-7834}\inst{\ref{aff95}}
\and A.~Secroun\orcid{0000-0003-0505-3710}\inst{\ref{aff79}}
\and E.~Sefusatti\orcid{0000-0003-0473-1567}\inst{\ref{aff33},\ref{aff32},\ref{aff34}}
\and S.~Serrano\orcid{0000-0002-0211-2861}\inst{\ref{aff96},\ref{aff97},\ref{aff98}}
\and C.~Sirignano\orcid{0000-0002-0995-7146}\inst{\ref{aff14},\ref{aff15}}
\and G.~Sirri\orcid{0000-0003-2626-2853}\inst{\ref{aff37}}
\and L.~Stanco\orcid{0000-0002-9706-5104}\inst{\ref{aff15}}
\and J.~Steinwagner\orcid{0000-0001-7443-1047}\inst{\ref{aff67}}
\and P.~Tallada-Cresp\'{i}\orcid{0000-0002-1336-8328}\inst{\ref{aff46},\ref{aff47}}
\and A.~N.~Taylor\inst{\ref{aff51}}
\and I.~Tereno\inst{\ref{aff60},\ref{aff99}}
\and R.~Toledo-Moreo\orcid{0000-0002-2997-4859}\inst{\ref{aff100}}
\and F.~Torradeflot\orcid{0000-0003-1160-1517}\inst{\ref{aff47},\ref{aff46}}
\and I.~Tutusaus\orcid{0000-0002-3199-0399}\inst{\ref{aff27}}
\and L.~Valenziano\orcid{0000-0002-1170-0104}\inst{\ref{aff1},\ref{aff2}}
\and T.~Vassallo\orcid{0000-0001-6512-6358}\inst{\ref{aff68},\ref{aff33}}
\and G.~Verdoes~Kleijn\orcid{0000-0001-5803-2580}\inst{\ref{aff101}}
\and A.~Veropalumbo\orcid{0000-0003-2387-1194}\inst{\ref{aff21},\ref{aff39},\ref{aff102}}
\and Y.~Wang\orcid{0000-0002-4749-2984}\inst{\ref{aff103}}
\and J.~Weller\orcid{0000-0002-8282-2010}\inst{\ref{aff68},\ref{aff67}}
\and G.~Zamorani\orcid{0000-0002-2318-301X}\inst{\ref{aff1}}
\and E.~Zucca\orcid{0000-0002-5845-8132}\inst{\ref{aff1}}
\and C.~Burigana\orcid{0000-0002-3005-5796}\inst{\ref{aff104},\ref{aff2}}
\and V.~Scottez\inst{\ref{aff3},\ref{aff105}}
\and A.~Spurio~Mancini\orcid{0000-0001-5698-0990}\inst{\ref{aff106},\ref{aff107}}
\and M.~Viel\orcid{0000-0002-2642-5707}\inst{\ref{aff32},\ref{aff33},\ref{aff35},\ref{aff34},\ref{aff108}}}
										   
\institute{INAF-Osservatorio di Astrofisica e Scienza dello Spazio di Bologna, Via Piero Gobetti 93/3, 40129 Bologna, Italy\label{aff1}
\and
INFN-Bologna, Via Irnerio 46, 40126 Bologna, Italy\label{aff2}
\and
Institut d'Astrophysique de Paris, 98bis Boulevard Arago, 75014, Paris, France\label{aff3}
\and
Oskar Klein Centre for Cosmoparticle Physics, Department of Physics, Stockholm University, Stockholm, SE-106 91, Sweden\label{aff4}
\and
Institut d'Astrophysique de Paris, UMR 7095, CNRS, and Sorbonne Universit\'e, 98 bis boulevard Arago, 75014 Paris, France\label{aff5}
\and
Instituto de F\'isica Te\'orica UAM-CSIC, Campus de Cantoblanco, 28049 Madrid, Spain\label{aff6}
\and
CERCA/ISO, Department of Physics, Case Western Reserve University, 10900 Euclid Avenue, Cleveland, OH 44106, USA\label{aff7}
\and
Dipartimento di Fisica e Scienze della Terra, Universit\`a degli Studi di Ferrara, Via Giuseppe Saragat 1, 44122 Ferrara, Italy\label{aff8}
\and
Istituto Nazionale di Fisica Nucleare, Sezione di Ferrara, Via Giuseppe Saragat 1, 44122 Ferrara, Italy\label{aff9}
\and
School of Physics and Astronomy, Queen Mary University of London, Mile End Road, London E1 4NS, UK\label{aff10}
\and
Department of Physics and Astronomy, University of the Western Cape, Bellville, Cape Town, 7535, South Africa\label{aff11}
\and
Department of Physics, P.O. Box 64, 00014 University of Helsinki, Finland\label{aff12}
\and
Helsinki Institute of Physics, Gustaf H{\"a}llstr{\"o}min katu 2, University of Helsinki, Helsinki, Finland\label{aff13}
\and
Dipartimento di Fisica e Astronomia "G. Galilei", Universit\`a di Padova, Via Marzolo 8, 35131 Padova, Italy\label{aff14}
\and
INFN-Padova, Via Marzolo 8, 35131 Padova, Italy\label{aff15}
\and
INAF-Osservatorio Astronomico di Padova, Via dell'Osservatorio 5, 35122 Padova, Italy\label{aff16}
\and
European Space Agency/ESTEC, Keplerlaan 1, 2201 AZ Noordwijk, The Netherlands\label{aff17}
\and
Institute Lorentz, Leiden University, Niels Bohrweg 2, 2333 CA Leiden, The Netherlands\label{aff18}
\and
Institute for Theoretical Particle Physics and Cosmology (TTK), RWTH Aachen University, 52056 Aachen, Germany\label{aff19}
\and
Institute of Cosmology and Gravitation, University of Portsmouth, Portsmouth PO1 3FX, UK\label{aff20}
\and
INAF-Osservatorio Astronomico di Brera, Via Brera 28, 20122 Milano, Italy\label{aff21}
\and
Dipartimento di Fisica, Universit\`a degli Studi di Torino, Via P. Giuria 1, 10125 Torino, Italy\label{aff22}
\and
INFN-Sezione di Torino, Via P. Giuria 1, 10125 Torino, Italy\label{aff23}
\and
INAF-Osservatorio Astrofisico di Torino, Via Osservatorio 20, 10025 Pino Torinese (TO), Italy\label{aff24}
\and
CEA Saclay, DFR/IRFU, Service d'Astrophysique, Bat. 709, 91191 Gif-sur-Yvette, France\label{aff25}
\and
Institut f\"ur Theoretische Physik, University of Heidelberg, Philosophenweg 16, 69120 Heidelberg, Germany\label{aff26}
\and
Institut de Recherche en Astrophysique et Plan\'etologie (IRAP), Universit\'e de Toulouse, CNRS, UPS, CNES, 14 Av. Edouard Belin, 31400 Toulouse, France\label{aff27}
\and
Universit\'e St Joseph; Faculty of Sciences, Beirut, Lebanon\label{aff28}
\and
Departamento de F\'isica, FCFM, Universidad de Chile, Blanco Encalada 2008, Santiago, Chile\label{aff29}
\and
Universit\'e Paris-Saclay, CNRS, Institut d'astrophysique spatiale, 91405, Orsay, France\label{aff30}
\and
School of Mathematics and Physics, University of Surrey, Guildford, Surrey, GU2 7XH, UK\label{aff31}
\and
IFPU, Institute for Fundamental Physics of the Universe, via Beirut 2, 34151 Trieste, Italy\label{aff32}
\and
INAF-Osservatorio Astronomico di Trieste, Via G. B. Tiepolo 11, 34143 Trieste, Italy\label{aff33}
\and
INFN, Sezione di Trieste, Via Valerio 2, 34127 Trieste TS, Italy\label{aff34}
\and
SISSA, International School for Advanced Studies, Via Bonomea 265, 34136 Trieste TS, Italy\label{aff35}
\and
Dipartimento di Fisica e Astronomia, Universit\`a di Bologna, Via Gobetti 93/2, 40129 Bologna, Italy\label{aff36}
\and
INFN-Sezione di Bologna, Viale Berti Pichat 6/2, 40127 Bologna, Italy\label{aff37}
\and
Dipartimento di Fisica, Universit\`a di Genova, Via Dodecaneso 33, 16146, Genova, Italy\label{aff38}
\and
INFN-Sezione di Genova, Via Dodecaneso 33, 16146, Genova, Italy\label{aff39}
\and
Department of Physics "E. Pancini", University Federico II, Via Cinthia 6, 80126, Napoli, Italy\label{aff40}
\and
INAF-Osservatorio Astronomico di Capodimonte, Via Moiariello 16, 80131 Napoli, Italy\label{aff41}
\and
INFN section of Naples, Via Cinthia 6, 80126, Napoli, Italy\label{aff42}
\and
Instituto de Astrof\'isica e Ci\^encias do Espa\c{c}o, Universidade do Porto, CAUP, Rua das Estrelas, PT4150-762 Porto, Portugal\label{aff43}
\and
Faculdade de Ci\^encias da Universidade do Porto, Rua do Campo de Alegre, 4150-007 Porto, Portugal\label{aff44}
\and
INAF-IASF Milano, Via Alfonso Corti 12, 20133 Milano, Italy\label{aff45}
\and
Centro de Investigaciones Energ\'eticas, Medioambientales y Tecnol\'ogicas (CIEMAT), Avenida Complutense 40, 28040 Madrid, Spain\label{aff46}
\and
Port d'Informaci\'{o} Cient\'{i}fica, Campus UAB, C. Albareda s/n, 08193 Bellaterra (Barcelona), Spain\label{aff47}
\and
INAF-Osservatorio Astronomico di Roma, Via Frascati 33, 00078 Monteporzio Catone, Italy\label{aff48}
\and
Dipartimento di Fisica e Astronomia "Augusto Righi" - Alma Mater Studiorum Universit\`a di Bologna, Viale Berti Pichat 6/2, 40127 Bologna, Italy\label{aff49}
\and
Instituto de Astrof\'isica de Canarias, Calle V\'ia L\'actea s/n, 38204, San Crist\'obal de La Laguna, Tenerife, Spain\label{aff50}
\and
Institute for Astronomy, University of Edinburgh, Royal Observatory, Blackford Hill, Edinburgh EH9 3HJ, UK\label{aff51}
\and
Jodrell Bank Centre for Astrophysics, Department of Physics and Astronomy, University of Manchester, Oxford Road, Manchester M13 9PL, UK\label{aff52}
\and
European Space Agency/ESRIN, Largo Galileo Galilei 1, 00044 Frascati, Roma, Italy\label{aff53}
\and
ESAC/ESA, Camino Bajo del Castillo, s/n., Urb. Villafranca del Castillo, 28692 Villanueva de la Ca\~nada, Madrid, Spain\label{aff54}
\and
Universit\'e Claude Bernard Lyon 1, CNRS/IN2P3, IP2I Lyon, UMR 5822, Villeurbanne, F-69100, France\label{aff55}
\and
Institute of Physics, Laboratory of Astrophysics, Ecole Polytechnique F\'ed\'erale de Lausanne (EPFL), Observatoire de Sauverny, 1290 Versoix, Switzerland\label{aff56}
\and
Institut de Ci\`{e}ncies del Cosmos (ICCUB), Universitat de Barcelona (IEEC-UB), Mart\'{i} i Franqu\`{e}s 1, 08028 Barcelona, Spain\label{aff57}
\and
Instituci\'o Catalana de Recerca i Estudis Avan\c{c}ats (ICREA), Passeig de Llu\'{\i}s Companys 23, 08010 Barcelona, Spain\label{aff58}
\and
UCB Lyon 1, CNRS/IN2P3, IUF, IP2I Lyon, 4 rue Enrico Fermi, 69622 Villeurbanne, France\label{aff59}
\and
Departamento de F\'isica, Faculdade de Ci\^encias, Universidade de Lisboa, Edif\'icio C8, Campo Grande, PT1749-016 Lisboa, Portugal\label{aff60}
\and
Instituto de Astrof\'isica e Ci\^encias do Espa\c{c}o, Faculdade de Ci\^encias, Universidade de Lisboa, Campo Grande, 1749-016 Lisboa, Portugal\label{aff61}
\and
Department of Astronomy, University of Geneva, ch. d'Ecogia 16, 1290 Versoix, Switzerland\label{aff62}
\and
INAF-Istituto di Astrofisica e Planetologia Spaziali, via del Fosso del Cavaliere, 100, 00100 Roma, Italy\label{aff63}
\and
Universit\'e Paris-Saclay, Universit\'e Paris Cit\'e, CEA, CNRS, AIM, 91191, Gif-sur-Yvette, France\label{aff64}
\and
Space Science Data Center, Italian Space Agency, via del Politecnico snc, 00133 Roma, Italy\label{aff65}
\and
FRACTAL S.L.N.E., calle Tulip\'an 2, Portal 13 1A, 28231, Las Rozas de Madrid, Spain\label{aff66}
\and
Max Planck Institute for Extraterrestrial Physics, Giessenbachstr. 1, 85748 Garching, Germany\label{aff67}
\and
Universit\"ats-Sternwarte M\"unchen, Fakult\"at f\"ur Physik, Ludwig-Maximilians-Universit\"at M\"unchen, Scheinerstrasse 1, 81679 M\"unchen, Germany\label{aff68}
\and
Institute of Theoretical Astrophysics, University of Oslo, P.O. Box 1029 Blindern, 0315 Oslo, Norway\label{aff69}
\and
Jet Propulsion Laboratory, California Institute of Technology, 4800 Oak Grove Drive, Pasadena, CA, 91109, USA\label{aff70}
\and
Felix Hormuth Engineering, Goethestr. 17, 69181 Leimen, Germany\label{aff71}
\and
Technical University of Denmark, Elektrovej 327, 2800 Kgs. Lyngby, Denmark\label{aff72}
\and
Cosmic Dawn Center (DAWN), Denmark\label{aff73}
\and
Universit\'e Paris-Saclay, CNRS/IN2P3, IJCLab, 91405 Orsay, France\label{aff74}
\and
Max-Planck-Institut f\"ur Astronomie, K\"onigstuhl 17, 69117 Heidelberg, Germany\label{aff75}
\and
NASA Goddard Space Flight Center, Greenbelt, MD 20771, USA\label{aff76}
\and
Department of Physics and Astronomy, University College London, Gower Street, London WC1E 6BT, UK\label{aff77}
\and
Department of Physics and Helsinki Institute of Physics, Gustaf H\"allstr\"omin katu 2, 00014 University of Helsinki, Finland\label{aff78}
\and
Aix-Marseille Universit\'e, CNRS/IN2P3, CPPM, Marseille, France\label{aff79}
\and
Universit\'e de Gen\`eve, D\'epartement de Physique Th\'eorique and Centre for Astroparticle Physics, 24 quai Ernest-Ansermet, CH-1211 Gen\`eve 4, Switzerland\label{aff80}
\and
NOVA optical infrared instrumentation group at ASTRON, Oude Hoogeveensedijk 4, 7991PD, Dwingeloo, The Netherlands\label{aff81}
\and
Universit\"at Bonn, Argelander-Institut f\"ur Astronomie, Auf dem H\"ugel 71, 53121 Bonn, Germany\label{aff82}
\and
INFN-Sezione di Roma, Piazzale Aldo Moro, 2 - c/o Dipartimento di Fisica, Edificio G. Marconi, 00185 Roma, Italy\label{aff83}
\and
Aix-Marseille Universit\'e, CNRS, CNES, LAM, Marseille, France\label{aff84}
\and
Dipartimento di Fisica e Astronomia "Augusto Righi" - Alma Mater Studiorum Universit\`a di Bologna, via Piero Gobetti 93/2, 40129 Bologna, Italy\label{aff85}
\and
Department of Physics, Institute for Computational Cosmology, Durham University, South Road, Durham, DH1 3LE, UK\label{aff86}
\and
Universit\'e Paris Cit\'e, CNRS, Astroparticule et Cosmologie, 75013 Paris, France\label{aff87}
\and
Institut de F\'{i}sica d'Altes Energies (IFAE), The Barcelona Institute of Science and Technology, Campus UAB, 08193 Bellaterra (Barcelona), Spain\label{aff88}
\and
School of Mathematics, Statistics and Physics, Newcastle University, Herschel Building, Newcastle-upon-Tyne, NE1 7RU, UK\label{aff89}
\and
DARK, Niels Bohr Institute, University of Copenhagen, Jagtvej 155, 2200 Copenhagen, Denmark\label{aff90}
\and
Centre National d'Etudes Spatiales -- Centre spatial de Toulouse, 18 avenue Edouard Belin, 31401 Toulouse Cedex 9, France\label{aff91}
\and
Institute of Space Science, Str. Atomistilor, nr. 409 M\u{a}gurele, Ilfov, 077125, Romania\label{aff92}
\and
Departamento de Astrof\'isica, Universidad de La Laguna, 38206, La Laguna, Tenerife, Spain\label{aff93}
\and
Consejo Superior de Investigaciones Cientificas, Calle Serrano 117, 28006 Madrid, Spain\label{aff94}
\and
Universit\"at Innsbruck, Institut f\"ur Astro- und Teilchenphysik, Technikerstr. 25/8, 6020 Innsbruck, Austria\label{aff95}
\and
Institut d'Estudis Espacials de Catalunya (IEEC),  Edifici RDIT, Campus UPC, 08860 Castelldefels, Barcelona, Spain\label{aff96}
\and
Satlantis, University Science Park, Sede Bld 48940, Leioa-Bilbao, Spain\label{aff97}
\and
Institute of Space Sciences (ICE, CSIC), Campus UAB, Carrer de Can Magrans, s/n, 08193 Barcelona, Spain\label{aff98}
\and
Instituto de Astrof\'isica e Ci\^encias do Espa\c{c}o, Faculdade de Ci\^encias, Universidade de Lisboa, Tapada da Ajuda, 1349-018 Lisboa, Portugal\label{aff99}
\and
Universidad Polit\'ecnica de Cartagena, Departamento de Electr\'onica y Tecnolog\'ia de Computadoras,  Plaza del Hospital 1, 30202 Cartagena, Spain\label{aff100}
\and
Kapteyn Astronomical Institute, University of Groningen, PO Box 800, 9700 AV Groningen, The Netherlands\label{aff101}
\and
Dipartimento di Fisica, Universit\`a degli studi di Genova, and INFN-Sezione di Genova, via Dodecaneso 33, 16146, Genova, Italy\label{aff102}
\and
Infrared Processing and Analysis Center, California Institute of Technology, Pasadena, CA 91125, USA\label{aff103}
\and
INAF, Istituto di Radioastronomia, Via Piero Gobetti 101, 40129 Bologna, Italy\label{aff104}
\and
ICL, Junia, Universit\'e Catholique de Lille, LITL, 59000 Lille, France\label{aff105}
\and
Department of Physics, Royal Holloway, University of London, TW20 0EX, UK\label{aff106}
\and
Mullard Space Science Laboratory, University College London, Holmbury St Mary, Dorking, Surrey RH5 6NT, UK\label{aff107}
\and
ICSC - Centro Nazionale di Ricerca in High Performance Computing, Big Data e Quantum Computing, Via Magnanelli 2, Bologna, Italy\label{aff108}}    

   \date{Received \textit{Month} \textit{Day} 2024; accepted \textit{Month} \textit{Day}, 2024}

  \abstract{
  A primary target of the \Euclid space mission is to constrain early-universe physics by searching for deviations from a primordial Gaussian random field. A significant detection of primordial non-Gaussianity would rule out the simplest models of cosmic inflation and transform our understanding of the origin of the Universe.
  This paper forecasts how well field-level inference of galaxy redshift surveys can constrain the amplitude of local primordial non-Gaussianity ($\fnll$), within a Bayesian hierarchical framework, in the upcoming \Euclid data. We design and simulate mock data sets and perform Markov chain Monte Carlo analyses using a full-field forward modelling approach.
  By including the formation history of the cosmic matter field in the analysis, the method takes into account all available probes of primordial non-Gaussianity, and goes beyond statistical summary estimators of $\fnll$. Probes include, for example, two-point and higher-order statistics, peculiar velocity fields, and scale-dependent galaxy biases. Furthermore, the method simultaneously handles systematic survey effects, such as selection effects, survey geometries, and galaxy biases.
  The forecast shows that the method can reach precision levels of up to $\sigma \left( \fnll \right) = 2.3$ (68.3\% confidence interval, and at the grid resolution $\Delta L = 62.5\,\Mpch$) with \Euclid data. We also provide data products, including realistic $N$-body simulations with nonzero values of $\fnll$ and maps of adiabatic curvature fluctuations.
  The results underscore the feasibility and advantages of field-level inference to constrain $\fnll$ in galaxy redshift surveys. Our approach consistently captures all the information available in the large-scale structure to constrain $\fnll$, and resolves the degeneracy between early-universe physics and late-time gravitational effects, while mitigating the impact of systematic and observational effects.
  }

   \keywords{large-scale structure of the Universe --
                cosmological parameters --
                initial conditions -- inflation -- Methods: data analysis -- Methods: statistical
               }
\titlerunning{Field-level inference of primordial non-Gaussianity}
\authorrunning{Andrews et al.}
   \maketitle

\textcolor{white}{-}
\newpage
\hfill
\section{Introduction}
One of the major tasks of modern cosmology is to understand the origin of the cosmic structure and the nature of the physical processes that governed the beginning of our Universe \citep{2004PhR...402..103B,biagetti_hunt_2019,achucarro_inflation_2022}. The current canonical mechanism, cosmic inflation, generates quantum fluctuations from one or more quantum fields. These fields drove an epoch of quasi-exponential cosmic expansion at the beginning of the Universe \citep{starobinsky_new_1980,guth_inflationary_1981}. Standard inflationary theory predicts primordial fluctuations as adiabatic, (almost) Gaussian, and (nearly) scale-invariant. Examples include single-field slow-roll models with quantum vacuum fluctuations as initial conditions, which induce tiny departures from Gaussianity \citep{1990PhRvD..42.3936S,Gangui_1993tt,Acquaviva_2002ud,2003JHEP...05..013M,Creminelli_2004yq,byrnes_scale-dependent_2010,Creminelli_2011rh,Baldauf_2011bh}. To test these predictions, ongoing and upcoming cosmological surveys aim to further constrain early-universe physics by searching for deviations from a primordial Gaussian random field \citep{lsst_science_collaboration_lsst_2009, dore_cosmology_2014, euclid_physics, euclid_collaboration_euclid_2020}. A significant detection of such a signal would radically transform our understanding of the early universe, as it would hint toward more complex inflationary models, involving multiple fields \citep[see e.g., ][]{2010AdAst2010E..72C,2010CQGra..27l4010K,alvarez_testing_2014,CORE_2016ymi,2018arXiv181208197C,meerburg_primordial_2019}.

The deviations from a primordial Gaussian random field are described by primordial non-Gaussianity (PNG). Potential sources of PNG include nonlinearity of gravity during inflation, inflaton self-interactions, and additional, yet unknown, light or heavy quantum fields, with various models predicting higher levels of PNG with respect to the standard single-field slow-roll models \citep[][]{1993ApJ...403L...1F, Gangui_1993tt,2003JHEP...05..013M,2004PhR...402..103B,2010AdAst2010E..72C,byrnes_review_2010,2015arXiv150308043A,meerburg_primordial_2019,2022JCAP...08..083C,2023arXiv231104882G}. To lowest order, local PNG is parameterised by the nonlinearity parameter $\fnll$ (see Eq.~\ref{eq:fnl_pert} for the definition). The perturbation of PNG induces a global rescaling of the primordial gravitational potential, leading to a multitude of effects and probes that can be used to measure $\fnll$ \citep{2000ApJ...542....1S,2001PhRvD..63f3002K,2001MNRAS.325..412V,2004PhRvD..69j3513S,2009astro2010S.158K,2009JCAP...02..017B,2010AdAst2010E..72C,biagetti_hunt_2019}. A subset of these probes has been used in observations of the cosmic microwave background \citep[CMB,][]{Planck_2013wtn,Planck:2013jfk,Ade:2015lrj,planck_collaboration_planck_2016,planck_2018,planck_collaboration_planck_2019_X,planck_collaboration_planck_2019_IX}, and the cosmic large-scale structures \citep[LSS,][]{castorina_redshift-weighted_2019,mueller2021clustering,damico_limits_2022,cabass_constraints_2022} to constrain PNG.
\begin{figure}[!]
	\begin{center}
    \includegraphics[width=1.0\columnwidth]{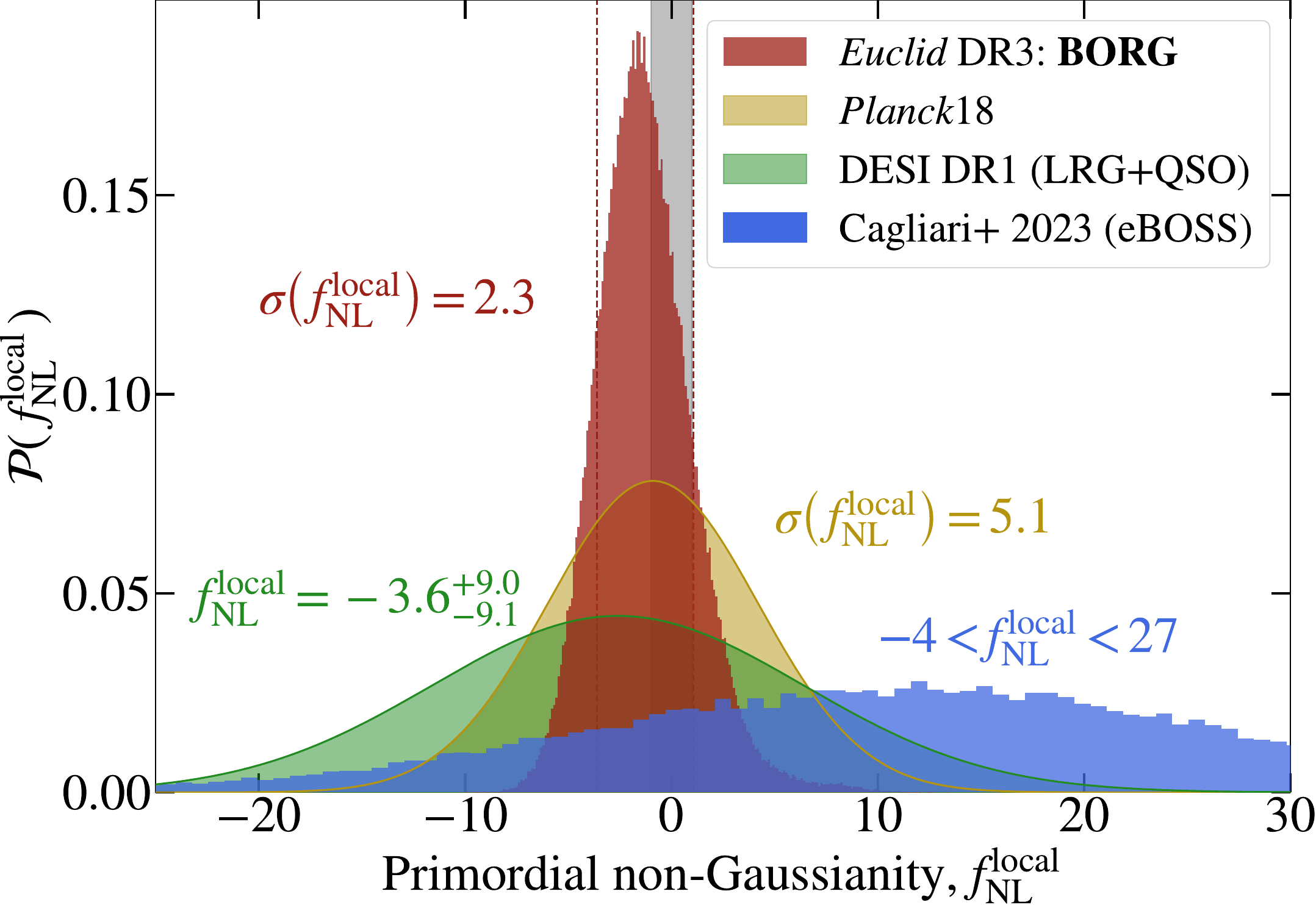}
    \end{center}
	\caption{The inference power of \borg{} on $\fnll$ in mock \euclid{} data, compared to the inferred value of {\it Planck} \citep{planck_collaboration_planck_2019_IX}\new{, and current state-of-the-art constrants from the large-scale structure} \citep{2023arXiv230915814C,2024arXiv241117623C}. The \borg{} results are based on the main results in this paper and indicate that field-level inference will be able to provide state-of-the-art constraints on $\fnll$ with \Euclid data.}
    \label{fig:money_plot}
\end{figure}
\begin{figure*}
	\centering
    \includegraphics[trim={0.75cm 17cm 3.5cm 0.75cm},clip,width=2\columnwidth]{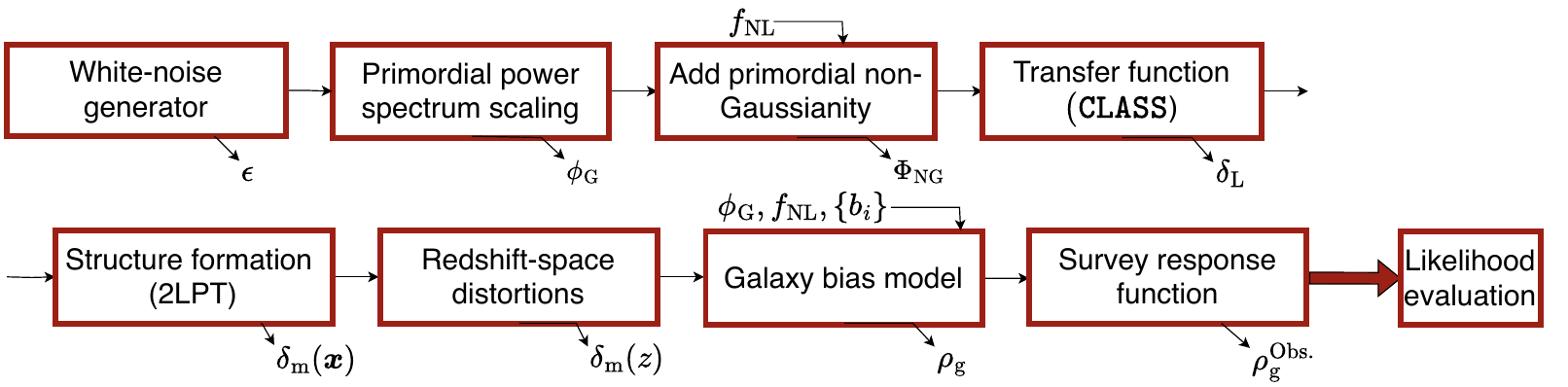}
	\caption{Flow chart illustrating the forward model implemented into \borg{} \citep{jasche_bayesian_2013,andrews_bayesian_2023}. The forward model connects a set of initial conditions to a model prediction. This output can then be compared to the data at the field level through a likelihood evaluation. The parameter under each box represents the output of the box and what is provided to the next step of the forward model. The parameters above some boxes represent the additional input of each computation, especially highlighting the inclusion of the $f_{\mathrm{NL}}^{\mathrm{local}}$ parameter.}
    \label{fig:flowchart}
\end{figure*}
 
To date, the tightest constraint on $\fnll$ (hereafter written as $\fnl$) has been obtained by CMB observations of the \textit{Planck} satellite, which finds $\fnl = -0.9 \pm 5.1$ at 68.3\% confidence interval \citep[CI,][]{planck_collaboration_planck_2019_IX}. Although the CMB is a powerful cosmological probe, its information content on PNG has been depleted, because large-scale temperature observations are expected to have reached the cosmic-variance limit\footnote{Some improvements may arise from small-scale polarisation measurements \citep{CMBPolStudyTeam_2008rgp,CMB_S4_2016ple,2020PhRvD.102b3521D,2021JCAP...04..050K}, but these are not expected to reach the science target of $\sigma(\fnl)=1$ \citep{meerburg_primordial_2019}.}.  In contrast, next-generation three-dimensional galaxy surveys can provide additional information by covering large cosmic volumes probing the largest scales of the cosmic matter distribution \citep{biagetti_hunt_2019,McQuinn_2021,achucarro_inflation_2022}.

Among the tightest constraints on local PNG from the LSS are the results of Dark Energy Spectroscopic Instrument (DESI), first data release, which finds $\fnl = -3.6^{+9.0}_{-9.1}$ \citep[with $p=1.6$ for the quasar sample][]{2024arXiv241117623C}. For quasars only, a well-studied data set is the SDSS--IV/eBOSS catalog \citep[DR 16, quasar sample][]{mueller2021clustering,2023arXiv230915814C}. For this data set, \citet{mueller2021clustering} find $|\fnl| < 21$ at 68.3\% CI (\new{$p=1.6$}), and \citet{2023arXiv230915814C} find $-4 < \fnl < +27$ at 68.3\% CI (\new{$p=1.0$}). Another example is \cite{castorina_redshift-weighted_2019}, who measured an earlier data release (DR14, quasar sample) to constrain $\fnl$ to  $-26< \fnl < +14$ at 68.3\% CI, (\new{$p=1.0$})\footnote{\new{For a discussion and definition of $p$, see Sect. \ref{sdb}}}.

On a side note, the two-dimensional CMB information can be cross-correlated with the three-dimensional cosmic LSS \citep{2014MNRAS.441L..16G,Euclid:2021qvm,mccarthy2022constraints,2023arXiv230507650K}. In this way, PNG can be measured without cosmic variance \citep{seljak_measuring_2009,Schmittfull_2018,ballardini_constraining_2019,barreira2023optimal,Karagiannis_2023, 2023arXiv230700058B}, with forecasts that demonstrate an improvement large enough to reduce the error to the aforementioned science goal \citep{M_nchmeyer_2019}.

So far, current LSS analyses have been based on statistical estimators that are sensitive to two- and three-point correlation functions. These estimators are thus sensitive to the large-scale effect on the power spectrum and the small-scale effect on the bispectrum. The scale-dependent bias effect depends on the initial bispectrum that, in the local model, has a large signal in squeezed configurations which correlates large scales with small scales responsible for halo formation \citep[][]{dalal_imprints_2008, 2008JCAP...04..014L,matarrese_effect_2008,Carbone_2008iz,Verde_2009hy,scoccimarro_large-scale_2012}. As a result, the scale-dependent bias effect is the most informative and crucial probe for measuring $\fnl$ with galaxy surveys \citep{2018MNRAS.474.2853U,mueller_optimising_2019, Karagiannis:2020dpq, Biagetti_2022, giri2023constraining,2023MNRAS.523..603R,2024arXiv240318789B,2024arXiv240317657P,2024arXiv240313985Y,2024arXiv240300490J}. However, it is heavily affected by large-scale contamination and systematic survey effects \citep{leistedt_constraints_2014,Rezaie_2021,mueller2021clustering,rezaie2023local,2024arXiv240600191C}, which require the application of data cleaning techniques to provide unbiased measurements of $\fnl$. On the other hand, the bispectrum of galaxies is sensitive to small-scale effects from the imprint of PNG onto the matter field \citep[][]{baldauf_primordial_2011,karagiannis_constraining_2018,friedrich_primordial_2019,Goldstein_2022,2023arXiv231012959G,chen2024reconstructing}. However, these primordial perturbations are intertwined with the effects from late-time structure formation, meaning that the signal of interest is non-trivial to disentangle in the data. Interestingly, the bispectrum can decipher the different shapes of PNG, for example, the local, equilateral, and orthogonal modes. We refer to the literature for more information on these modes \citep[][]{Babich_2004gb,scoccimarro_large-scale_2012,regan_universal_2012,Planck_2013wtn,schmidt_imprint_2015,planck_collaboration_planck_2016,karagiannis_constraining_2018,planck_collaboration_planck_2019_IX}.

Ongoing and upcoming galaxy redshift surveys such as \euclid{} \citep{laureijs_euclid_2011,euclid_physics,euclid_collaboration_euclid_2020,scaramella_euclid,2023arXiv230917287B,2024arXiv240513491E} are designed with the aim, among others, of probing early-universe physics by providing an unprecedented amount of data. However, data from these missions are expected to be affected by the systematic and observational effects of the survey \citep{Graham_2018}. Consequently, strategies for mitigating and modelling effects such as instrumentation noise and astrophysical contamination are crucial for handling the largest scales. Neglecting to address these factors can introduce biases in the results, particularly the constraints of $\fnl$. \citep{huterer2013,leistedt_constraints_2014,ho2015,jasche2017,2019A&A...625A..64J}.

To solve the open issues mentioned above and go beyond statistical summary estimators, we apply a Bayesian field-level inference approach to constrain PNG. This method uses a forward modelling approach to connect the primordial matter fluctuations of the early universe with the three-dimensional galaxy distribution at the field level \citep{jasche_bayesian_2013}. In this way, we leverage the complete formation history of the Universe in a holistic manner to constrain $\fnl$ in the observed data. With this novel approach, we provide a complementary and independent method for optimally measuring deviations from Gaussian initial conditions. In fact, this method allows us, among other things, to simultaneously:
\begin{itemize}
    \item naturally incorporate all probes of PNG in the LSS;\footnote{Examples include scale-dependent bias, higher-order statistics in the density field, mass distribution of tracers, velocity field imprints, etc.}
    \item disentangle the late-time effects of nonlinearities from early-universe signals of PNG;
    \item marginalise unknown nuisance parameters and large-scale foreground contamination;\footnote{See for example \citet{2019A&A...624A.115P,Lavaux2019}}
    \item account for survey systematic effects, for example, survey geometry and instrumentation noise.
\end{itemize}
In \citet{andrews_bayesian_2023}, we developed a field-level inference method to measure $\fnl$ in galaxy surveys. We forecasted that for a simplified Stage IV survey, the method can achieve up to $\sigma(\fnl) = 5.7$, with $\sigma(\fnl)$ denoting the 68.3\% CI.

In this paper, we improve on our previous work in several aspects. We apply a more advanced forward model, which goes beyond the forward model used in \cite{andrews_bayesian_2023}. This includes a more realistic structure formation model and additional higher-order bias terms, to account for finer small-scale physics in galaxy formation. In addition, more realistic survey specifications of \euclid{} are incorporated, leading to more realistic survey aspects than in the previous work. This includes, for example, a more accurate survey mask, a radial selection function, and galaxy number counts. The combination of these modifications leads to generally improved forecasts of $\fnl$ inference of spectroscopic data from the \Euclid space telescope. Our preliminary results indicate that our method can constrain $\fnl$ to the order of $\sigma(\fnl) = 2.3$ (68.3\% CI) for a realistic \Euclid survey, as can be seen in Fig.~\ref{fig:money_plot}.

The paper is structured as follows. In Sect.~\ref{method}, we provide an overview of the method and the algorithmic design choices. Emphasis has been placed on the galaxy bias model employed, the scale-dependent bias model, and on how the adiabatic curvature fluctuation maps were generated. In Sect.~\ref{data}, we provide descriptions of the data generation setup and the data sets generated. Specifications for \Euclid mock data and the objectives of the \Euclid mission with respect to PNG are included here. We finalise the section with the list of tests included in this project. We present the results in Sect.~\ref{results}, which cover convergence results, results on $\fnl$, as well as test-specific results. Finally, we summarise in Sect.~\ref{discussion}, and discuss future work in Sect. \ref{future_work}.

\newpage
\section{Method}
\label{method}
The primary goal of this paper is to forecast how well field-level inference can jointly constrain $\fnl$ and cosmic initial conditions in spectroscopic \Euclid{} mock data. In this context, field-level inference uses the entirety of the 3D cosmic matter field and its formation history to optimally extract the available information from the data to constrain cosmology, such as primordial fluctuations or cosmological parameters \citep{jasche_bayesian_2013,leclercq_accuracy_2021,baumann2021power,2021JCAP...03..058N,2024arXiv240303220N}. This is achieved by constructing a data model that forward models an arbitrary set of initial conditions, so that the corresponding predicted galaxy field can be directly compared with the data at the field level \citep{jasche_bayesian_2010,jasche_bayesian_2013,2017JCAP...12..009S,kostic2021machinedriven,2022MNRAS.509.3194P,andrews_bayesian_2023,2023arXiv230404785P,2023arXiv230709504B,2024MNRAS.527.1244S}. To test the performance of field-level inference, we set up a series of mock data sets, all emulating the survey features of the \euclid{} mission, including selection effects, window function, and galaxy number counts. Then, we analyse the mock data sets by running Markov chain Monte Carlo (MCMC) analyses in a Bayesian hierarchical framework. The MCMC runs produce plausible values of the primordial matter fluctuations, $\fnl$, and marginalised nuisance parameters, all conditioned on the observed data. These outputs constitute the main results to make the said forecasts on $\fnl$ and the cosmic initial conditions.

In this section, we provide a more detailed description of the field-level inference algorithm used (Sect. \ref{method_borg}). Next we describe the applied PNG model (Sect. \ref{subsec:fnl}), and the galaxy bias model used (Sect. \ref{galaxy_bias}). In the following, we provide a brief description of the running of MCMC analyses. Finally, we outline the generation of inferred 3D maps of adiabatic curvature fluctuations (Sect. \ref{zeta_maps}), which is a new data product presented for the first time, to our knowledge, in this paper.

\subsection{Overview of \textup{\borg{}}}
\label{method_borg}

\begin{figure}
	\centering
    \includegraphics[width=1\columnwidth]{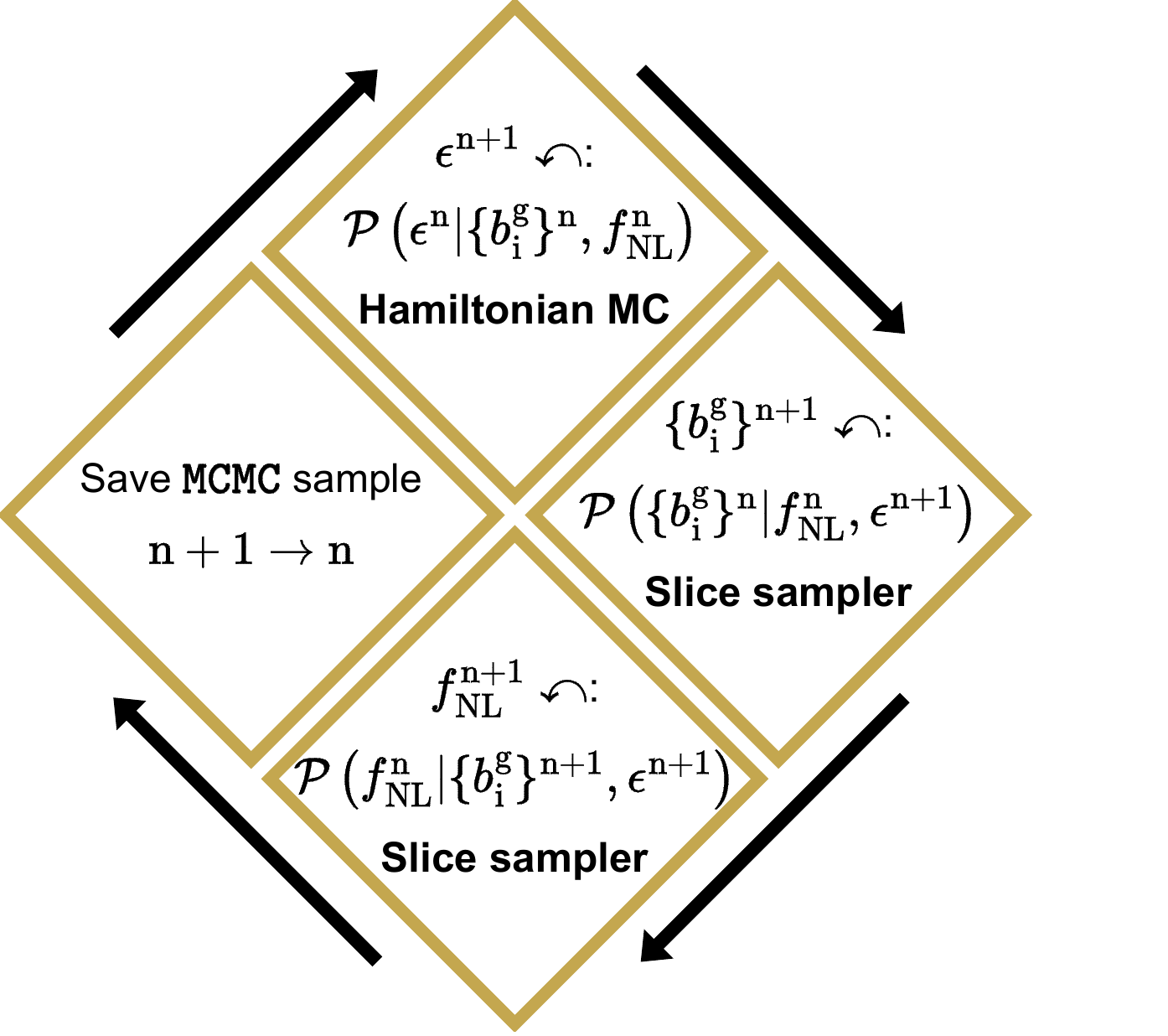}
	\caption{Flow chart depicting the sampling scheme in \borg{}. The sampling scheme can be divided into four major parts: Sampling the 3D initial conditions, sampling the galaxy bias parameters (one set for each tracer catalogue), sampling the $\fnl$ parameter, and finally saving the data products and restarting the cycle. Each sub-box depicts the conditional posterior from which the sample is drawn, and the sampling technique that is used.} 
	\label{fig:sample_scheme}
\end{figure}

The Bayesian Origin Reconstruction from Galaxies (\borg) algorithm is a Bayesian hierarchical inference framework and is designed for the analysis of cosmic structure in cosmological surveys through forward modelling of three-dimensional galaxy fields \citep[]{jasche_bayesian_2013,2015JCAP...01..036J,2016MNRAS.455.3169L,2019A&A...625A..64J,Lavaux2019}. The forward model in \borg{} aims to recreate the physical process in which the galaxies were formed and observed, as close as possible as the underlying physical process. In other words, the data model connects the three-dimensional primordial matter field to the observed distribution of galaxies, effectively reformulating the inverse problem of inferring the initial conditions into a statistical forward problem. In practice, we still have to simplify the model providing the local galaxy abundances through statistical mapping between the matter field and the galaxy distribution. Thus, the objective of \borg{} is, given the assumed forward model to explore the joint posterior distribution of initial conditions (denoted as $\epsilon$), cosmological parameters, and nuisance parameters, as constrained by the observed data \citep[]{jasche_bayesian_2013,andrews_bayesian_2023}. The problem can be formulated in the form of a joint posterior distribution $\mathcal{P}_{\rm{post}}\left(\epsilon, \fnl, \left\{b_i^\textrm{g} \right\}|N_{\textrm{g}}^\textrm{O}\right)$:
\begin{align}
&\mathcal{P}_{\rm{post}}\left(\epsilon, \fnl, \left\{b_i^\textrm{g} \right\}|N_{\textrm{g}}^\textrm{O}\right) \propto \nonumber \\ & \quad \quad \quad \quad \quad \mathcal{P}_{f}\left(\fnl\right) \, \mathcal{P}_{\epsilon}\left(\epsilon\right)\, \mathcal{P}_{b}\left\{b_i^\textrm{g} \right\} \, \mathcal{P}_{\rm{like}}\left(N_{\textrm{g}}^\textrm{O}|\epsilon, \fnl, \left\{b_i^\textrm{g} \right\}\right) \, , 
\label{eq:full_posterior}
\end{align}
where $N_{\textrm{g}}^\textrm{O}$ is the observed data in form of galaxy counts, $\left\{b_i^\textrm{g} \right\}$ are the bias parameters, and $\mathcal{P}_{f},\mathcal{P}_{\epsilon},\mathcal{P}_{b}$ are the prior distributions. The prior on the white-noise field $\mathcal{P}_{\epsilon}\left(\epsilon\right)$ is a Gaussian with zero mean and unit standard variance. The likelihood distribution $\mathcal{P}_{\rm{like}}\left(N_{\textrm{g}}^\textrm{O}| \epsilon, \fnl, \left\{b_i^\textrm{g} \right\}\right)$ is defined by the data model.

The data model forward-evolves a set of initial conditions to the corresponding predicted galaxy field, in the form of galaxy number counts. The forward model starts by simulating the gravitational progression of the matter field over time through a structure formation model. This process yields a predicted realisation of the late-time dark matter field, which is populated using a galaxy bias model \citep[Sect. \ref{galaxy_bias}, ][]{jasche_bayesian_2013,2015JCAP...01..036J,2016MNRAS.455.3169L,2019A&A...625A..64J,Lavaux2019}. The window function and survey selection effects are accounted for by evaluating the survey footprint and radial selection functions at each voxel. We note that while the sky map in this work consists of a binary selection function in the form of the survey footprint, \borg{} is able to account for more complex selection functions, e.g. those in the form of relative probabilities in each sky pixel \citep{Lavaux2019}. The resulting predicted galaxy field is compared to the observed galaxy field through a likelihood distribution. A schematic overview of the data model used in the \borg algorithm in this paper is given in Fig.~\ref{fig:flowchart}. One feature to be specifically pointed out in the data model is that the primordial perturbation with $\fnl$ enters the data model at a different step from the evaluation of the structure formation model. This allows the algorithm to break the degeneracy between gravitational nonlinearities and primordial signals \citep{baumann2021power,andrews_bayesian_2023}. To emphasise, a major advantage of relying on forward model analysis to infer PNG is that more information is available, beyond what is available for perturbation approaches \citep{leclercq_accuracy_2021,andrews_bayesian_2023,2024arXiv240303220N}.

To sample the joint posterior spanned by the data model, the algorithm relies on MCMC analysis. The complete multivariate distribution (Eq.~\ref{eq:full_posterior}) is handled using a Gibbs sampling approach. The sampling scheme consists of a mixture of Hamiltonian Monte Carlo \citep[HMC,][]{1987PhLB..195..216D,betancourt_2017} and slice sampling techniques \citep[]{hastings_1970,neal_slice_2003,2011hmcm.book..113N}. To reiterate, the joint posterior distribution includes the three-dimensional initial density fields, cosmological parameters (i.e., $\fnl$), and the galaxy bias parameters \new{for each galaxy catalogue (see Sect. \ref{euclid_specs})}. Exploring this joint posterior distribution allows \borg{} to effectively leverage all available information in the data for optimal parameter constraints, while also marginalising over nuisance parameters, given the data model and its resolution. Thus, through an iterative MCMC analysis, \borg{} performs a statistically rigorous analysis, allowing us to quantify the significance of inferred quantities \citep[]{jasche_bayesian_2013}. The sampling scheme is given in Fig.~\ref{fig:sample_scheme}.

For more details on the forward model and likelihood, we refer to the previous publication \citep{andrews_bayesian_2023}. For a description of the structure formation model, second-order Lagrangian perturbation theory (2LPT), we refer to similar work \citep{jasche_bayesian_2013,2015JCAP...01..036J,2016MNRAS.455.3169L,Lavaux2019, Tsaprazi_2022}.

\subsection{Model of local primordial non-Gaussianity $\fnl$}

\begin{figure}
	\centering
    \includegraphics[trim={0 2cm 0 0},clip,width=1\columnwidth]{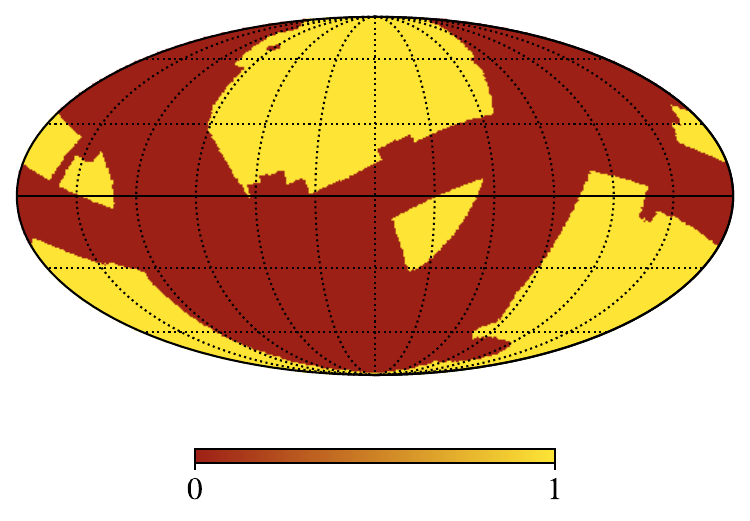}
	\caption{\new{\Euclid sky map.} This sky map illustrates the observed (yellow) and masked (red) regions for the \Euclid survey of this project. The survey mask is a result of the observation strategy of the \Euclid mission \citep{scaramella_euclid}. We point out that each tracer catalogue uses one single survey strategy but extends outwards at different redshifts (as illustrated in Fig.~\ref{fig:rad_sel}).}
	\label{fig:sky_map}
\end{figure}

\label{subsec:fnl}
Before applying the primordial perturbation with $\fnl$, we first convolve the Gaussian white noise field $\epsilon$ with the primordial transfer function $T_{\rm G}(k)$. The field $\epsilon$ is the set of initial conditions for the forward model shown in Fig.~\ref{fig:flowchart}. The transfer function $T_{\rm G}(k)$ scales the white noise field so that it has the properties of the primordial power spectrum. The choice of primordial power spectrum in this project follows the discrete case, and the resulting field $\phi_{\textrm{G}}$ has the following covariance:
\begin{align}
    \left\langle \hat{\phi}^{\phantom{*}}_{\textrm{g},\mvec{a}} \hat{\phi}^{*}_{\textrm{g},\mvec{b}} \right\rangle = V \, \delta^{\rm K}_{\mvec{a},-\mvec{b}} \, A_{\textrm{s}} \frac{2\pi^2}{k_\mvec{a}^3} \left(\frac{k_\mvec{a}}{k_\mathrm{pivot}}\right)^{n_s-1} \equiv \delta^{\rm K}_{\mvec{a},-\mvec{b}} \, T^2_{\rm G}\left(k_\mvec{a}\right),
    \label{eq:wn_to_phi}
\end{align}
with $V=L^3$ the volume of the data cube, $\mvec{a}$ and $\mvec{b}$ mesh indices, and $\delta^{\rm K}_{\mvec{a},\mvec{b}}$ the Kronecker delta. $A_{\textrm{s}}$ provides the amplitude of the post-inflationary gravitational potential in a radiation-dominated era, prior to the perturbation of $\fnl$. Thus, the gravitational potential $\phi_{\rm g}$, which is evaluated at $a\to0$, corresponds to the physical initial conditions of our forward model. It also contains 3D information on the primordial matter fluctuations. The primordial gravitational potential in real space is then calculated with an inverse Fourier transform, with which we compute the perturbed primordial gravitational potential $\Phi_{\rm NG}$. Generally, PNG is the deviation of the primordial gravitational potential from Gaussian statistics. To the lowest order, these deviations are parameterised by the quantity $\fnl$. Although the deviation can have different shapes \citep{karagiannis_constraining_2018}, in this paper we focus on forecasting the constraining power of the local form (or squeezed shape). We parameterise $\fnl$ through the Bardeen potential \citep{1990NuPhB.335..197H,Kofman_1991qx,1990PhRvD..42.3936S,Gangui_1993tt,2000MNRAS.313..141V,Wang_1999vf,2001PhRvD..63f3002K}
\begin{align}
    \Phi_{\rm NG} = \phi_{\rm g} + \fnl \, \left(\phi_{\rm g}^2 - \left\langle \phi_{\rm g}^2 \right\rangle \,\right) \, ,
    \label{eq:fnl_pert}
\end{align}
where $\phi_{\rm g}$ is a field described by Gaussian statistics, evaluated at the Lagrangian position. We consider $\fnl$ to be a constant parameter, independent of scale and any other parameter, including time. It is also the lowest order of PNG. We leave the incorporation of the higher order expansion of PNG for future implementations \citep{Jeong_2009,roth_can_2012,leistedt_constraints_2014}.

\subsection{Galaxy bias model}
\label{galaxy_bias}

\begin{figure}
	\centering
    \includegraphics[width=1\columnwidth]{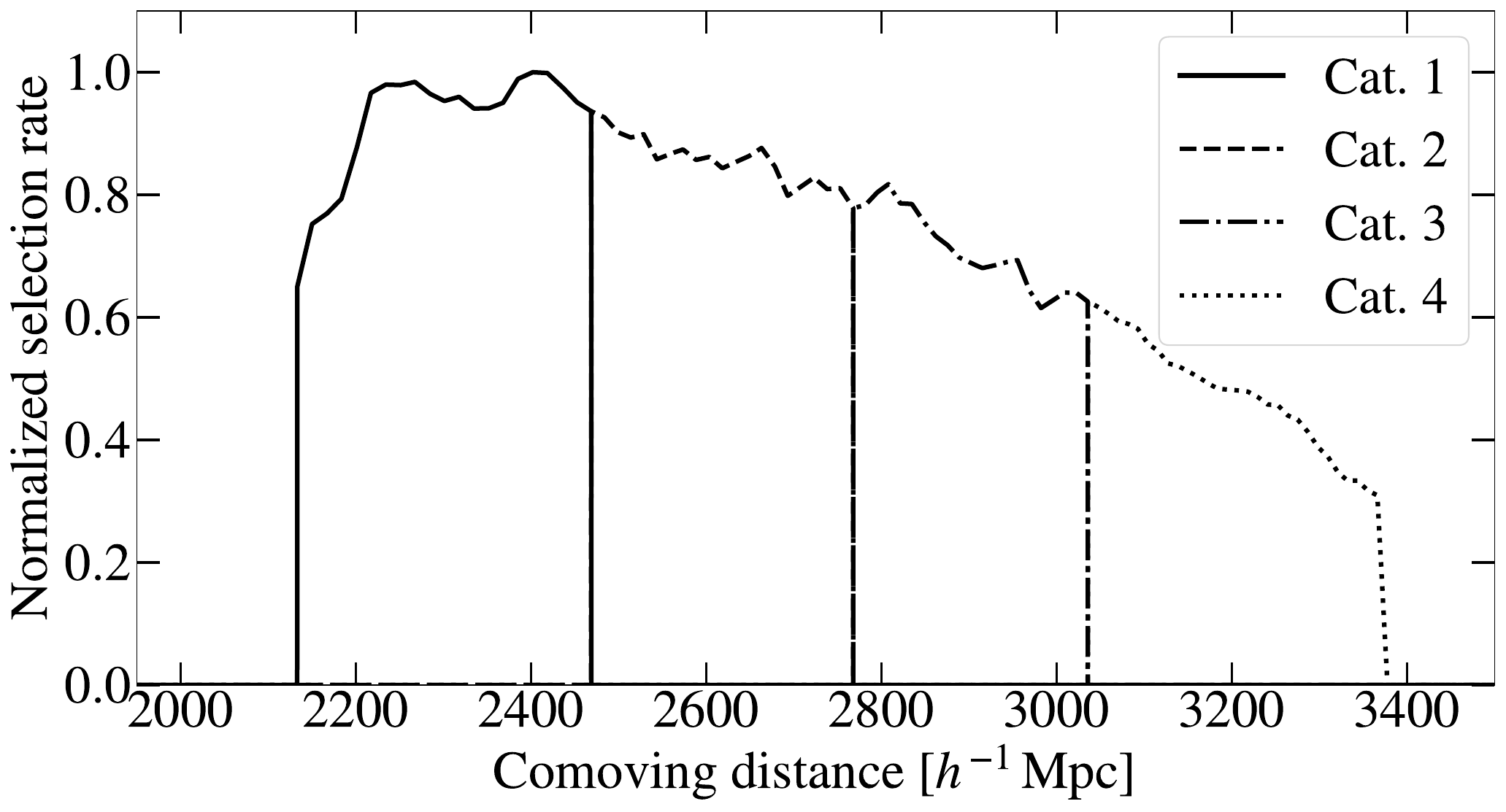}
	\caption{\new{\Euclid radial selection function.} This plot displays the normalised radial selection, $\mathrm{d}N(z)/(\mathrm{d} \Omega \, \mathrm{d}z)$, for the four galaxy catalogues in this project. Notice how the tracer catalogues do not overlap but rather cover separate regions in the mock universe.}
	\label{fig:rad_sel}
\end{figure}

In galaxy survey analysis, galaxy formation is typically described as a functional relationship of the dark matter field and bias parameters \citep{Assassi_2015,desjacques_large-scale_2018}. Specifically, we treat galaxies as `biased' tracers of the dark matter field, meaning that they share similar clustering statistics and properties. However, the true relationship is unknown and constitutes one of the most important unresolved problems in LSS cosmology \citep[for an exhaustive review, see][]{desjacques_large-scale_2018}. Thus, while noting that the forecasts are highly dependent on the estimated relationship, we assume a next-to-linear order bias model with scale-dependent bias components \citep{Assassi_2015,Barreira_2020b}. The motivation is that since we are still at relatively large scales ($>60\,\Mpch$), a relationship between the dark matter field and the galaxy field can be described as a linear function. By further including additional terms beyond the linear bias, we allow the model to account for nonlinear features as well.

The complete bias model for a given tracer population $g$ adopted in this paper is defined as
\begin{align}
\rho_{\rm g}(z, \vec{q}) = \left\langle N_{\rm g}^{\rm O} \, \right\rangle \, \Bigg[ 1 + b_1 \, \delta_{\rm m}(z)  +  \frac{b_2}{2} \delta_{\rm m}^2(z) + b_{K} \, K^2(z) \nonumber \\ \quad \quad + \, b_\phi \, \fnl \, \phi_{\rm g}(\vec{q}) + \bpd \, \fnl \, \delta_{\rm m}(z) \, \phi_{\rm g} (\vec{q})  \Bigg] \, ,
\end{align}
where $\left\langle N_{\rm g}^{\rm O} \, \right\rangle$ is the mean number of observed galaxies, $b_1$ is the linear bias, $b_2$ is the second-order bias, $b_{K}$ is the tidal field bias, and $K^2(z)$=$\rm{tr}$$\left [ K^2_{ij} (z) \right ]$, where $K_{ij}(z) \equiv \left ( \partial_i \partial_j/\nabla^2 - \delta^{\rm K}_{i,j}/3 \right ) \delta_{\rm m}(z)$ is the long-wavelength tidal field \citep{2021JCAP...10..063L,2021JCAP...08..029B}.\footnote{$K_{ij}$ is computed in Fourier space, where the differential operators correspond to multiplications by components of the wavevector $k$.} Scale-dependent bias terms $b_\phi$ and $\bpd$ are defined and further discussed in Sect.~\ref{sdb}. We note that $\phi(\vec{q})$ is evaluated in Lagrangian space, while $\delta_{\rm m}(z)$ is evaluated in a space that includes redshift-space distortions, which we call redshift space. In this project, we analyse the data at constant redshifts and thus do not account for effects that arise due to observing the galaxies on their light cones. We leave the inclusion of light-cone effects when inferring PNG to a future publication.

We assume a fixed noise level for the Gaussian distribution of galaxies, set to $\sigma_{\rm g}^2 = \left\langle N_{\rm g}^{\rm O} \right\rangle$ \citep{andrews_bayesian_2023}. In an upcoming publication, we will assess both the noise-level assumption and the likelihood model for describing the distribution of galaxies at the voxel level. Additionally, we will explore how the constraints on $\fnl$ depend on these choices. The ground truth values of the galaxy parameters (used to generate the mock data) can be found in Table~\ref{tab:euclid_specs}.

We mention that there is also the possibility of using effective field theory (EFT) to model the galaxy bias formalism and likelihood in a field-level inference approach \citep{2019JCAP...01..042S,2020JCAP...11..008S,2021JCAP...04..032S,2022JCAP...08..007B,2023arXiv231003741T,2023JCAP...10..069S,2024arXiv240701524B,2024arXiv240910937S,2024arXiv241104513S}. For a review of the galaxy bias problem, the interested reader is referred to the literature \citep{desjacques_large-scale_2018}.

\subsubsection{Scale-dependent bias model}
\label{sdb}

In the model of local PNG considered, the primordial perturbation gives rise to a scale-dependent imprint on the biased tracer populations. This is due to the coupling of short- and long-wavelength modes in a nonzero $\fnl$ universe \citep{dalal_imprints_2008,Slosar_2008,matarrese_effect_2008,Carbone_2008iz,Verde_2009hy}. Thus, the PNG adds a scale-dependent contribution to the galaxy bias relation between galaxies and the underlying primordial gravitational field, which scales as $\propto k^{-2}$ \citep{dalal_imprints_2008,Slosar_2008,matarrese_effect_2008,Carbone_2008iz,Verde_2009hy}, which is the most prominent on the largest scales. In this paper, we adopt the universal mass function \citep{barreira_local_2021,Lucie_Smith_2023,2023arXiv231110088F,2023arXiv231212405A} for a non-zero $\fnl$ universe. This allows us to model the bias parameter of the primordial gravitational potential $b_{\phi}$ as a function of the linear bias $b_1$
\begin{align}
b_{\phi} = 2 \, \delta_{\rm c} \, \left(b_1 - p\right) \, ,
\label{eq:b_phi}
\end{align}
with $\delta_{\rm c} = 1.686$ being the spherical critical overdensity in an Einstein--de Sitter universe \citep{Percival_2005vm}, and $p$ a tracer-dependent parameter. For this forecast, we fix $p$ to $0.55$ or $1$ for each tracer population \citep{Barreira_2020,cabass_constraints_2022}. Additionally, we incorporate the bias for the cross-field term $\bpd$, adopting the parameterisation of \citet{barreira_predictions_2021,cabass_constraints_2022}\footnote{\new{We note that alternative parameterisations of $\bpd$ exist in the literature, for example \citet{Moradinezhad_Dizgah_2021} and \citet{damico_limits_2022}.}},
\begin{align}
\bpd = b_{\phi} - b_1 + 1 + \delta_{\rm c} \, \left [b_2 - \frac{8}{21} \, \left(b_1 - 1\right) \right ] \, .
\label{eq:b_phidelta}
\end{align}
The problem of accurately modelling $b_\phi$ and $\bpd$ remains an unresolved challenge within the cosmological community \citep{biagetti_hunt_2019, Barreira_2022, achucarro_inflation_2022, barreira2023optimal,sullivan2023learning,2023arXiv231212405A,2023arXiv231110088F,2024arXiv240701391D,2024arXiv241018039S,Kvasiuk_2024gbz}. Improving the precision of the models for these bias parameters is crucial for robust and unbiased inference of $\fnl$ \citep{Moradinezhad_Dizgah_2021,barreira_local_2021,Lazeyras_2023}. Further model-driven investigations of the treatment of $b_\phi$ and $\bpd$ will result in more robust and comprehensive models that better capture the underlying physical processes. In this paper, we adopt the universal mass approximation as a practical choice \citep{Barreira_2020b,barreira_local_2021}. However, we emphasise that this assumption is not fundamental to the method itself, but as advances are achieved in the modelling of $b_\phi$ and $\bpd$, the forward model will be revised accordingly. For a more in-depth discussion of the problem, see \cite{Moradinezhad_Dizgah_2021} and \cite{barreira_local_2021}.

\subsection{Running the MCMC analysis}
\label{mcmc}

\begin{table}[]
\caption{The detailed specifications of the spectroscopic \Euclid mock data used in this paper. These specifications form the basis for the four tracer catalogues generated for all of the runs in this paper. The galaxy bias parameters used in this galaxy bias model are the linear bias ($b_1$), second-order bias ($b_2$), and tidal field bias ($b_{K}$). Scale-dependent bias parameters $b_\phi$ and $\bpd$ have values derived based on Eqs.~\eqref{eq:b_phi} and ~\eqref{eq:b_phidelta}, with $p=0.55$. The values of the galaxy bias parameters are based on table 1 in \cite{yanky}, and the number densities are based on table 2 in \cite{euclid_collaboration_euclid_2020}. The galaxy density, $n_{\mathrm{gal}}$, is given in units of $10^{-4} \, \left(h^3\,\text{Mpc}^{-3}\right)$.}
\begin{tabular}{llllllll}
\hline
 Cat. & $b_1$ & \hspace{-0.05cm}$b_2$  & \hspace{-0.05cm}$b_{K}$ & \hspace{-0.05cm}${b_\phi}^*$ & \hspace{-0.05cm}${\bpd}^*$ & \hspace{-0.05cm}$z_{\mathrm{min}}$/$z_{\mathrm{max}}$ & \hspace{-0.05cm}$n_{\mathrm{gal}}$ \\ \hline
1         & \hspace{-0.05cm}$1.30$  &  \hspace{-0.05cm}$-0.74$   & \hspace{-0.05cm}$-0.17$  & \hspace{-0.05cm}$2.53$ & \hspace{-0.05cm}$0.79$ &    \hspace{-0.05cm}$0.9$/$1.1$    & \hspace{-0.05cm}$6.86$       \\
2         & \hspace{-0.05cm}$1.38$  &  \hspace{-0.05cm}$-0.70$   & \hspace{-0.05cm}$-0.22$  & \hspace{-0.05cm}$2.80$ & \hspace{-0.05cm}$0.99$ &    \hspace{-0.05cm}$1.1$/$1.3$    & \hspace{-0.05cm}$5.58$       \\
3         & \hspace{-0.05cm}$1.46$  &  \hspace{-0.05cm}$-0.66$   & \hspace{-0.05cm}$-0.26$  & \hspace{-0.05cm}$3.13$ & \hspace{-0.05cm}$1.23$ &    \hspace{-0.05cm}$1.3$/$1.5$    & \hspace{-0.05cm}$4.21$       \\
4         & \hspace{-0.05cm}$1.54$  &  \hspace{-0.05cm}$-0.60$   & \hspace{-0.05cm}$-0.31$  & \hspace{-0.05cm}$3.33$ & \hspace{-0.05cm}$1.44$ &    \hspace{-0.05cm}$1.5$/$1.8$    & \hspace{-0.05cm}$2.61$       \\ \hline
\end{tabular}
{\tiny * \textit{Derived values}}
\label{tab:euclid_specs}
\end{table}

The \borg{} algorithm performs a large-scale MCMC to explore the joint posterior distribution of Eq.~\eqref{eq:full_posterior}, given the mock data sets. We briefly touch on the details of the MCMC analysis performed in this paper. We follow the prescription as in \citet{ramanah_cosmological_2019} and \citet{andrews_bayesian_2023}, which provide more details.

To ensure that the sampler can sample from the target posterior distribution, we initialise the initial conditions $\epsilon$ at a randomly chosen point, set at one-tenth of the overall amplitude. The bias parameters are initialised at nine-tenths of their ground truth values and $\fnl$ is shifted by $+5$. The prior distribution on $\fnl$ is a Gaussian distribution with $\mu_{\fnl}=0,\,\sigma(\fnl)=100$. This design is motivated by the choice to have a broad and non-informative prior.

We start the burn-in phase of the MCMC runs by exclusively sampling the initial conditions $\epsilon$ with $\fnl$ and the bias parameters kept constant to their starting values. This first step continues until the amplitude of the initial conditions fluctuates around the prior expectation. This is monitored by the power spectra estimated from Markov samples. For an example of such a plot, see the results of similar work \citep[e.g.,][]{2019A&A...630A.151P,ramanah_cosmological_2019,2022MNRAS.509.3194P,andrews_bayesian_2023}. Next, we continue the run with sampling bias parameters, \new{where each catalogue is assigned its own set,} which are allowed to converge and stabilise. Finally, we include the sampling of $\fnl$. To ensure the inclusion of only post-burn-in samples, additional $5000$ samples are generated before including MCMC samples in the analysis. At this point, \borg{} explores the parameter space of plausible large-scale structure realisations, spanned by $\epsilon$, $\fnl$, and the bias parameters.

The MCMC chains are run until convergence, as determined by the Gelman--Rubin statistic $\hat{R}$ \citep{gr_test_1992}. The Gelman--Rubin statistic is calculated by dividing the sequential samples of the chains into $M$ distinct sets, each with an equal number of samples, where $M$ typically ranges from 4 to 8. The sets are separated by $N$ discarded samples, where $N$ denotes the number of samples required to decorrelate $\fnl$, to ensure statistical independence. More details on the autocorrelation length of $\fnl$ for each run can be found in Appendix \ref{app_tests_of_algo}. When the threshold $|\hat{R}| \leq 1.05$ is reached, the chain is considered to have converged.

It should be noted that a proper Gelman--Rubin test assumes independent MCMC chains. In our case, a single MCMC chain has been split into several chains for the evaluation of the Gelman--Rubin test, due to limitations in computational resources. By doing this, we acknowledge the risks associated with this, for example, the sampler getting stuck in a local minimum or biasing our results. However, from investigating the convergence in the other diagnostic results (e.g., correlation lengths, corner plots, and estimates on uncertainty of uncertainty), we deem these risks to be negligible.

\subsection{Generating 3D maps of adiabatic curvature fluctuations}
\label{zeta_maps}

Our field-level inference method, in addition to providing measurements of $\fnl$, infers the primordial gravitational potential. For a given inference of the primordial gravitational potential, we can generate a 3D map of the adiabatic curvature fluctuations, or $\mathcal{R}$ maps in short \citep{planck_collaboration_planck_2016}. From Eqs.~\eqref{eq:wn_to_phi} and \eqref{eq:fnl_pert}, we compute $\Phi_{\rm NG}(\vec{q})$, and then relate this to the corresponding $\mathcal{R}$ field\footnote{This expression comes from the exact formula $\mathcal{R}_k = -[(5+3w) / (3+3w)] \Phi_k$, evaluated at super-Hubble scales for a radiation-dominated universe in Fourier space, with $w$ being the equation of state of the dominant energy form.}
\begin{align}
\mathcal{R}(\vec{q}) = -\frac{3}{2} \Phi_{\rm NG}\left(\vec{q}\right)   \, ,
\label{eq:acf_computation}
\end{align}
where $\vec{q}$ represents the vector in Lagrangian space \citep{2001PhRvD..63f3002K,2002PhRvD..66f3008O,2002PhLB..524....5L,sasaki_2006}. Since \borg runs an MCMC analysis to sample the posterior distribution of $\Phi_{\rm NG}(\vec{q})$, we can also estimate the corresponding uncertainties over the chain of samples. The resulting maps are provided in Sect.~\ref{zeta_results}, together with a more detailed prescription and the choice of data set. For a 2D reconstruction of the adiabatic curvature fluctuation, see section 6.3 in \citet{planck_collaboration_planck_2016}.

\section{Data and data generation}
\label{data}
\subsection{The Euclid Wide Survey}

The Euclid Wide Survey has among its goals to probe the expansion history and evolution of our Universe \citep{scaramella_euclid_2014, scaramella_euclid}. To perform this task, \euclid{} will observe a region of $15\,000$ square degrees, over a redshift range of $0.9<z<1.8$. Over the next six years, it will observe up to 30 million spectroscopic redshift galaxies with high precision ($\sigma_z \approx 0.001$), which can be used for galaxy clustering studies. 

Among several cosmological measurements, inferring PNG is one of the primary objectives of the \Euclid mission. More specifically, we focus solely on the local shape of PNG (Eq.~\ref{eq:fnl_pert}). The earliest goals were set at $\sigma(\fnl) \approx 2$ (68.3\% CI). This forecast is based on statistical information from two-point correlation functions measured by spectroscopic redshift galaxies and for a ground truth value of $f_{\rm NL}^{\mathrm{gt}} = 0$ \citep{laureijs_euclid_2011}. However, more recent Fisher forecasts estimate that $\sigma(\fnl)$ around $4$ to $5$ (68.3\% CI) is achievable, given the accuracy of the spectroscopic redshift measurements and updated galaxy counts \citep{euclid_physics}. These results include marginalisation over the galaxy bias parameters, nuisance parameters, and other cosmological parameters \citep{giannantonio_constraining_2012, euclid_physics}. We acknowledge that the constraining power of the forecasts is highly dependent on the measured linear galaxy bias $b_1$ and the relationship $b_\phi(b_1)$, due to the scale-dependent bias effect (as discussed in Sect.~\ref{sdb}). That being said, the constraints on $\fnl \times b_\phi$ will be largely insensitive to this uncertainty \citep{Barreira_2022}. \new{Unless unspecified, we use the value of $p=0.55$}.

\subsection{\euclid{} specifications}
\label{euclid_specs}

\begin{table}[]
\caption{An overview of the runs included in this project. The runs are designed to fulfil the tests outlined in Sect.~\ref{run_overview}. Run \#3 is the main run of the paper (bolded in the table), which uses the most realistic settings for a future \Euclid study. The computed resolutions and values of $k_{\rm max}$ are given in $\Mpch$ and $\hMpc$, respectively. The ground truth value used to generate the mock data is denoted as $f_{\rm NL}^{\mathrm{gt}}$. For the resulting inferred $\fnl$ values with uncertainties, see Table~\ref{tab:fnl_table}.}
\begin{tabular}{lllll}
\hline
Run \# & Resolution  & $k_{\rm max}$     & $f_{\rm NL}^{\mathrm{gt}}$ & Note \\ \hline
1 & 250\phantom{.0} & 0.025 & 0 & Low resolution    \\
2 & 125\phantom{.0} & 0.05\phantom{0} & 0 & Medium resolution \\
\textbf{3} & \textbf{62.5} & \textbf{0.1\phantom{0}\phantom{0}} & \textbf{0}                               & \textbf{High resolution} \\
4 & 62.5 & 0.1\phantom{0}\phantom{0} & 5                               & Different $f_{\mathrm{NL}}^{\mathrm{gt}}$    \\
5 & 125\phantom{.0} & 0.05\phantom{0} & 0 & Fixed bias \\
6 & 125\phantom{.0} & 0.05\phantom{0} & 0 & $p=1$ \\
7 & 125\phantom{.0} & 0.05\phantom{0} & 0 & Sample $b_{\phi}, \bpd$   \\ \hline
\end{tabular}
\label{tab:overview}
\end{table}

We base the mock data sets on the forecast specifications of \Euclid, meaning that we make use of survey features of the \Euclid mission to generate the mock data. Examples include sky completeness coverage, radial selection effects, bias parameter values, and galaxy counts. 

In Fig.~\ref{fig:sky_map}, we illustrate the sky completeness map used. We emphasise that the same completeness map is used for all four galaxy catalogues, and has a total sky coverage of roughly $15\,000$ square degrees. Also, while the sky completeness map used in this forecast is the survey geometry footprint of the \Euclid{} survey \citep{scaramella_euclid}, \borg{} has the ability to use more complex sky completeness masks in its data model \citep{2019A&A...625A..64J, Lavaux2019,andrews_bayesian_2023}.

The radial selection functions are plotted in Fig.~\ref{fig:rad_sel}. The observed number counts are split into four different galaxy populations, each population corresponding to a catalogue. This design choice is based on the redshift binning of the \euclid{} forecasts \citep{euclid_collaboration_euclid_2020}. Here, it can be seen that each subsequent catalogue reaches further into the observable universe, covering a nonoverlapping redshift range $0.9< z < 1.8$. The number counts decrease towards higher redshifts \citep[]{euclid_collaboration_euclid_2020}. 

The values of the galaxy bias parameters and number densities are listed in Table~\ref{tab:euclid_specs}. These bias values have been chosen as in table 1 of \citet{yanky}, which constitutes the current estimate of the bias values as a function of redshift. The number densities are based on table 3 of \citet{euclid_collaboration_euclid_2020}. The values for the scale-dependent bias parameters are derived with Eqs.~\eqref{eq:b_phi} and ~\eqref{eq:b_phidelta}.

\subsection{Mock data generation}

\begin{figure*}
	\centering
    \includegraphics[width=2\columnwidth]{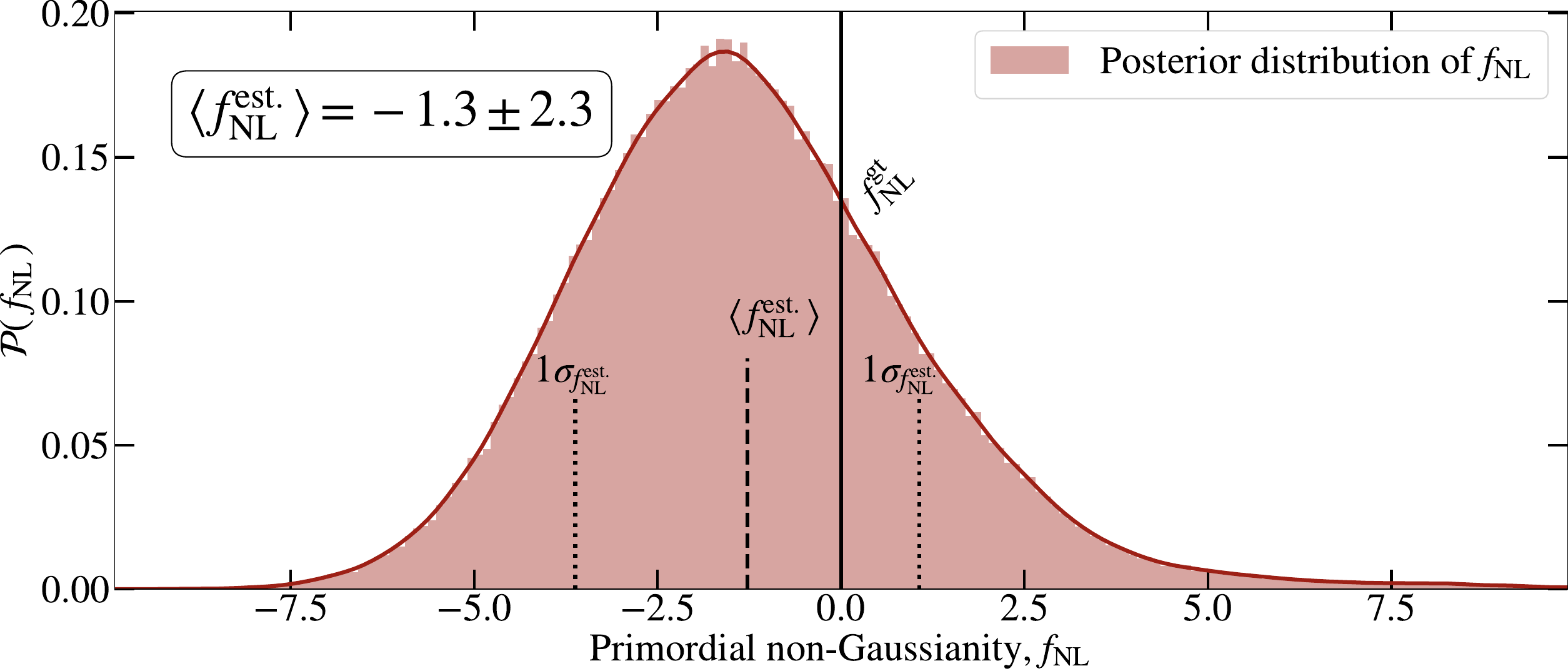}
	\caption{Field-level results for inferring $\fnl$ for a high-resolution run (Run \#3). The figure illustrates that the method can find a unimodal marginalised distribution of $\fnl$ that best explains the data, with the ground truth value $f_{\rm NL}^{\mathrm{gt}}$ (solid line) within the 68.3\% CI of the estimated value. The value in the box is the maximum of the distribution in the range of 68.3\%. We point out that these results are based on the universal mass function, which means that the scale-dependent bias parameters are fixed to the expressions in Eq.~\eqref{eq:b_phi} and Eq.~\eqref{eq:b_phidelta}.}
	\label{fig:trace_pdf_N128}
\end{figure*} 

To generate mock data, we follow similar procedures as described in previous works \citep[]{jasche_fast_2010, jasche_bayesian_2013,ramanah_cosmological_2019,andrews_bayesian_2023}. In general, these mock data sets are generated by running the forward model on a set of randomly drawn initial conditions. \new{Thus, the same forward model is used for both mock data generation and the inference process, maintaining no model mismatch.} We provide the details below, in a step-by-step description.

\begin{enumerate}
    \item The evaluation of the physics forward model was prepared for a cubic Cartesian box of side length $L=8000\,\Mpch$ and $N_{\rm grid}=32$, $64$, or $128$, yielding grid resolutions in the range of $\Delta L = 250\,\Mpch$, $\Delta L = 125\,\Mpch$, and $\Delta L = 62.5\,\Mpch$, respectively.
    \item  A random three-dimensional field $\epsilon$, with zero mean and unit standard deviation, was generated. Given this white-noise field, a primordial density field was computed by applying the primordial power spectrum (Eq.~\ref{eq:wn_to_phi}), perturbing it with the $\fnl$ parameter (Eq.~\ref{eq:fnl_pert}), and then apply the cosmological transfer function as provided by \texttt{CLASS} \citep{Lesgourgues_2014}. This produced the linear matter field $\delta_{\rm L}$, which was the starting point for the gravitational structure formation model.
    \item  To reduce the sample variance of the particle distribution, we oversampled the initial density by a factor of 2 per dimension, resulting in a total number of $(2N)^3$ simulation particles. Particles were then evolved to the present epoch, using 2LPT, and were assigned to a three-dimensional Cartesian grid via the Cloud-In-Cell (CIC) kernel to yield the present-day three-dimensional density field $\delta_{\rm m}$. In addition, redshift-space distortions were added to transform the particles from the rest frame to the redshift frame.  
    \item  To emulate a biased galaxy distribution, we applied a next-to-linear order, scale-dependent galaxy bias (as described in Sect.~\ref{galaxy_bias}) to the forward-modelled density field. The output is the galaxy field, wherein the galaxy counts in each voxel are characterised by a Gaussian distribution. Detailed specifications are organised in Table~\ref{tab:euclid_specs}, including the parameter choices for the galaxy bias model.  
    \item  Finally, the radial selection functions and the survey geometry were applied to the simulated galaxy field to emulate the observational effects of the survey.
\end{enumerate}

We use the set of best-fit cosmological parameters ($\Omega_{\mathrm{m}}=0.3153$, $\Omega_{\Lambda}=0.6847 $, $\Omega_{\mathrm{b}}=0.0493$, $h=0.6736$, $A_{\rm s}=2.1\times10^{-9}$, $n_{\rm s}=0.9649$) from \textit{Planck} \citep[][]{planck_2018} to calculate the cosmological power spectrum and transfer functions. A summary of the specifications, including detailed parameter choices for the galaxy bias model, is provided in Table~\ref{tab:euclid_specs}. We provide a rendering of one of the mock data sets in Appendix \ref{appendix_cosmic_fields}. We note that by relying on 2LPT to model structure formation we forego including small-scale physics that capture information at higher-order correlation functions beyond the bispectrum. To accurately and completely capture these higher-order effects at small-scales, we will in future publications rely on $N$-body solvers, e.g. tCOLA \citep{tassev_solving_2013}, particle mesh \citep{2019A&A...625A..64J}, or field-level emulators \citep{2023arXiv231209271D}. \new{Since the data analysis in this paper is self-consistent with the mock data generation -- using the same forward model for both inference and mock data creation -- it captures all the information contained in the mock data sets.}

\subsection{Overview of runs}
\label{run_overview}

In this section, we provide a complete list of the runs that we performed for this paper. Although the main aim of this paper is to forecast $\fnl$ measurements with mock \Euclid galaxy surveys, we are also interested in investigating the performance of the method in various configurations. To achieve this, we vary the specifications in the data and the analysis (e.g., resolution, marginalisation, and parameters values). The main questions for each test, together with the design of the runs, are outlined in the following list. For a concise overview of the runs in this project, we refer to Table~\ref{tab:overview}.

\begin{enumerate}
    \item \textbf{Resolution study} \newline 
    The imprint of PNG affects the full cosmic matter field, both at large and small scales. To test the method's constraining power as a function of scale, we set up three different runs. The first at coarser grid resolution $\big($$\Delta L = 250\,\Mpch$, Run \#1$\big)$, one at medium grid resolution $\big($$\Delta L = 125\,\Mpch$, Run \#2$\big)$, and one at finer grid resolution $\big($$\Delta L = 62.5\,\Mpch$, Run \#3$\big)$. By increasing the voxel resolution, we allow the algorithm to have more degrees of freedom in describing the 3D cosmic matter field. Therefore, we expect that the method can use more small-scale information in the LSS to constrain $\fnl$. We also perform these runs as a benchmark as we adjust other parameters, for example, changing the sampling scheme or parameter values.
    
    \item \textbf{$\bm{f_{\rm NL}=0}$, $\bm{f_{\rm NL}=5}$} \newline 
    We aim to investigate whether the algorithm's constraining power depends on the fiducial amplitude of PNG. To test this, we generated two identical mock data sets that differ only by their values of $\fnl$: one with $\fnl=0$ (Run \#3) and the other with $\fnl=5$ (Run \#4), both with finer grid resolutions. In this way, we test the algorithm roughly in the upper 68.3\% CI of the \textit{Planck} 2018 measurement and the other in null detection \citep[]{planck_collaboration_planck_2019_IX}. This also allows us to check if the algorithm can accurately retrieve a nonzero ground truth value of $f_{\rm NL}^{\mathrm{gt}}$.
    \item \textbf{Idealised} \newline
    We aim to examine how marginalising the galaxy bias parameters affects the constraint on $\fnl$. To do this, we analyse the same mock data as in the medium resolution case, but this time with the bias parameters fixed to their ground truth values (Run \#5). This setup represents an ideal scenario where we have complete and perfect knowledge of the galaxy bias relationship. By analysing mock data in this idealised case, we investigate the nuances of parameter interactions and their implications for $\fnl$ constraints. For a list of ground truth values, see Table~\ref{tab:euclid_specs}. 
    
    \item \textbf{$p = 1$} \newline
    We aim to investigate how much the algorithm's constraining power depends on the amplitude of the scale-dependent bias effect. To test this, we generate and analyse mock data that have $p$ (Eq.~\ref{eq:b_phi}) set to $1$ (Run \#6) instead of $0.55$ (which is the default for the other runs). Since this value of $p$ is larger, there is a weaker scale-dependent bias amplitude in this analysis, meaning that the expectation is that the inferred uncertainty in $\fnl$ will be larger.

    \item \textbf{Sampling $b_{\phi}, \, \bpd$} \newline
    As described in Sect.~$\ref{sdb}$, PNG gives rise to a scale-dependent bias effect that is modelled by $b_{\phi}$ and $\bpd$. In the other runs, we fix $b_{\phi}$ and $\bpd$ to the expressions of the universal mass function (Eqs.~\ref{eq:b_phi} and ~\ref{eq:b_phidelta}). In this test, we want to test whether the algorithm is able to jointly sample $\fnl$, $b_{\phi}$ and $\bpd$, in the presence of priors. Therefore, in Run \#7, we include the sampling of $b_{\phi}$ and $\bpd$, in addition to the initial conditions $\epsilon$, $\fnl$, and the other bias parameters. In the mock data, the ground truth values of $b_{\phi}$ and $\bpd$ are set to the universal expressions of the mass function. \replace{The priors for $b_{\phi}$ and $\bpd$ are Gaussians centred on universal mass function expressions (as a function of $b_1$ and $b_2$). The standard deviation of these priors is $\sigma(b_\phi) = \sigma(\bpd) = 0.4$.}{The priors for $b_{\phi}$ and $\bpd$ are Gaussians centred on universal mass function expressions (as a function of $b_1$ and $b_2$), with standard deviations set to 40\% of their values: $\mathcal{P}_{\phi}\left(b_\phi\right) = \mathcal{G}\left(b^{\rm UMA}_{\phi},0.4b^{\rm UMA}_{\phi}\right)$, and $\mathcal{P}_{\phi\delta}\left(\bpd\right) = \mathcal{G}\left(b_{\phi,\delta}^{\rm UMA},0.4b_{\phi,\delta}^{\rm UMA}\right)$, where $b^{\rm UMA}_{\phi}$ and $b_{\phi,\delta}^{\rm UMA}$ are the expressions in Eqs.~\eqref{eq:b_phi} and ~\eqref{eq:b_phidelta}. We note that $b_1$ and $b_2$ used to evaluate $b^{\rm UMA}_{\phi}$ and $b_{\phi,\delta}^{\rm UMA}$ correspond to the values in the current state of the MCMC chain, rather than the ground-truth values of $b^{\rm UMA}_{\phi}$ and $b_{\phi,\delta}^{\rm UMA}$.} We are mainly interested in evaluating the performance of the method, in terms of $\sigma(\fnl)$, and investigating possible correlations between bias parameters $b_{\phi}$ and $\bpd$, and $\fnl$. If successful, this run further highlights the flexibility of the field-level inference approach in adjusting the galaxy bias model, and to sample bias parameters arising due to primordial effects.
\end{enumerate}

\section{Results}
\label{results}

\begin{figure*}
	\centering
    \includegraphics[width=2.0\columnwidth]{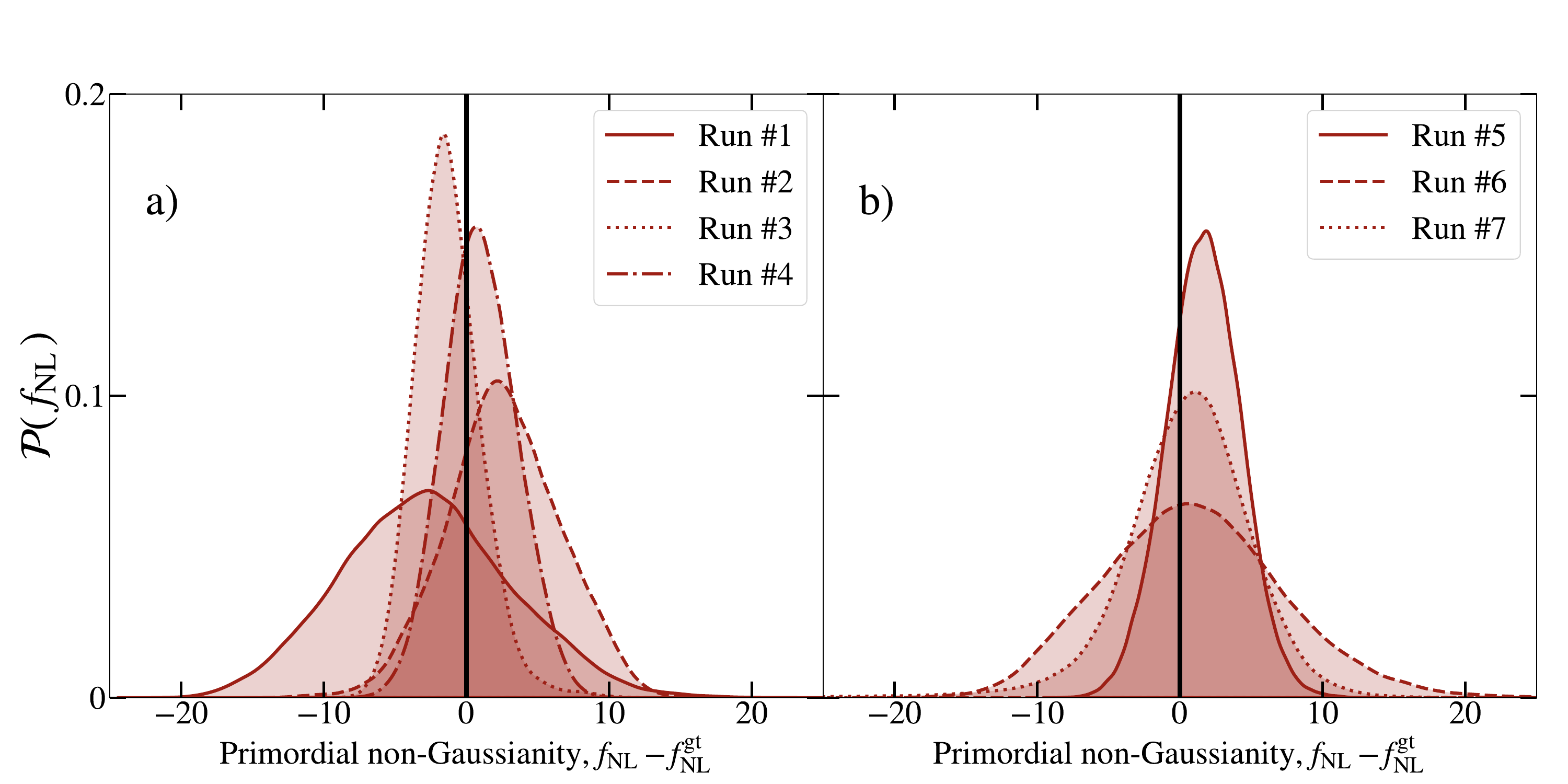}
	\caption{Field-level results for inferring $\fnl$ for all runs. The first four runs, Runs \#1-\#4, are included in panel a). The last three runs, Runs \#5-\#7, are included in panel b). The marginal distribution of Run \#4 has been shifted to be relative to the ground truth $f_{\rm{NL}}^{\rm{gt}}=0$.}
	\label{fig:all_pdfs}
\end{figure*}

The main goal of this work is to infer the marginal posterior distribution of $\fnl$ for each generated mock data set. In this way, we provide both forecasts of how well \borg{} can constrain $\fnl$ with the Euclid Wide Survey, and test the inference power under a variety of configurations. The results have been summarised in Table~\ref{tab:fnl_table}. For each run, we obtain an ensemble of samples, each containing plausible values of $\fnl$, the initial conditions, and the nuisance parameters, given the mock data. From these ensembles, we calculate the ensemble mean and uncertainty of $\fnl$ and include these in the table. We also compute the uncertainty of the uncertainty, which quantifies the margin of error. For completeness, we also include the ground truth $f_{\mathrm{NL}}^{\mathrm{gt}}$ values, the resolution, and the corresponding $k_{\rm max}$. We highlight that each inferred ensemble mean $\left\langle \fnl \right\rangle$ is within the 68.3\% CI of the ground truth $f_{\rm NL}^{\mathrm{gt}}$ value.

\begin{table}[]
\caption{The inferred $\fnl$ for each run. The table illustrates how different exchanges in the physics model and setup affect the constraint power in the data to infer $\fnl$. Examples include an increase in resolution or a change of the structure formation model. We note that all inferred values of $\fnl$ are within the 68.3\% CI of the ground truth $f_{\rm NL}^{\mathrm{gt}}$ values. This suggests that the method can reliably infer $\fnl$ from the data. We have also included the uncertainty of the uncertainty estimates, denoted as $\mathrm{std}\left[\sigma(\fnl)\right]$, using the batch means method \citep{fishman_1997}. As a reminder, for Runs \#1--\#6 we assume the universal mass function, while for Run \#7 we sample the scale-dependent bias parameters (with a prior centred as in Eq.~\ref{eq:b_phi} and Eq.~\ref{eq:b_phidelta}). The computed resolutions and values of $k_{\rm max}$ are given in $\Mpch$ and $\hMpc$, respectively.}
\begin{tabular}{lllllll}
\hline
Run & $f_{\mathrm{NL}}^{\mathrm{gt}}$     & $\left\langle \fnl \right\rangle$  & $\sigma(\fnl)$ & $\mathrm{std}\left[\sigma(\fnl)\right]$ & $k_{\rm max}$     &  Resol. \\  \hline
1         & 0 & $-2.8$  & 6.0 & 0.12 & 0.025 &  250\phantom{.}\phantom{0}  \\
2         & 0 & $\phantom{-}2.6$  & 4.0 & 0.54 & 0.05\phantom{0} &  125\phantom{.0}  \\
3         & 0 & $-1.3$  & 2.3 & 0.23 & 0.1\phantom{0}\phantom{0} &  \phantom{0}62.5 \\
4         & 5 & $\phantom{-}6.0$  & 2.5 & 0.22 & 0.1\phantom{0}\phantom{0} & \phantom{0}62.5  \\
5         & 0 & $\phantom{-}1.7$  & 2.6 & 0.07 & 0.05\phantom{0} &  125\phantom{.0}  \\
6         & 0 & $\phantom{-}0.9$  & 6.3 & 0.82 & 0.05\phantom{0} &  125\phantom{.0}  \\
7         & 0 & $\phantom{-}0.4$  & 5.0 & 0.59 & 0.05\phantom{0} &  125\phantom{.0}  \\ \hline
\end{tabular}
\label{tab:fnl_table}
\end{table}

The main run of this project, Run \#3, constitutes the most realistic \euclid{} mock data. With it, we infer $\fnl$ at a voxel resolution of $62.5 \, \Mpch$ $\left(k_{\rm max} = 0.1 \, \hMpc\right)$. We illustrate the inferred posterior distribution in Fig.~\ref{fig:trace_pdf_N128}, with which we find $\fnl = -1.3 \pm 2.3$. This marginal posterior distribution contains the entirety of the information available in the data, given the physics model and resolution. However, we emphasise that this distribution function is not a pure Gaussian, as seen, for example, by the elongated tails. To make this point clear, we compute the skewness and excess kurtosis (with Fisher's definition), which we find to be 0.6 and 1.3, respectively. This is an example of the flexibility of the algorithm, as it is able to infer non-Gaussian posterior distributions.

Furthermore, for a visual comparison between the runs, we include Fig. \ref{fig:all_pdfs}, with all the inferred marginal posterior distributions of $\fnl$. To maintain legibility, we have divided the seven posteriors into two subplots.

\subsection{Further tests}

Below, we present the results for the tests outlined in Sect. \ref{run_overview}. In Sect. \ref{zeta_results}, we describe and present the results of generating maps of adiabatic curvature fluctuations.

\subsubsection{Resolution study}

We discuss the results of the resolution investigation, in which we test the constraining power of \borg{} as a function of the available small scales. This is tested by comparing Runs \#1--3. As can be seen, by increasing the resolution from $32^3$ to $64^3$, we improve the results by roughly 33\%. Furthermore, by increasing the resolution from $64^3$ to $128^3$, we improve the constraining power by approximately $30$\%. We note that we are still at the mildly linear regime ($k_{\rm max} = 0.1 \, \hMpc$), but we use a forward model that can account for nonlinearities in the data. This means that we can, in principle, further increase the resolution while still being able to model the emerging nonlinear small-scale physics. However, due to the computational cost, we leave this investigation to a future project.

\subsubsection{$\fnl=0$, $\fnl=5$}

We outline the results for the performance of the algorithm for a nonzero ground truth value of $f_{\rm NL}^{\mathrm{gt}}$. We generate two different mock data sets with the same white noise and galaxy bias parameters but with different $\fnl$ values. These two are Run \#3 $\left(f_{\mathrm{NL}}^{\mathrm{gt}} = 0\right)$ and Run \#4 $\left(f_{\mathrm{NL}}^{\mathrm{gt}} = 5\right)$. The results show that the constraints for the two runs are similar, which indicates that the algorithm's performance is largely independent of the underlying $\fnl$ value. Moreover, the uncertainty of the uncertainty $\mathrm{std}\left[\sigma(\fnl)\right]$ is similar for the two runs, providing additional confirmation of this. The results indicate that our algorithm does not have a strong bias or performance issue related to ground truth $f_{\rm NL}^{\mathrm{gt}}$, which means that we expect the same sensitivity for a null signal or a primordial signal. We leave the exploration of sensitivity to large ground truth values of $f_{\rm NL}^{\mathrm{gt}}$ to a future project, which previous work has shown can influence the estimated uncertainty \citep{Creminelli_2006gc,Liguori_2007sj}.

\subsubsection{Effect of fixing bias parameters}

We compare the performance of two runs, where we marginalise over galaxy bias parameters (Run \#2) for one and keep them fixed in the other (Run \#5), given the same mock data set, \new{at medium grid resolution $\Delta L = 125\,\Mpch$}. The results show that marginalising galaxy bias parameters increases the constraints on $\sigma(\fnl)$ from $2.6$ to $4.0$. This indicates that in cases where we can forego marginalising over galaxy bias parameters, for example, when the galaxy biasing relationship is known, we can expect improvements up to roughly $35$\%.

\subsubsection{Amplitude of scale-dependent bias effect}

We discuss the results of weakening the effect of the scale-dependent bias in the model, by changing the parameter $p$. We perform two runs at the same resolution, but with differing values of $p$ from $0.55$ (Run \#2) to $1$ (Run \#6). The results show that by increasing $p$ we reduce the constraining power in $\fnl$. In fact, the change in $p$ from $0.55$ to $1$ results in a decrease in the constraints of roughly 40\%. This outcome highlights the dependence of the algorithm forecast on changes in scale-dependent bias parameters.

\begin{figure*}
\centering
\includegraphics[width=1.9\columnwidth]{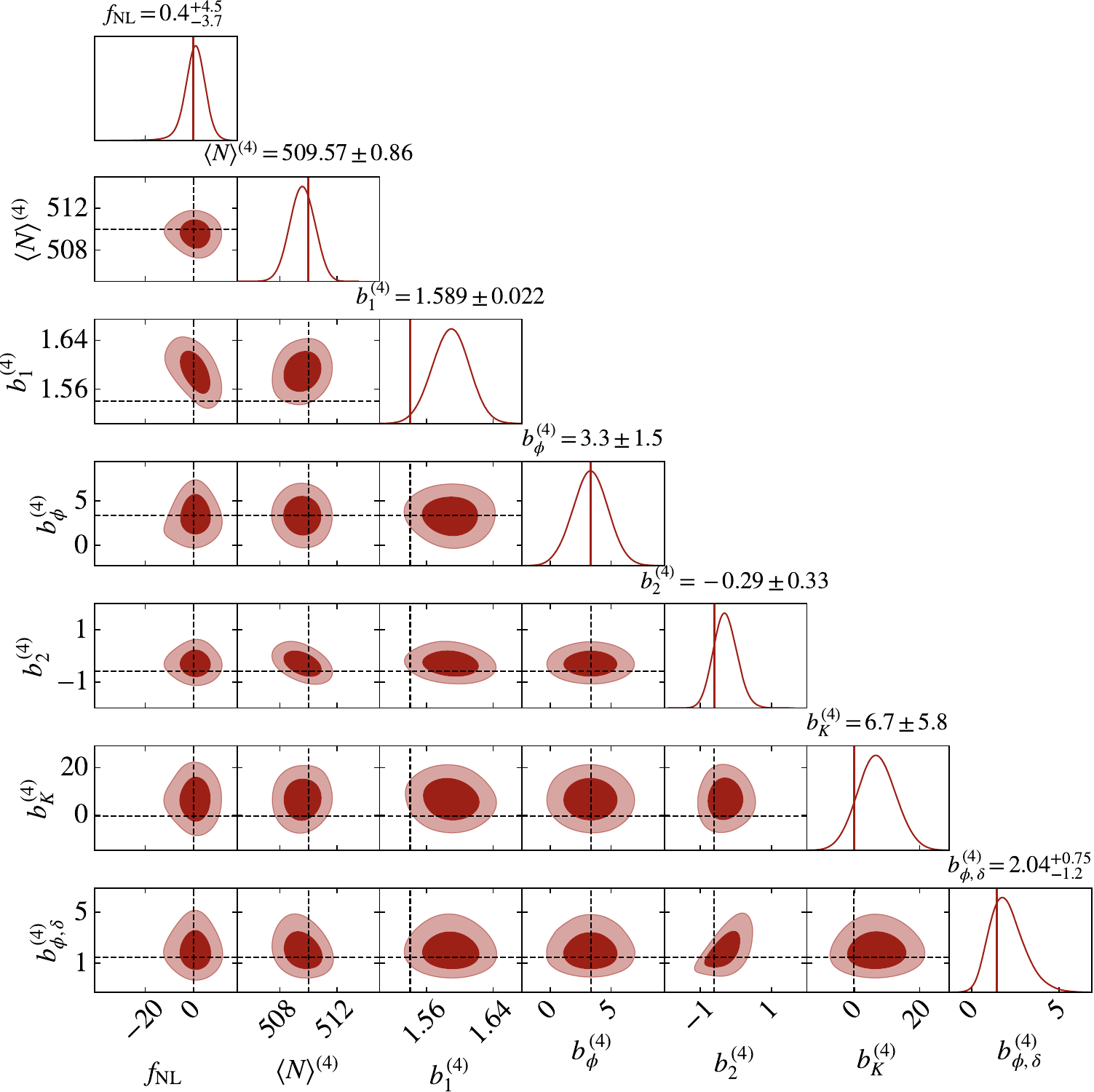}
\caption{Field-level results for inferring $\fnl$, when simultaneously also sampling $b_{\phi}$ and $\bpd$. Corner plot for $\fnl$ and bias parameters, for Run \#7, catalogue 4. Although the priors (described in Sect.~\ref{run_overview}) keep the inferred value of $\fnl$ centering around the expected value of $0$, the possible degeneracies with $b_\phi$ and $\bpd$ are still explored. Thus, while the results indicate that our field-level inference method can jointly sample $\fnl$ together with $b_\phi$ and $\bpd$ (in the presence of priors), more work is needed to stabilise the region of explored $\fnl$ values. \replace{The priors on $b_{\phi}$ and $\bpd$ is centered around their Gaussians centred on universal mass function expressions, with standard deviations $\sigma(b_\phi) = \sigma(\bpd) = 0.4$}{The priors on $b_{\phi}$ and $\bpd$ is centered around their Gaussians centred on universal mass function expressions, with standard deviations at 40\% of that value: $\mathcal{P}_{\phi}\left(b_\phi\right) = \mathcal{G}\left(b^{\rm UMF}_{\phi},0.4b^{\rm UMF}_{\phi}\right)$, and $\mathcal{P}_{\phi\delta}\left(\bpd\right) = \mathcal{G}\left(b_{\phi,\delta}^{\rm UMF},0.4b_{\phi,\delta}^{\rm UMF}\right)$, where $b^{\rm UMF}_{\phi}$ and $b_{\phi,\delta}^{\rm UMF}$ are the expressions in Eqs.~\eqref{eq:b_phi} and ~\eqref{eq:b_phidelta}}. The corner plots for the other catalogues and for Run \#3 can be found in Appendix \ref{app_tests_of_algo}.}
\label{fig:pyramid_6_4}
\end{figure*}

\begin{figure*}[h!]
	\centering
    \includegraphics[width=2.\columnwidth]{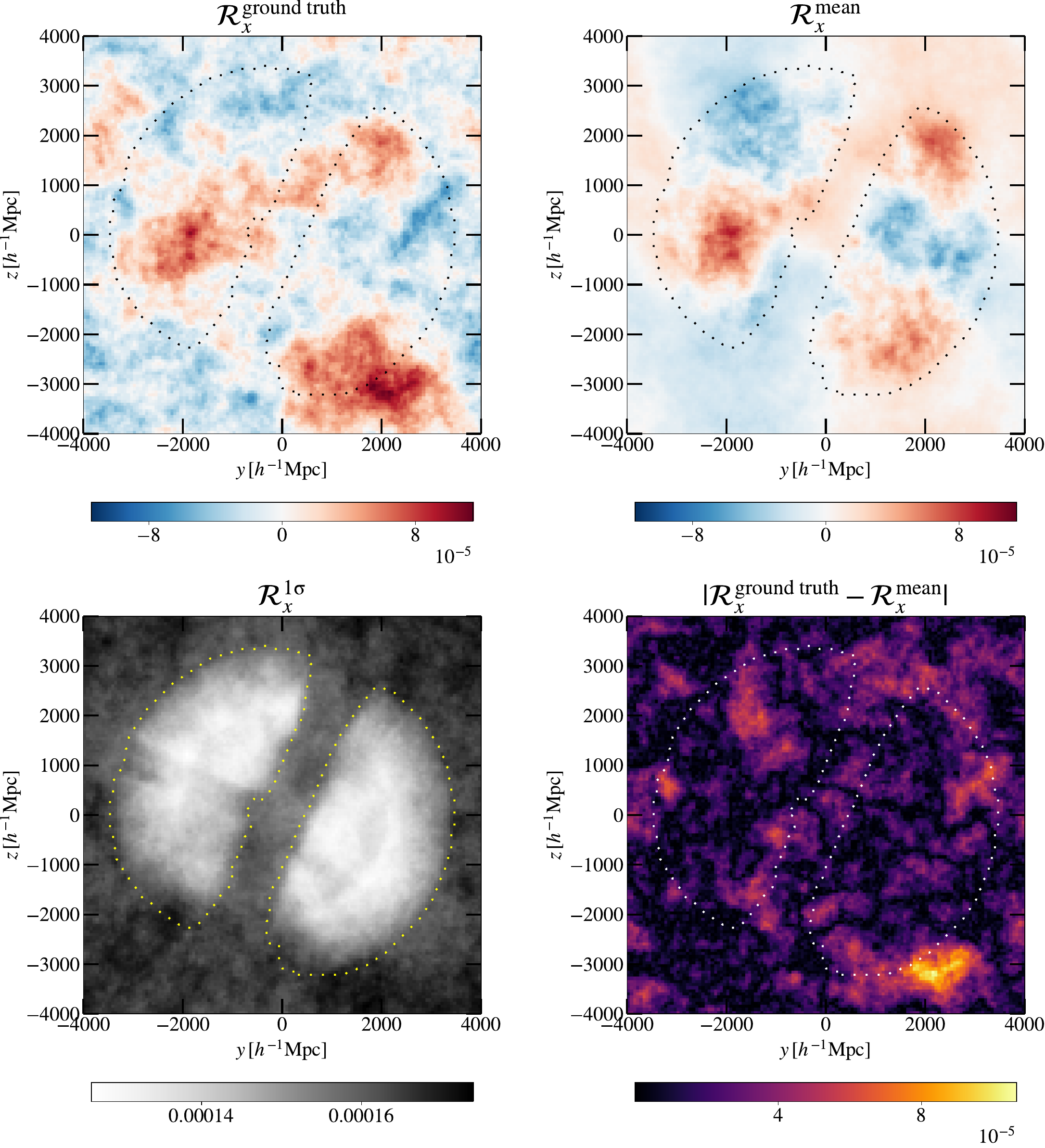}
	\caption{
	Illustrations of the inferred adiabatic curvature fluctuations. For each saved sample, we have a set of initial conditions $\epsilon$ that produce a plausible set of model predictions, constrained by the data. Each set of initial conditions corresponds to a field of adiabatic curvature fluctuations (Eqs.~\ref{eq:wn_to_phi} and ~\ref{eq:acf_computation}), which are the input to the structure formation model. By computing these fluctuations for a subset of the chain, the method can provide an expected estimate of the fluctuation of the adiabatic curvature along with uncertainty. We highlight the edge of the survey window with the dotted lines, meaning that voxels outside of the inner regions in the final observed field contain no observations. In the top left plot, we have included the ground truth field of adiabatic curvature fluctuations used to generate the mock data, averaged over the $x$ direction. In the top right plot, we have the expectation value of the adiabatic curvature fluctuations averaged over the $x$ direction. In the bottom left plot, we have the corresponding uncertainty averaged over the $x$ direction. Lastly, in the bottom right plot, we include the absolute difference between the mean inferred field and the ground truth field.}
	\label{fig:acf_map}
\end{figure*}

\subsubsection{Sampling of $b_\phi$ and $\bpd$}
With this investigation, we want to test the flexibility of \borg{} to include the joint sampling of the scale-dependent bias parameters, including priors. This is relevant since it is still not fully understood how to model the scale-dependent bias effect \citep{barreira2023optimal}. 
We present the results of Run \#7, in which we include the sampling of $b_\phi$ and $\bpd$, in addition to the initial conditions, $\fnl$, and the other parameters of the galaxy bias. The results show that \borg{} is still able to provide unbiased constraints on $\fnl$, but that the constraints degrade by approximately 20\%. In Fig.~\ref{fig:pyramid_6_4}, we provide a visualisation of the corner plot of the sampled $\fnl$ and galaxy bias parameters of the fourth catalogue in the Markov chain. The corner plots for the other catalogues and the full correlation matrix are presented and discussed in Appendix \ref{app_tests_of_algo}. Although the results show promise, further developments \new{and choice of priors} \citep[for example the ones presented in][]{2023arXiv231110088F,2023arXiv231212405A} will be investigated in a future publication. 

\subsection{Maps of adiabatic fluctuations}
\label{zeta_results}

In addition to measuring $\fnl$, our field-level inference method infers the initial 3D conditions of the data. With these sets of plausible initial conditions, we generate maps of the 3D adiabatic curvature fluctuations of the post-inflationary universe. To generate maps of adiabatic curvature fluctuations, we apply Eq.~\eqref{eq:acf_computation} to a subset of the inferred samples of Run \#3. The subset consists of every tenth sample from the Markov chain, starting from the first sample after the burn-in phase has concluded. By analysing the posterior ensemble, we can compute the average $\mathcal{R}$ field and the corresponding variance. We mention that we also include the inferred values of $\fnl$ in the evaluation of the perturbed primordial gravitational potential $\Phi_{\rm NG}$. These estimated statistical 3D fields constitute a novel data product that the method enables.

In Fig.~\ref{fig:acf_map}, we render the 2D projections (in the $x$ direction) of the resulting 3D maps. We also include the ground truth $\mathcal{R}$ field of the mock data and the absolute residual between the inferred and ground truth field. The upper-left panel illustrates the ground truth map of $\mathcal{R}$ provided by the mock data. The upper-right panel illustrates the average inferred $\mathcal{R}$ values. The bottom-left panel illustrates the uncertainty in the inferred adiabatic curvature fluctuations, whereas the bottom-right panel depicts the residual between the ground truth and the inferred. To highlight the effect of the selection function, the edges of the survey are marked with dotted lines, with voxels outside the edge completely masked in Eulerian space. The voxels inside the edges are regions of the data that are unmasked or only partially masked. However, we emphasize that \borg{} is capable of extrapolating information into unobserved voxels using the physics-informed data model \citep{jasche_bayesian_2013,2015arXiv151202242L,2015JCAP...01..036J,2017JCAP...06..049L}. In summary, these maps offer a comprehensive representation of plausible adiabatic curvature fluctuations and capture the statistical properties and (at least the three-point) correlation functions of these fluctuations.

\subsubsection{Additional diagnostics of the runs}

Before concluding, we briefly mention the additional diagnostics and results of the runs provided in the appendix. In Appendix \ref{app_tests_of_algo} we present correlation matrices, correlation lengths, and corner plots of $\fnl$ and the bias parameters for Run \#3 and \#7. In Appendix \ref{appendix_cosmic_fields}, we highlight the generated mock data and illustrate the inferred final density fields and compare them with their ground truth representations. We present further tests of the adiabatic curvature fluctuations, including plots, in Appendix \ref{appendix_acf}.

\section{Discussion}
\label{discussion}

The detection of PNG would have profound implications for our understanding of the early universe and the inflationary paradigm. Local PNG can serve as a powerful probe to test the single-field inflation hypothesis \citep{1993ApJ...403L...1F, Gangui_1993tt,2001PhRvD..63f3002K,2003JHEP...05..013M,2004PhR...402..103B,2010AdAst2010E..72C,biagetti_hunt_2019,meerburg_primordial_2019,2023arXiv231104882G}. In this context, the local $\fnl$ parameter provides a convenient parameterisation to quantify the lowest order of PNG. Projects like \Euclid aim at constraining $\fnl$ with high-precision redshift surveys \citep{laureijs_euclid_2011,euclid_physics}. To forecast the detectable level of PNG in \Euclid{} data, we apply a field-level inference method on \Euclid-like spectroscopic mock data to constrain $\fnl$. This approach allows us to naturally and jointly use all of the information available in the data to provide measurements of PNG and the primordial matter fluctuations.

To generate and infer $\fnl$ from the mock data, we use \borg{}, which is a Bayesian hierarchical field-level inference algorithm designed to analyse galaxy redshift surveys. The data model for this project, illustrated in Fig.~\ref{fig:flowchart}, forward models the primordial matter fluctuations to a predicted observation of the 3D galaxy field, so that the initial conditions can be directly inferred from the data. The data model captures all the effects of the perturbation of the primordial gravitational potential with $\fnl$ and the scale-dependent bias effect on the final observable. We point out that we still assume the universal mass approximation \citep[]{Moradinezhad_Dizgah_2019,barreira_predictions_2021} and lowest-order $\fnl$ contributions. The data model will be updated as more progress is made in modelling the scale-dependent bias effect. Our approach can handle a variety of observational and systematic effects in the forward model, providing robust inferences of PNG and the early universe \citep{jasche2017,2019A&A...624A.115P,Lavaux2019}. Moreover, the explicit physics-informed data model allows the method to maintain interpretability \citep{jasche_bayesian_2013,2019A&A...624A.115P} and perform posterior predictive tests of the results \citep{2019A&A...625A..64J,Lavaux2019}.

In this study, we use \Euclid forecast specifications to generate mock data, such as the number of galaxies, number of catalogues, observed volume, and redshift binning, among others. To generate the mock data, we employ a forward model on a white-noise field, including effects such as structure formation effects, redshift selection functions, sky masks, etc., allowing us to simulate realistic observations. Throughout the study, we generate and analyse mock data under a variety of conditions, for example, tests with different ground truth $f_{\rm NL}^{\mathrm{gt}}$ values, and with different resolutions. The investigations paint a comprehensive picture of the impact of various factors on the inference process.

This study demonstrates the successful sampling of the parameter $\fnl$ using mock \Euclid data. All the runs carried out for this analysis result in inferred values of $\fnl$ within the 68.3\% CI of the ground truth $f_{\rm NL}^{\mathrm{gt}}$ value. These constraints demonstrate the reliability and robustness of the inference methodology. In particular, at a resolution of $62.5 \, \Mpch$, we achieve a constraining power of $\sigma(\fnl) = 2.3$ (68.3\% CI). The resolution study carried out in this paper demonstrates the ability of our method to use information on small scales to constrain $\fnl$. In addition, we present detailed maps of primordial adiabatic curvatures, or $\mathcal{R}$ maps, and their corresponding uncertainties, providing a comprehensive rendering of the inferred initial conditions. In general, these results highlight the potential of using field-level inference to constrain primordial physics with \Euclid data.

\hfill{}
\newpage
\hfill{}
\newpage
\hfill{}
\newpage{}
We forecast how well \borg{} can constrain $\fnl$ in mock \textit{Euclid} spectroscopic data sets. We highlight our major findings below and thereafter provide a summary.

\begin{itemize}
\item The field-level inference framework successfully handles \Euclid-like specifications, accommodating the large survey volume, the number of galaxies, and various physical effects and noise properties. The data model used includes the generation of primordial gravitational potential, perturbation with local PNG, running a structure formation computation, applying a bias model, and evaluating the likelihood of galaxy formation.
\item Our primary finding is that, at a resolution of $62.5\,\Mpch$ $\left(k_{\rm max}=0.1 \, \hMpc\right)$, our method achieves an uncertainty of $\sigma(\fnl) = 2.3$ for $\fnl = 0$ at 68.3\% CI, \new{assuming the universal mass function and $p=0.55$}.
\item Our method can infer the parameter $\fnl$ with high precision at multiple resolutions, different fiducial $\fnl$ values, and sampling configurations, demonstrating the consistency and versatility of the methodology.
\item As we increase the resolution, our algorithm provides tighter constraints on $\fnl$, confirming the results in \citet[][see figure 7]{andrews_bayesian_2023}. Thus, we expect additional improvements by further increasing the resolution, especially since our data model is capable of handling nonlinearities beyond the considered $k_{\rm max}$.
\item When assuming known bias parameters, an improvement of $\sim$$35$\% is achieved in the estimation of $\fnl$, compared to the test in which the bias parameters are sampled and marginalised over.
\item When the amplitude of the scale-dependent bias effect is weakened, \new{from $p=0.55$ to $p=1$, the constraints on the inferred $\fnl$ decrease by approximately $38\%$ (68.3\% CI)}.
\item We perform a run by also sampling the scale-dependent bias parameters $b_{\phi}$ and $\bpd$, with priors centred on Eqs.~\eqref{eq:b_phi} and ~\eqref{eq:b_phidelta}. The run provides unbiased constraints on $\fnl$, but with an increase in $\sigma(\fnl)$ by approximately 20\% (68.3\% CI).
\item At a resolution of $62.5\,\Mpch$, we generate maps of adiabatic curvature fluctuations from inferred initial conditions, offering valuable data products for conducting further investigations into the early universe.
\item We showcase additional convergence tests and data products, for example corner plots, correlation lengths, and inferred cosmic fields, in Appendices \ref{app_tests_of_algo}--\ref{appendix_cosmic_fields}.
\end{itemize}
In conclusion, our method provides a complementary and independent approach to statistical summary estimators for constraining primordial physics in galaxy redshift surveys. By using the full formation history of the Universe in the data model, field-level inference opens up the possibility to perform optimal measurements of PNG up to the given resolution and data model. With these measurements, the scientific community will be able to significantly reduce the parameter space of plausible models for the inflationary universe. In this way, our method will contribute to the scientific success of the \Euclid mission by enabling it to excel in one of its primary research objectives.

\newpage

\section{Future work}
\label{future_work}

In this section, we provide a discussion on the further testing and development of the method for inferring primordial physics in galaxy redshift survey data, with a focus on the following aspects: (1) validation against the \Euclid flagship simulation \citep{2017ComAC...4....2P,2024arXiv240513495E}, (2) handling observational effects such as relativistic effects and foreground systematic effects, and (3) inclusion of additional inflationary model parameters and (4) cosmological parameters.

\subsection{Method validation with the \Euclid flagship simulation}

Before analysing the upcoming observed data, further tests of the adopted data model against more complex data is required, for example against the \Euclid flagship simulation. The \Euclid flagship simulation is a gravity-only dark matter simulation consisting of $12\,600^3$ dark matter particles within the size of $3780\,\Mpch$. With the method, one can use the simulated halo catalogue as a substitute for galaxies, and thus perform further tests on the data model and the performance of the algorithm in constraining $\fnl$. Examples include further testing of the structure formation model and bias model in \borg, and how well they can capture the results of the full simulation. Furthermore, analysing $N$-body simulations with nonzero $\fnl$ values is another way to validate the implemented PNG model \citep{Jung_2022, Coulton_2023, Jung_2023, jung2023quijotepng,2023arXiv231110088F,2024arXiv240210881H}. In addition, another insightful investigation is to  test how changes in the data model, such as the choice of structure formation model, impact the algorithm's inference power on $\fnl$. Successfully further testing the algorithm with $N$-body simulations is a crucial step before the subsequent application on the upcoming real data.

\subsection{Handling more complex observational effects}

Real observational data are subject to various survey systematic and observational effects. Future work will therefore have to incorporate and mitigate these effects in the analysis framework. One example is the consideration of general relativistic effects, which imprint an effect similar to PNG on large scales. Another is the modelling of light-cone effects, which account for the evolution of the observable universe over cosmic time \citep{Bruni_2011ta,2012PhRvD_85b3504J,PhysRevD.88.023515,Bertacca_2014dra,Bertacca_2014wga,Jeong_2014ufa,PhysRevD.90.123507,Yoo_2015uxa,Koyama_2018ttg,Desjacques_2018pfv,Umeh_2019qyd,Lavaux2019,Wang_2020ibf,Maartens_2021,Martinez_Carrillo_2021,2022JCAP...01..061C,Enr_quez_2022,shiveshwarkar2023postinflationary,2024arXiv240706301R,2024arXiv240700168A}. By properly handling these effects, the algorithm aims to break the degeneracy between these effects and PNG, in such a way that one can ensure that the method accurately captures the relevant information from observed data. Furthermore, foreground systematic effects, such as contamination from Galactic emissions or instrumental artefacts, can significantly bias the measured cosmological parameters if they are not taken into account \citep[]{leistedt_constraints_2014,Rezaie_2021,mueller2021clustering,rezaie2023local}. Lastly, tests incorporating photometric redshift data sets will also be conducted \citep{2012MNRAS.425.1042J,2023arXiv230103581T}. Developing robust methods to identify and marginalise these systematic effects is a key question in the \Euclid mission.

\subsection{Inclusion of additional inflationary model parameters}

Inflationary models offer a rich framework for understanding the early universe and its subsequent evolution. To further constrain the space of plausible inflationary models, the data model can be extended to include additional inflationary model parameters beyond the parameter $\fnl$. For example, the data model could include other parameters such as the local trispectrum non-Gaussianity parameter $g_{\rm NL}$ \citep{2002PhRvD..66f3008O,sasaki_2006,Jeong_2009,leistedt_constraints_2014,roth_can_2012,2023arXiv231215038S, pardede2023wideangle}, the running-of-the-scalar index $\alpha_s$ \citep{fedeli_primordial_2010, planck_collaboration_planck_2019_X,german_constraints_2020}, and other shapes of PNG, for example, the equilateral non-Gaussianity parameter $f_{\rm NL}^{\rm equi}$ \citep{Babich_2004gb,scoccimarro_large-scale_2012,regan_universal_2012,Planck_2013wtn,schmidt_imprint_2015,planck_collaboration_planck_2016,karagiannis_constraining_2018,planck_collaboration_planck_2019_IX,Karagiannis_2020,baumann2021power}. Such implementations would be subject to validation tests and mock data studies before making predictions on how well field-level inference methods can constrain such parameters in the large-scale structure. In short, by incorporating additional primordial parameters, the data model would be able to perform a more thorough exploration of the inflationary paradigm in the data, and thus allowing the method to further distinguish different inflationary models.

\subsection{Joint sampling of cosmological parameters}
Another test of interest is to explore the parameter space of the $\Lambda$CDM model together with non-Gaussian cosmic initial conditions. By doing so, one would be ale to assess the efficacy of field-level inference in the context of jointly sampling other cosmological parameters together with $\fnl$. Examples of such parameters include $\Omega_{\rm{m}}$, $w_0$ \citep{ramanah_cosmological_2019}, and $\sigma_8$ \citep{2021MNRAS.502.3035P,2022MNRAS.509.3194P,2023arXiv230404785P}. However, extending the analysis to include additional degrees of freedom exposes the algorithm to potential parameter degeneracies, which could degrade the constraints in $\fnl$. Based on previous work, we anticipate a marginal decline in performance, projected to be within the limit of 10\% \citep{Jung_2023,jung2023quijotepng}. Confirmation of these expectations will be made through future mock data tests, within the context of \euclid{} simulations and other data sets.

\begin{acknowledgements}
We thank Fabian Schmidt for valuable feedback on the manuscript. JJ acknowledges support by the Swedish Research Council (VR) under the project 2020-05143 -- `Deciphering the Dynamics of Cosmic Structure' and from the contract ASI/ INAF for the Euclid mission n.2018-23-HH.0. GL acknowledge financial support from the Centre National d’Etudes Spatiales (project GCEUCLID), and the Simons Foundation collaboration programme `Learning the Universe'. FF, MB, and DP acknowledge partial financial support from the contract ASI/ INAF for the Euclid mission n.2018-23-HH.0 and from the INFN InDark initiative.\\

The computations and data handling were enabled by resources provided by the National Academic Infrastructure for Supercomputing in Sweden (NAISS) and the Swedish National Infrastructure for Computing (SNIC) at Tetralith partially funded by the Swedish Research Council through grant agreements no. 2022-06725 and no. 2018-05973.\\

This research utilised the HPC facility supported by the Technical Division of the Department of Physics, Stockholm University.\\

MB acknowledges financial support from the INFN InDark initiative and from the COSMOS network ({\tt www.cosmosnet.it}) through the ASI (Italian Space Agency) 
Grants 2016-24-H.0 and 2016-24-H.1-2018, as well as 2020-9-HH.0 (participation in LiteBIRD phase A).\\

JV was supported by Ruth och Nils--Erik Stenbäcks Stiftelse and Research Council of Finland grant 347088.\\

FP acknowledges partial support from the INFN grant InDark and the Departments of Excellence grant L.232/2016 of the Italian Ministry of University and Research (MUR) and the FCT project with ref. number PTDC/FIS-AST/0054/2021.\\

We acknowledge the use of the following packages: \texttt{NumPy} \citep[]{harris2020array}, \texttt{Matplotlib} \citep[]{Hunter_2007}, \texttt{GetDist} \citep[]{lewis2019getdist}, and \texttt{HEALPix} \citep[]{gorski_healpix_2005}.\\

This work is done within the Aquila Consortium (\url{https://www.aquila-consortium.org/}).\\

\AckEC
\end{acknowledgements}

\bibliographystyle{aa_url}
\bibliography{paper}

\appendix

\section{Additional tests of the adiabatic curvature fluctuations}
\label{appendix_acf}

\begin{figure}[h]
\centering
\includegraphics[width=0.9\columnwidth]{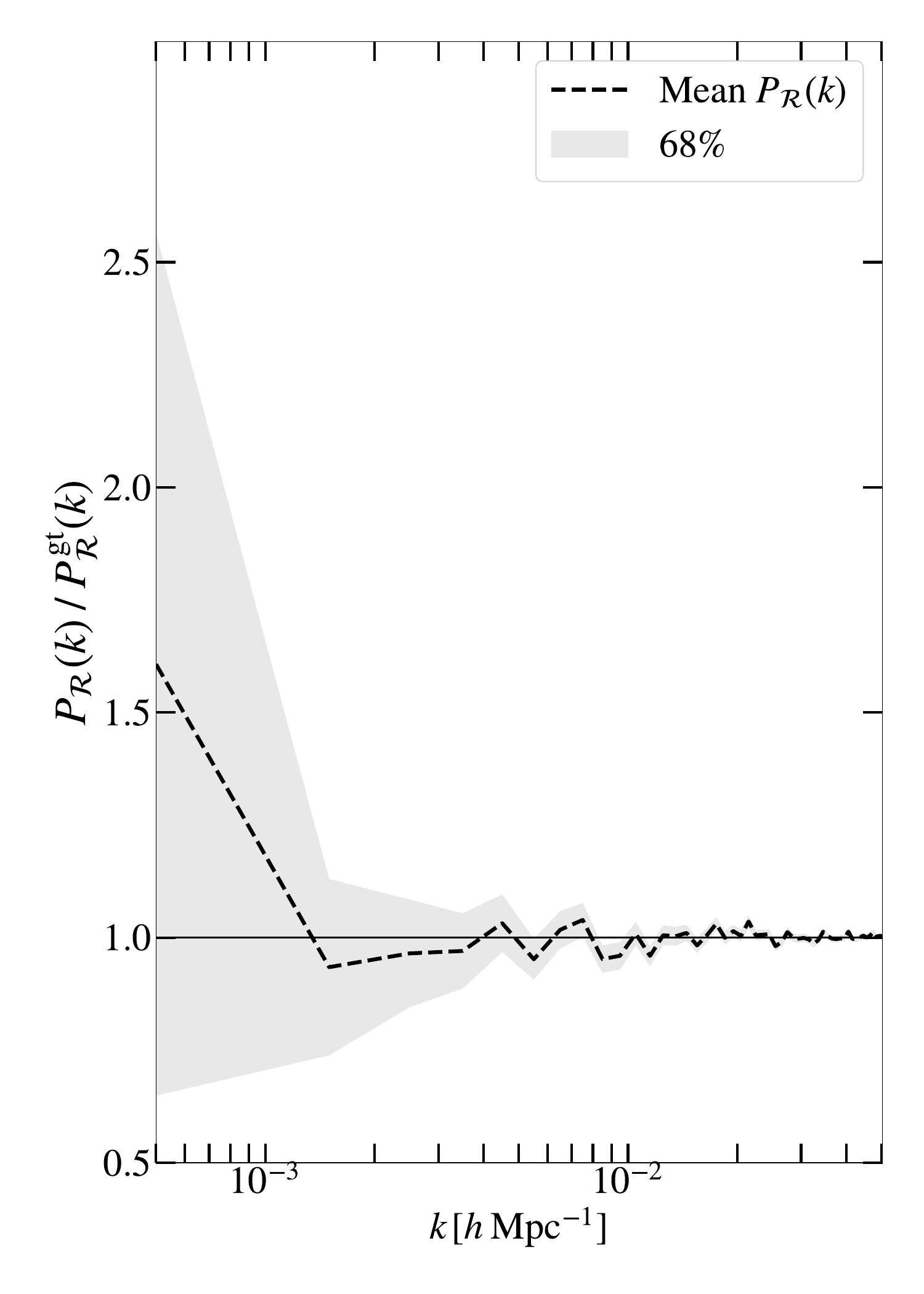}
\caption{Ensemble power spectra statistics of the inferred adiabatic curvature fluctuations, relative to the ground truth. The grey region is the 68\% scatter around the mean power spectrum of the ensemble.}
\label{fig:acf_pk_ensemble}
\end{figure}

To further test our generated data products, we compute how well they relate to the ground truth adiabatic curvature field. We compare the power spectra of the ensemble-predicted adiabatic curvature fluctuations with the ground truth. The results can be seen in Fig. \ref{fig:acf_pk_ensemble}. Furthermore, in each voxel, we have stored three values: i) the ground truth value (from the mock data itself), ii) the mean inferred estimate, and iii) the uncertainty of the estimate, which is given in terms of $\sigma(\mathcal{R})$. Thus, in the range of voxels, we evaluate whether the ground truth value is within the CIs 68.3\%, 95.4\%, or 99.7\%. Our results show that in the $128^3$ voxels, $69.7 \, \%$ are in the $1\sigma(\mathcal{R})$ range, $95.6 \, \%$ are in the $2\sigma(\mathcal{R})$ range and $99.8 \, \%$ are in the $3\sigma(\mathcal{R})$ range. The test also shows that roughly $49\,\%$ of the inferred voxels fall above the ground truth value, while $51\,\%$ of the inferred voxels fall below the ground truth value. Thus, our inferred $\mathcal{R}$ maps are representative of the ground truth $\mathcal{R}$ map, to the expected confidence.

\begin{figure}
\centering
\includegraphics[width=0.7\columnwidth]{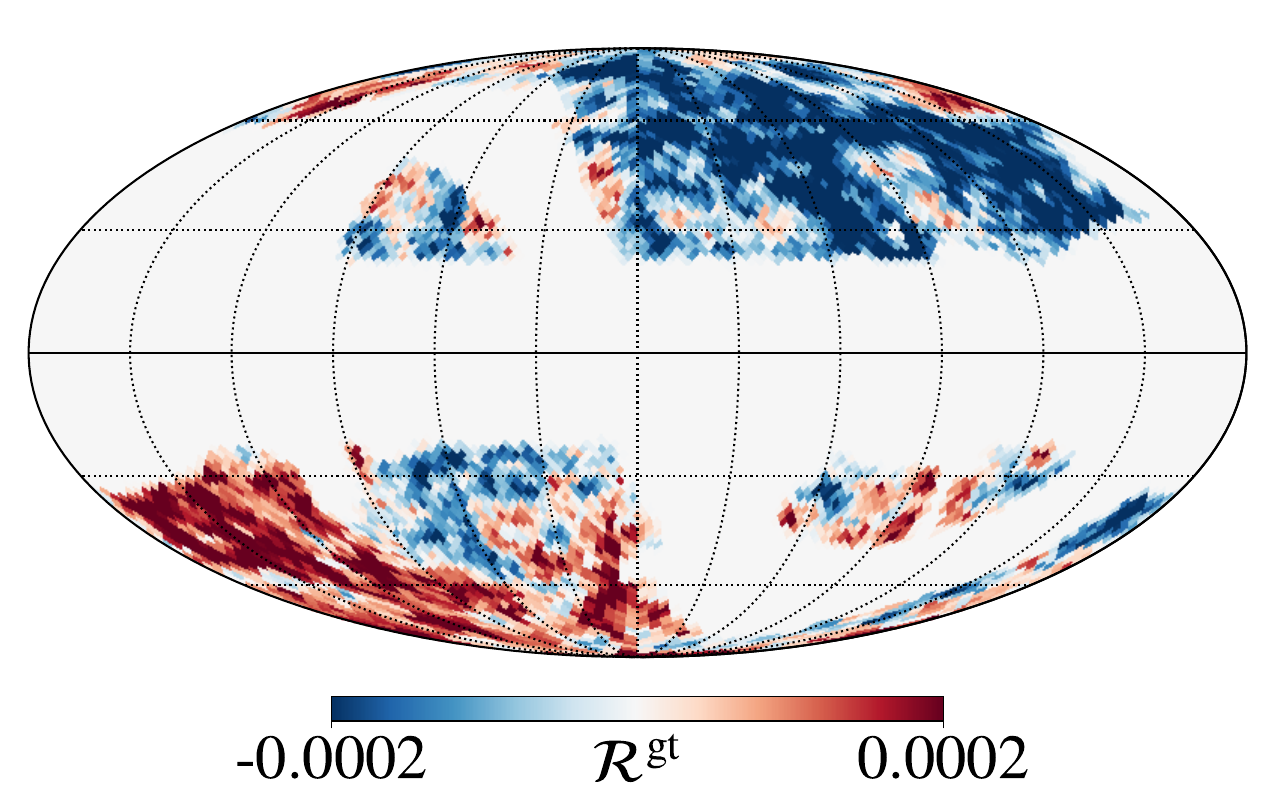}
\caption{Mollweide projection of the \textbf{ground truth} adiabatic curvature fluctuation map. The projection is computed for a distance of $r=2250 \, \Mpch$, for an observer placed in the centre of the cube, and multiplied by the window selection function.}
\label{fig:acf_projection_one}
\end{figure}

\begin{figure}
\centering
\includegraphics[width=0.7\columnwidth]{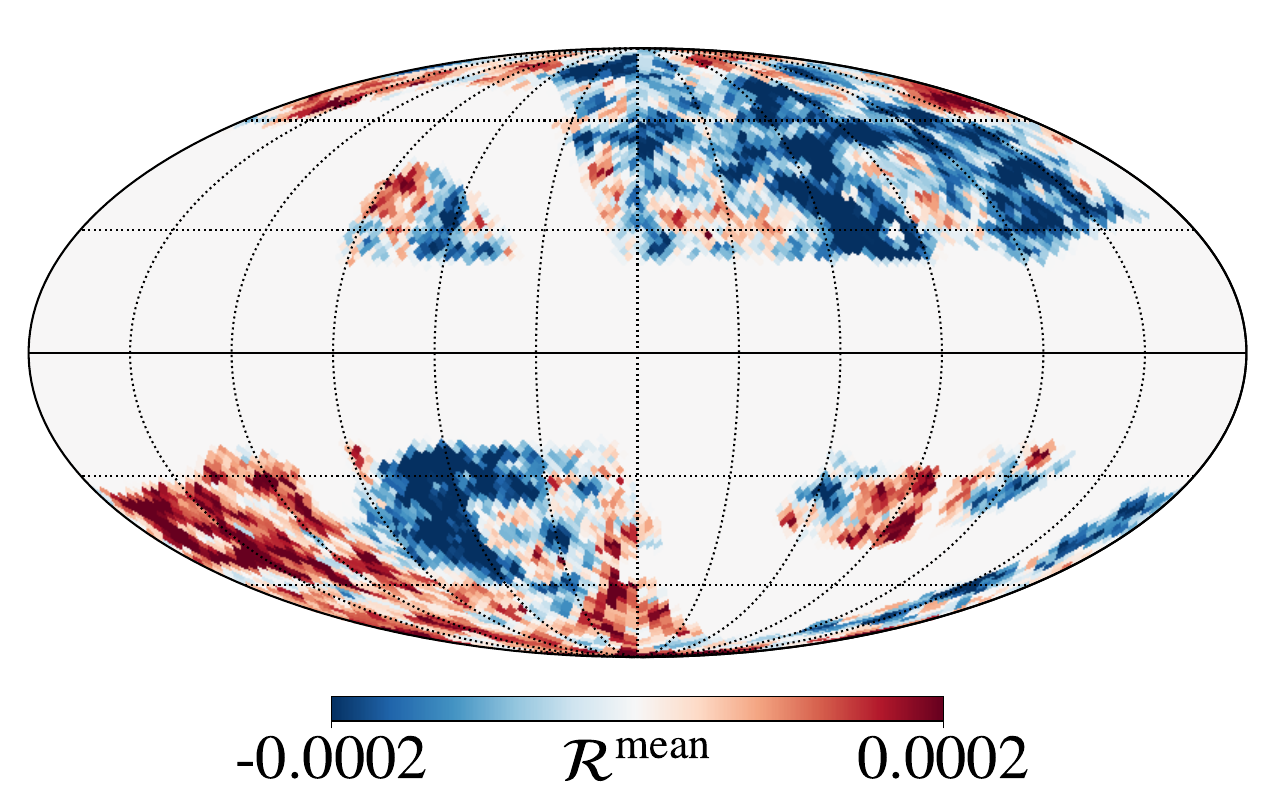}
\caption{Similar to Fig \ref{fig:acf_projection_one}, but for the \textbf{mean inferred} adiabatic curvature fluctuation map.}
\label{fig:acf_projection_two}
\end{figure}

\begin{figure}
\centering
\includegraphics[width=0.7\columnwidth]{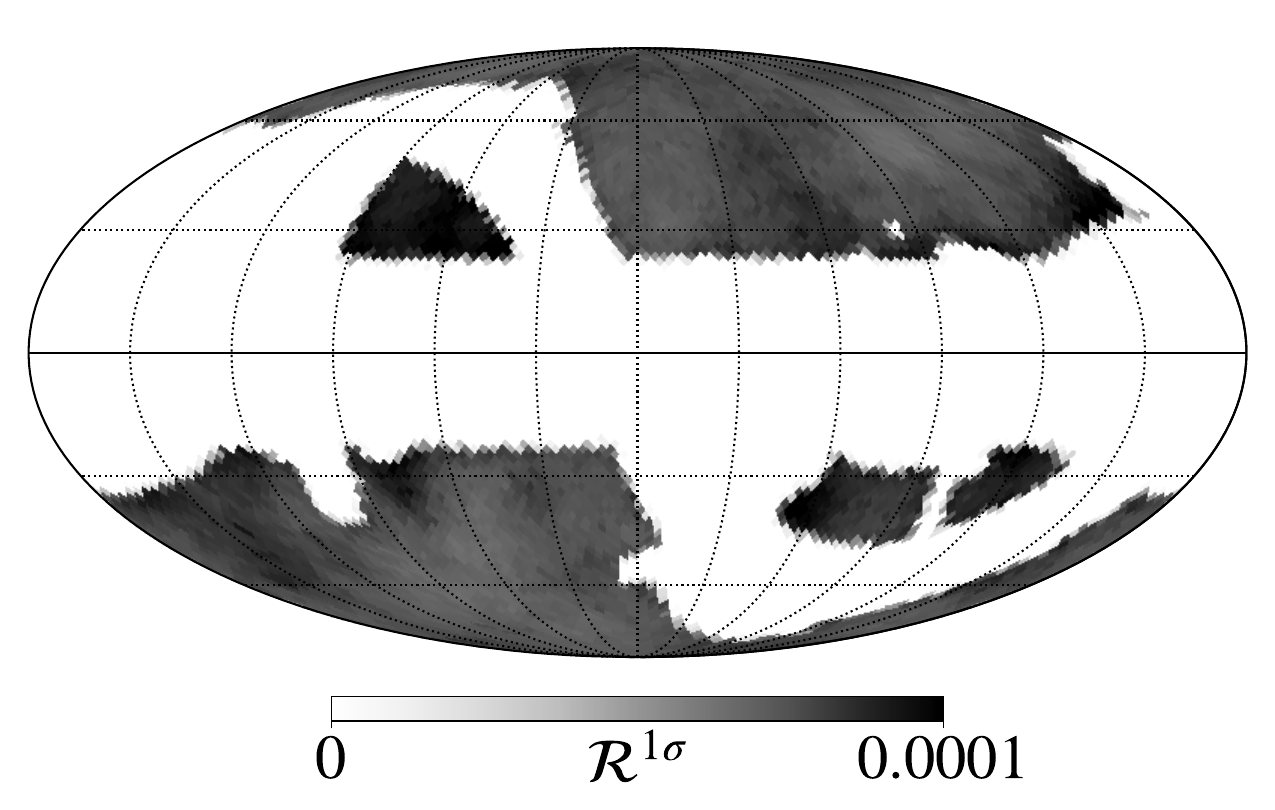}
\caption{Similar to Fig \ref{fig:acf_projection_one}, but for the \textbf{uncertainty of the inferred} adiabatic curvature fluctuation map.}
\label{fig:acf_projection_three}
\end{figure}

\begin{figure}
\centering
\includegraphics[width=0.7\columnwidth]{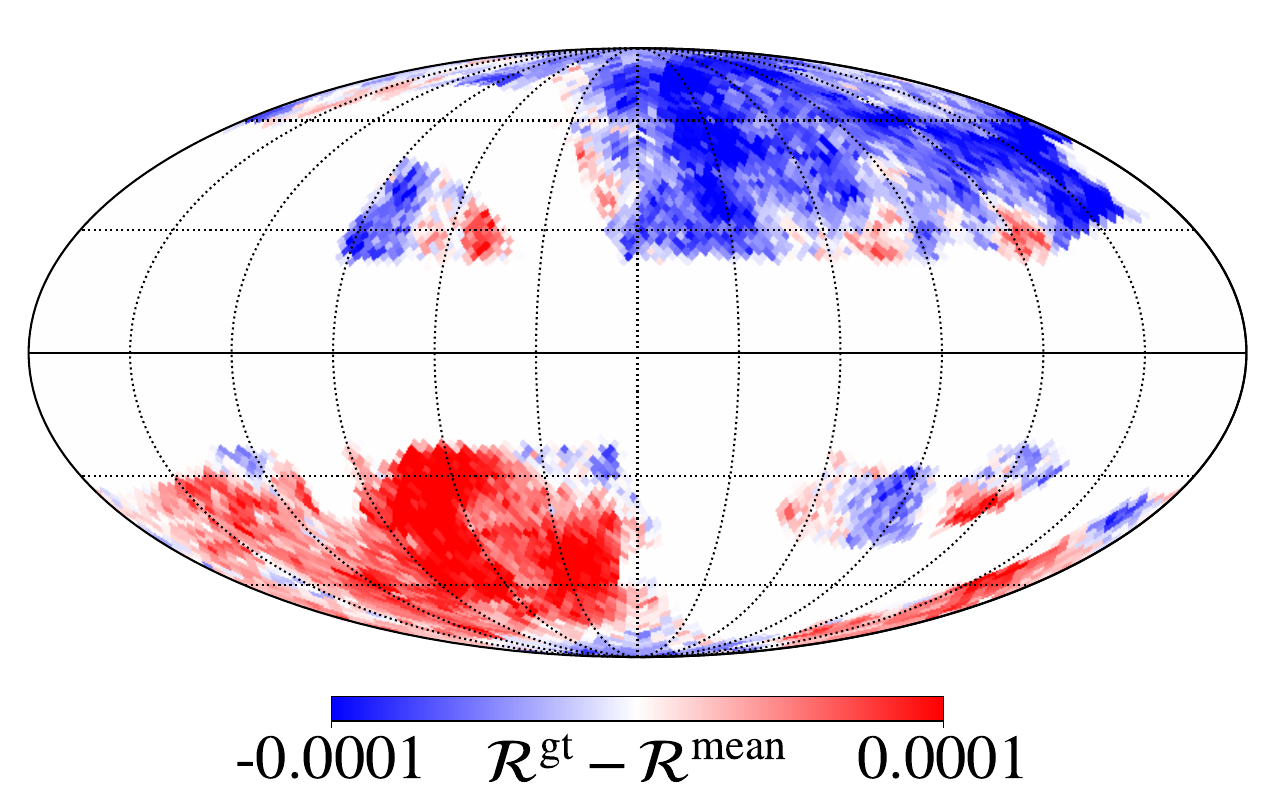}
\caption{Similar to Fig \ref{fig:acf_projection_one}, but for the \textbf{residual} adiabatic curvature fluctuation map, defined as $\mathcal{R}^{\rm ground \, truth} - \mathcal{R}^{\rm mean \, inferred}$.}
\label{fig:acf_projection_four}
\end{figure}

Figs.~\ref{fig:acf_projection_one}--~\ref{fig:acf_projection_four}: We present the ensemble statistics of the adiabatic curvature fluctuation maps, in the Mollweide projection, for a distance of $r=2250 \, \Mpch$. The fields shown are the averages of the ground truth, the mean inferred, the standard deviation of the inferred, and the residuals between the ground truth and inferred in a single direction. The fields are multiplied by the selection value in each direction, effectively setting the masked directions to zero.

\section{Tests of the algorithm}
\label{app_tests_of_algo}

This section presents additional plots to offer further insight into the algorithm's performance. The plots encompass autocorrelation lengths, correlation matrices, and corner plots that illustrate the relationships between $\fnl$ and the galaxy bias parameters.

\begin{figure*}
	\centering
    \includegraphics[width=2\columnwidth]{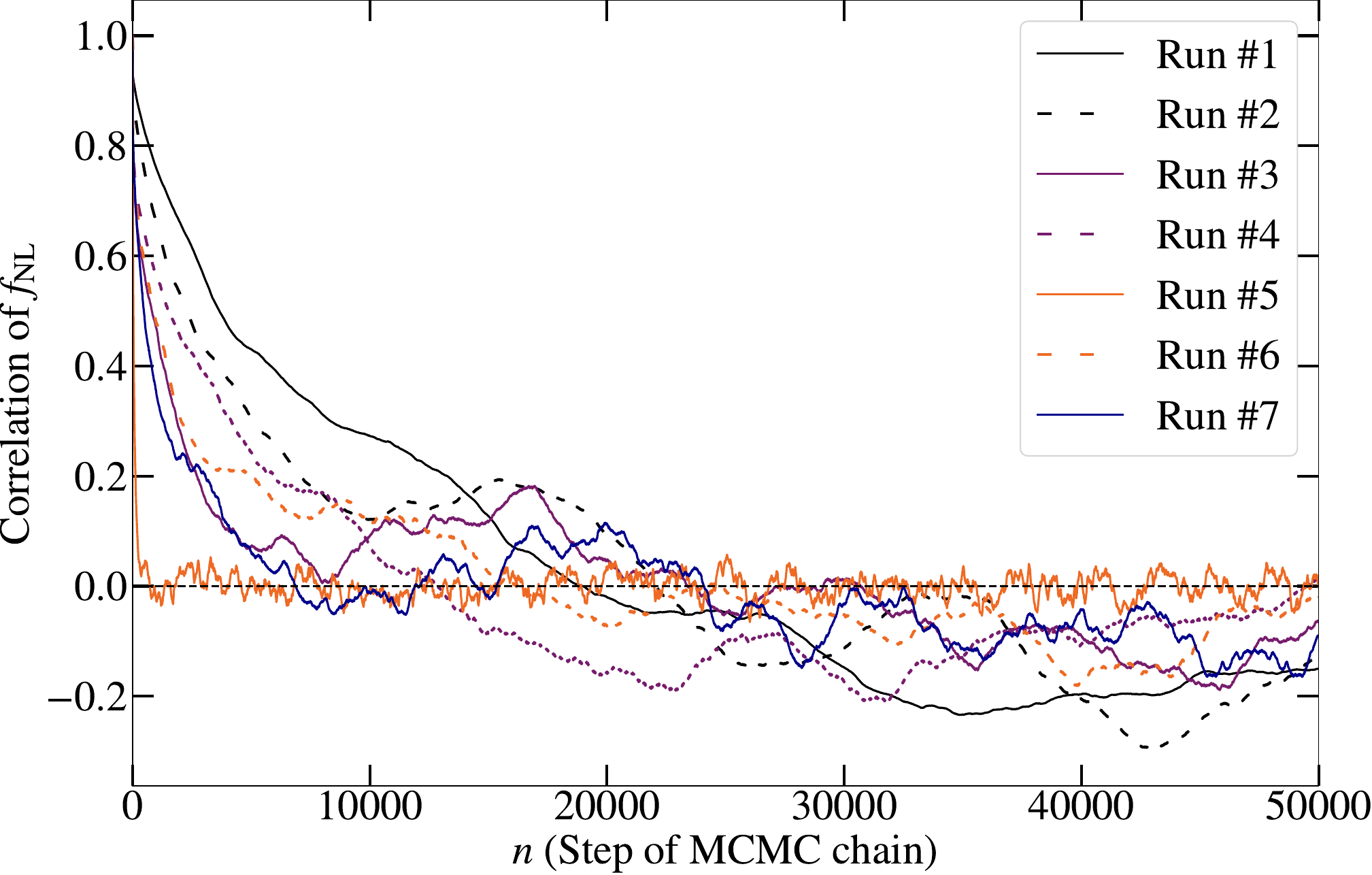}
	\caption{Correlation length of all chains, illustrating the rate at which samples in the various chains achieve independence. The typical correlation length for the chains is around $10\,000$ samples when bias parameters are also sampled.}
\label{fig:corr_lengths}
\end{figure*}

Figure~\ref{fig:corr_lengths} shows the autocorrelation lengths for $\fnl$ from all chains. The autocorrelation length is a metric used in MCMC algorithms to assess how quickly the samples in the chain become independent. Longer lengths indicate slower convergence, necessitating more iterations for reliable parameter estimates. On the contrary, shorter lengths denote faster convergence with reduced reliance on past samples. This metric is vital for evaluating MCMC efficiency, which impacts parameter estimation speed and computational requirements. In particular, the correlation length of $\fnl$ in the chains is approximately $10\,000$ samples for each chain, excluding Run \#5.

\begin{figure*}
	\centering
    \includegraphics[width=2\columnwidth]{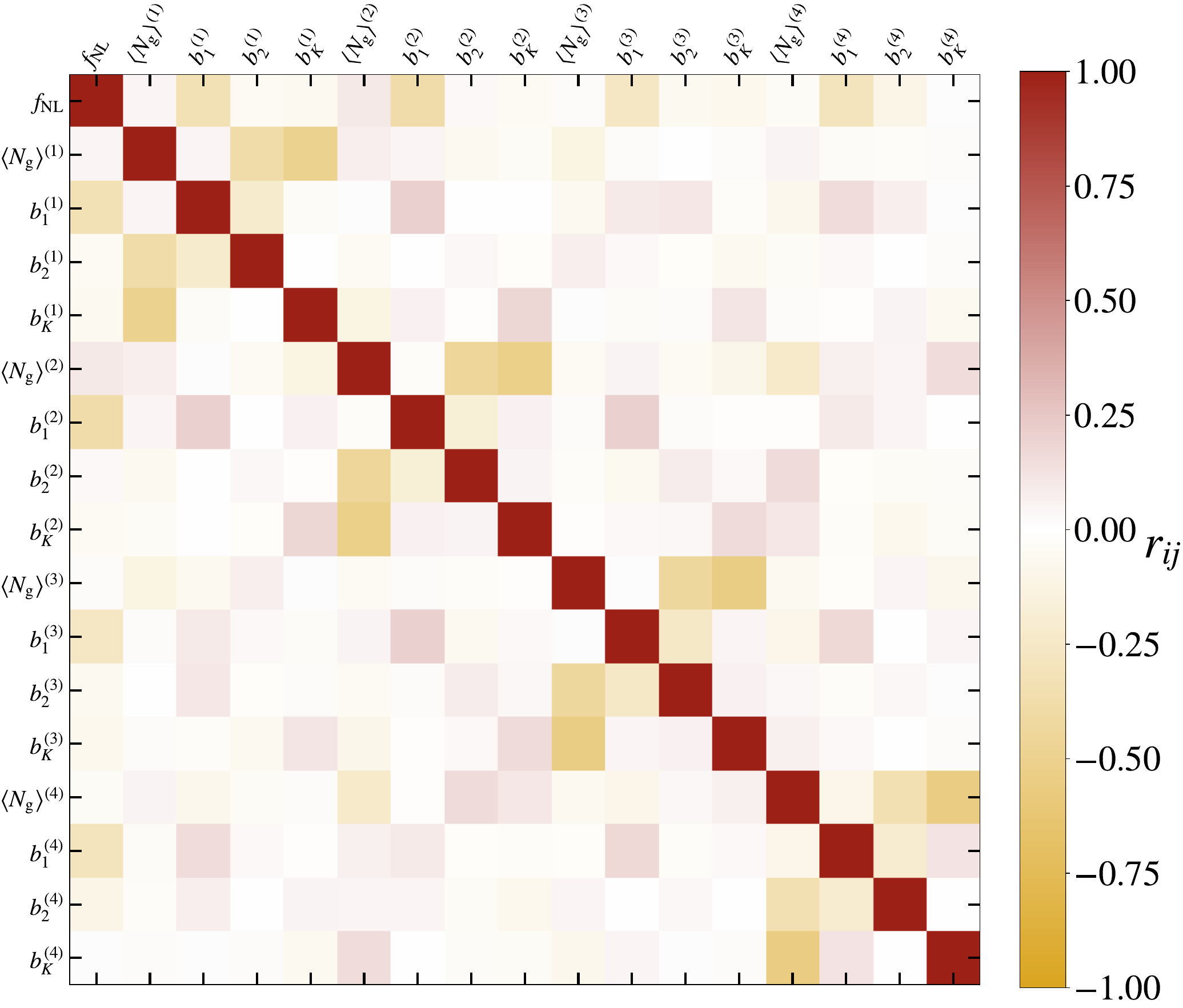}
	\caption{Correlation matrix of Run \#3 (the primary run). The correlation matrix illustrates the pairwise relationships among variables, with colour-coding indicating the strength and direction of correlations, aiding in the identification of patterns and dependencies within the data set. The results show little to no correlation, \replace{especially between $\fnl$ and the bias parameters}{except for a mild anti-correlation between $\fnl$ and the linear bias values}.}
	\label{fig:corr_mat_3}
\end{figure*}

\begin{figure*}
	\centering
    \includegraphics[width=2\columnwidth]{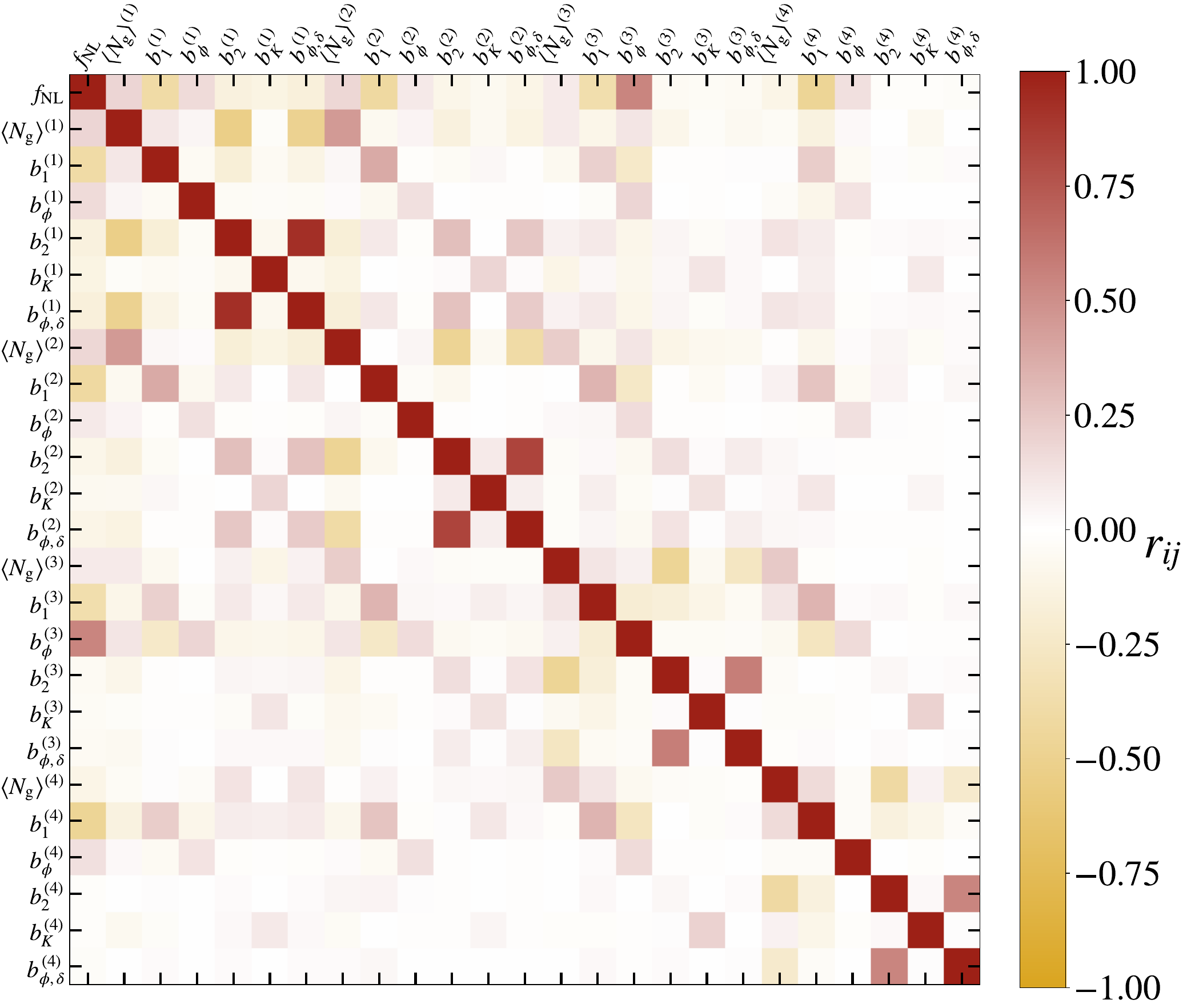}
	\caption{Correlation matrix of Run \#7 (which includes the sampling of the scale-dependent bias parameters, $b_\phi$ and $\bpd$). Colour coding indicates the strength and direction of the correlations, illustrating the little to no correlation between $\fnl$ and the \new{non-linear }bias parameters, including the scale-dependent bias parameters.}
	\label{fig:corr_mat_6}
\end{figure*}

The correlation matrix summarises the relationships between variables in a data set. It shows the magnitudes and direction of linear associations between pairs of variables. High positive values indicate strong positive correlations, while high negative values imply strong negative correlations. A correlation close to zero suggests a weak or no linear relationship. This matrix can also be used to identify potential degeneracies. The Pearson correlation coefficient $r_{ij}$ is defined as:
\begin{align}
r_{ij} = \frac{\text{cov}(X_i, X_j)}{\sigma_i \sigma_j} \, .
\label{eq:corr_mat_def}
\end{align}
Here, \( X_i \) and \( X_j \) are the bias parameters or $\fnl$, \( \text{cov}(X_i, X_j) \) is the covariance, and \( \sigma_i \) and \( \sigma_j \) are the standard deviations. As can be seen in Figs.~\ref{fig:corr_mat_3} and \ref{fig:corr_mat_6}, the galaxy bias parameters exhibit little or no correlations with $\fnl$.

\begin{figure*}
\centering
\includegraphics[width=1.9\columnwidth]{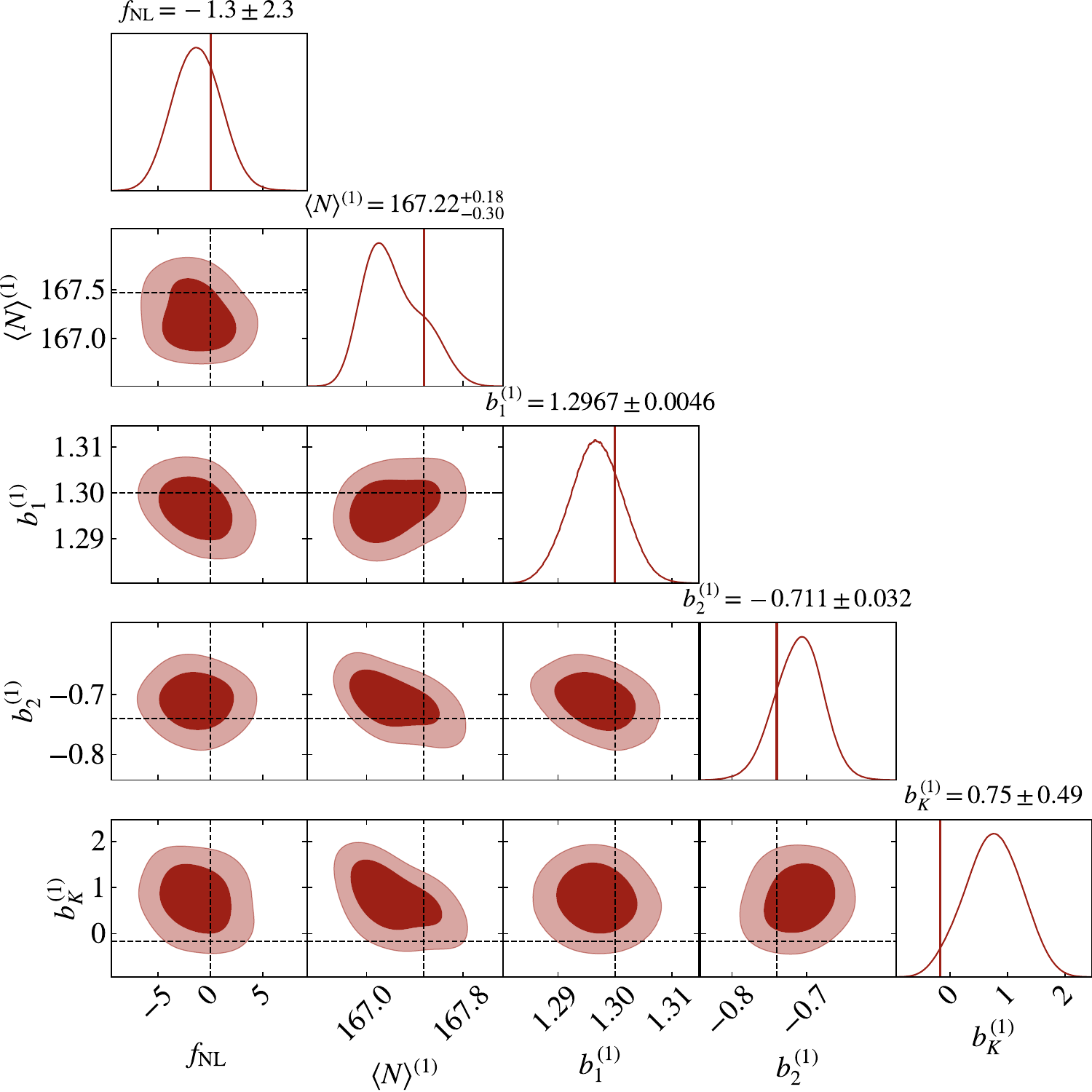}
\caption{Corner plot for $\fnl$ and bias parameters, for Run \#3, catalogue 1. The corner plot displays the joint distributions and marginal distributions of the variables in the multidimensional data set covered by $\fnl$ and the bias parameters. Each subplot captures the relationships between pairs of variables, offering an overview of the data set structure and dependencies. For the main run, there are few to no degeneracies in the bias parameters.}
\label{fig:pyramid_3_1}
\end{figure*}

\begin{figure*}
\centering
\includegraphics[width=1.9\columnwidth]{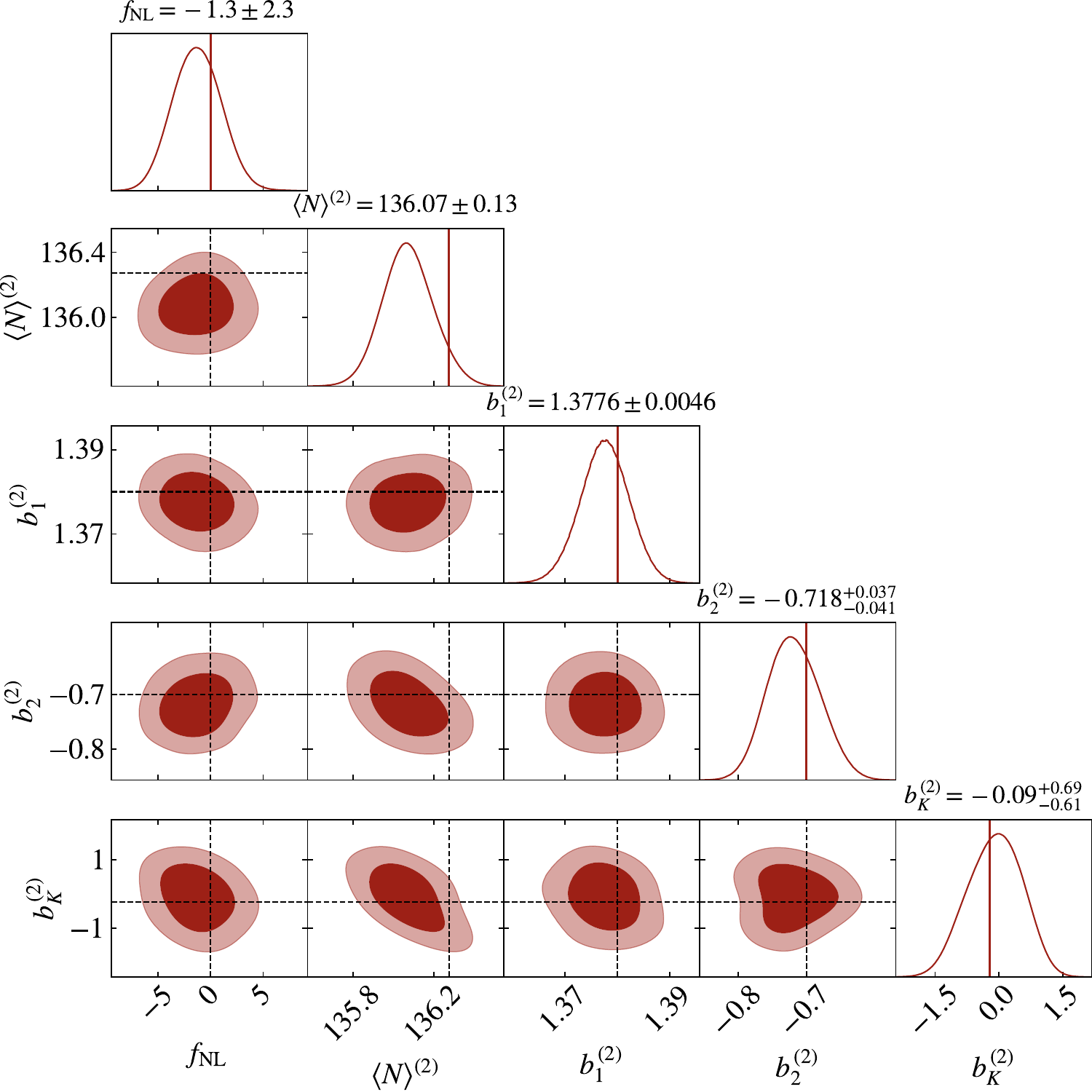}
\caption{Corner plot for $\fnl$ and bias parameters, for Run \#3, catalogue 2. Similar to Fig. \ref{fig:pyramid_3_1}, there are little to no degeneracies between $\fnl$ and the bias parameters.}
\label{fig:pyramid_3_2}
\end{figure*}

\begin{figure*}
\includegraphics[width=1.9\columnwidth]{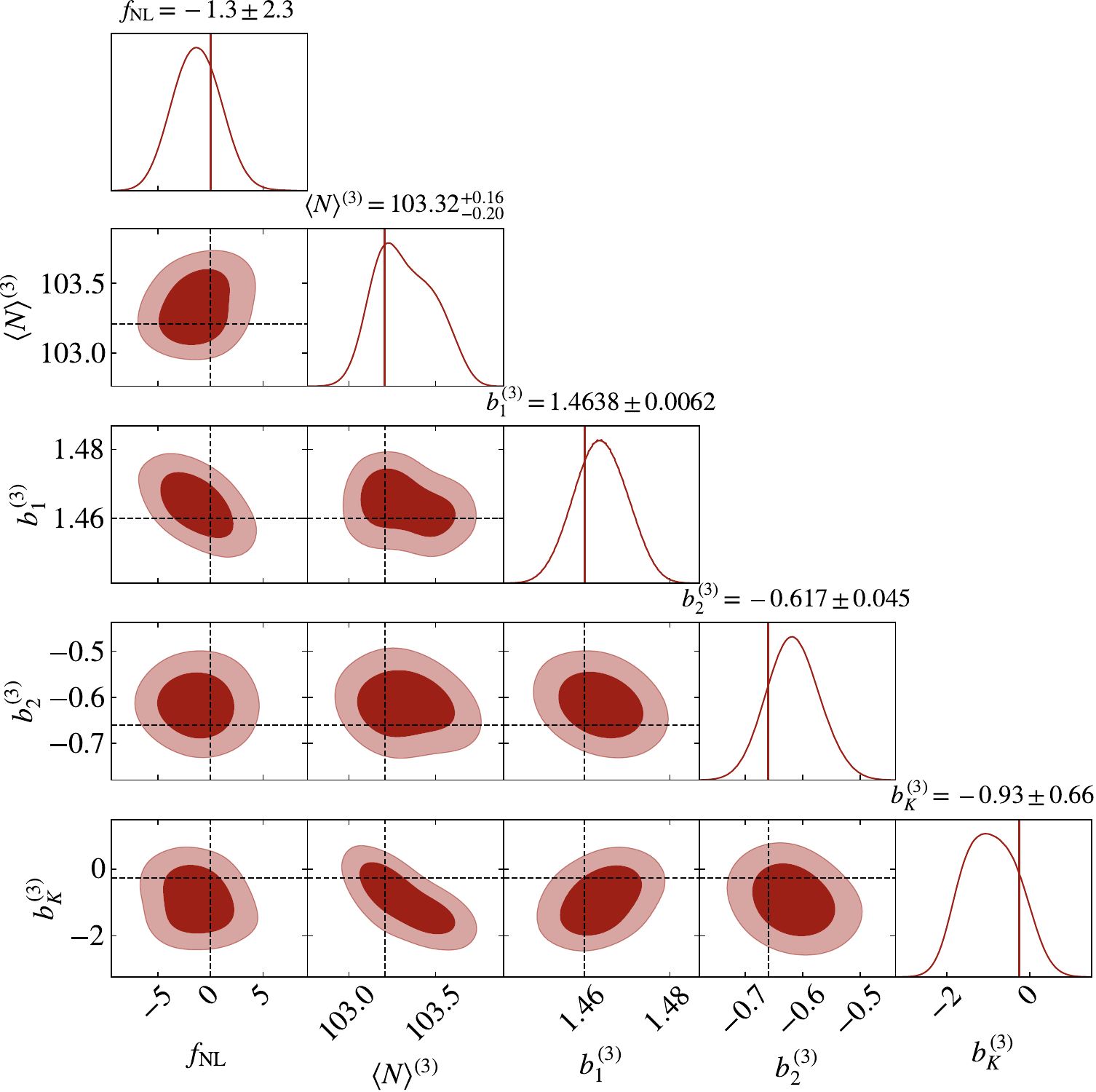}
\caption{Corner plot for $\fnl$ and bias parameters, for Run \#3, catalogue 3. Similar to Fig. \ref{fig:pyramid_3_1}, there are little to no degeneracies between $\fnl$ and the bias parameters.}
\label{fig:pyramid_3_3}
\end{figure*}

\begin{figure*}
\centering
\includegraphics[width=1.9\columnwidth]{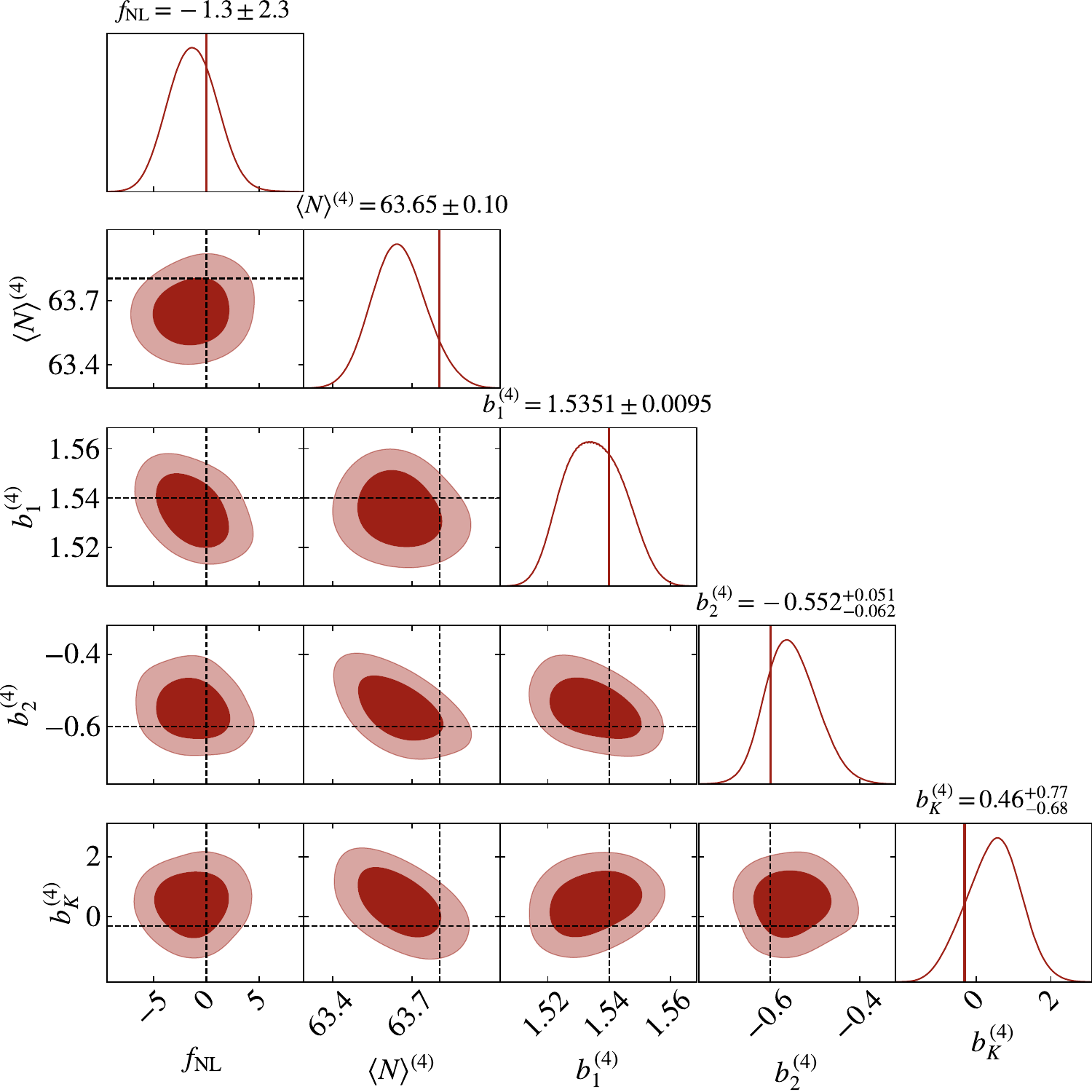}
\caption{Corner plot for $\fnl$ and bias parameters, for Run \#3, catalogue 4. Similarly to Fig. \ref{fig:pyramid_3_1}, there are few to no degeneracies between $\fnl$ and the bias parameters.}
\label{fig:pyramid_3_4}
\end{figure*}

\begin{figure*}
\centering
\includegraphics[width=1.9\columnwidth]{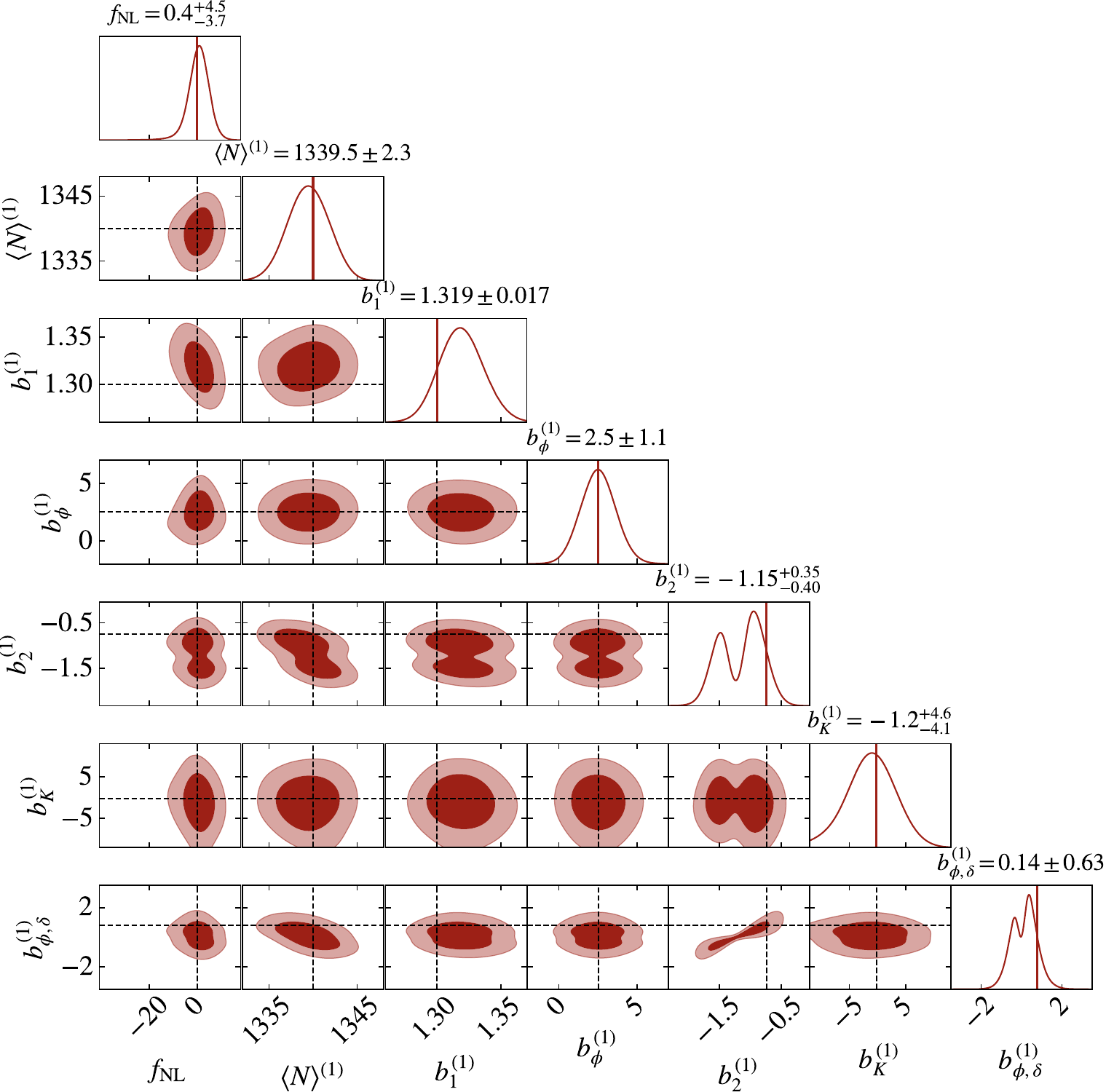}
\caption{Corner plot for $\fnl$ and bias parameters, for Run \#7, catalogue 1. Similar to Fig. \ref{fig:pyramid_3_1}, there are little to no degeneracies between $\fnl$ and the bias parameters.}
\label{fig:pyramid_6_1}
\end{figure*}

\begin{figure*}
\includegraphics[width=1.9\columnwidth]{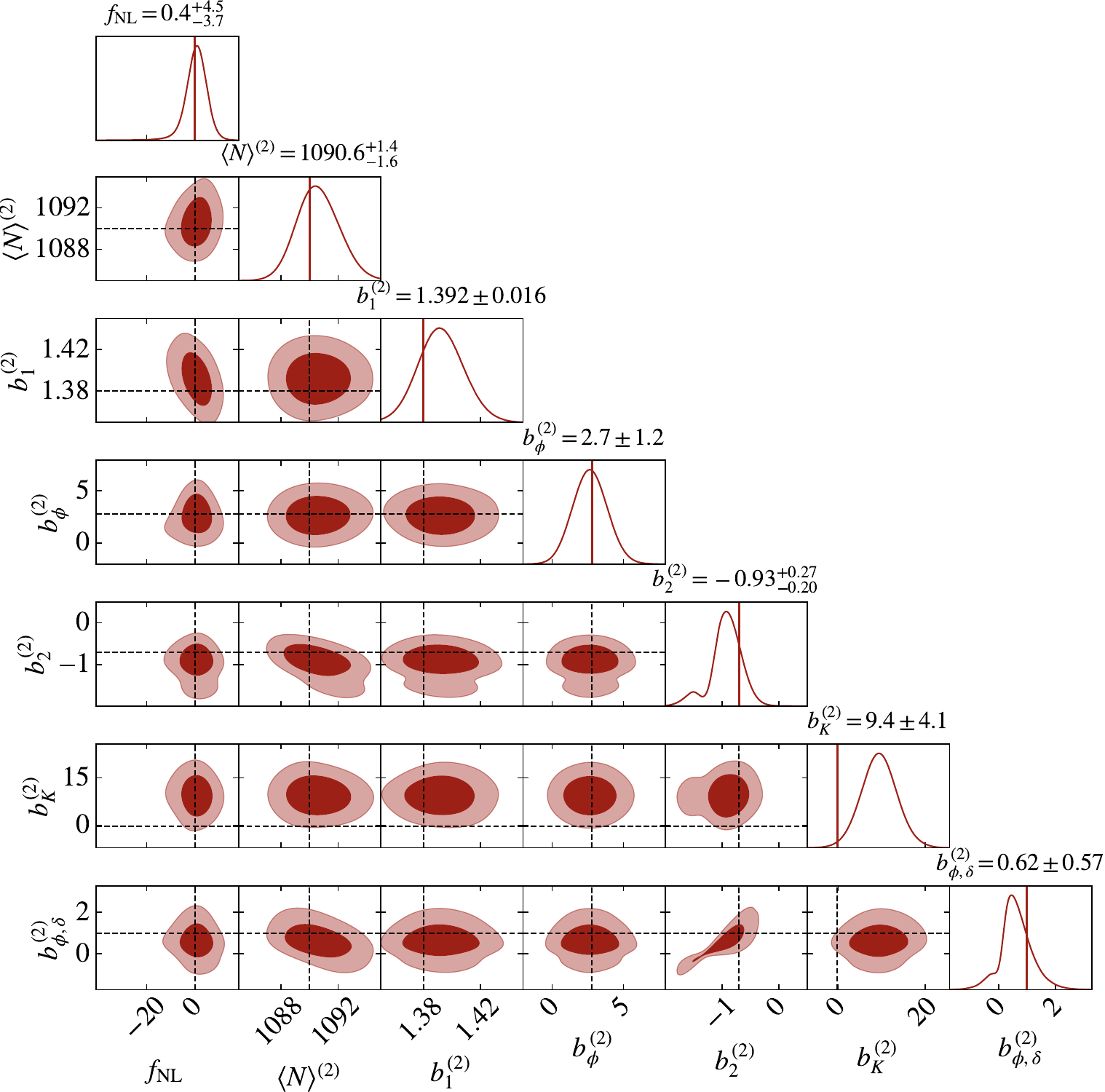}
\caption{Corner plot for $\fnl$ and bias parameters, for Run \#7, catalogue 2. Similar to Fig. \ref{fig:pyramid_3_1}, there are little to no degeneracies between $\fnl$ and the bias parameters.}
\label{fig:pyramid_6_2}
\end{figure*}

\begin{figure*}
\centering
\includegraphics[width=1.9\columnwidth]{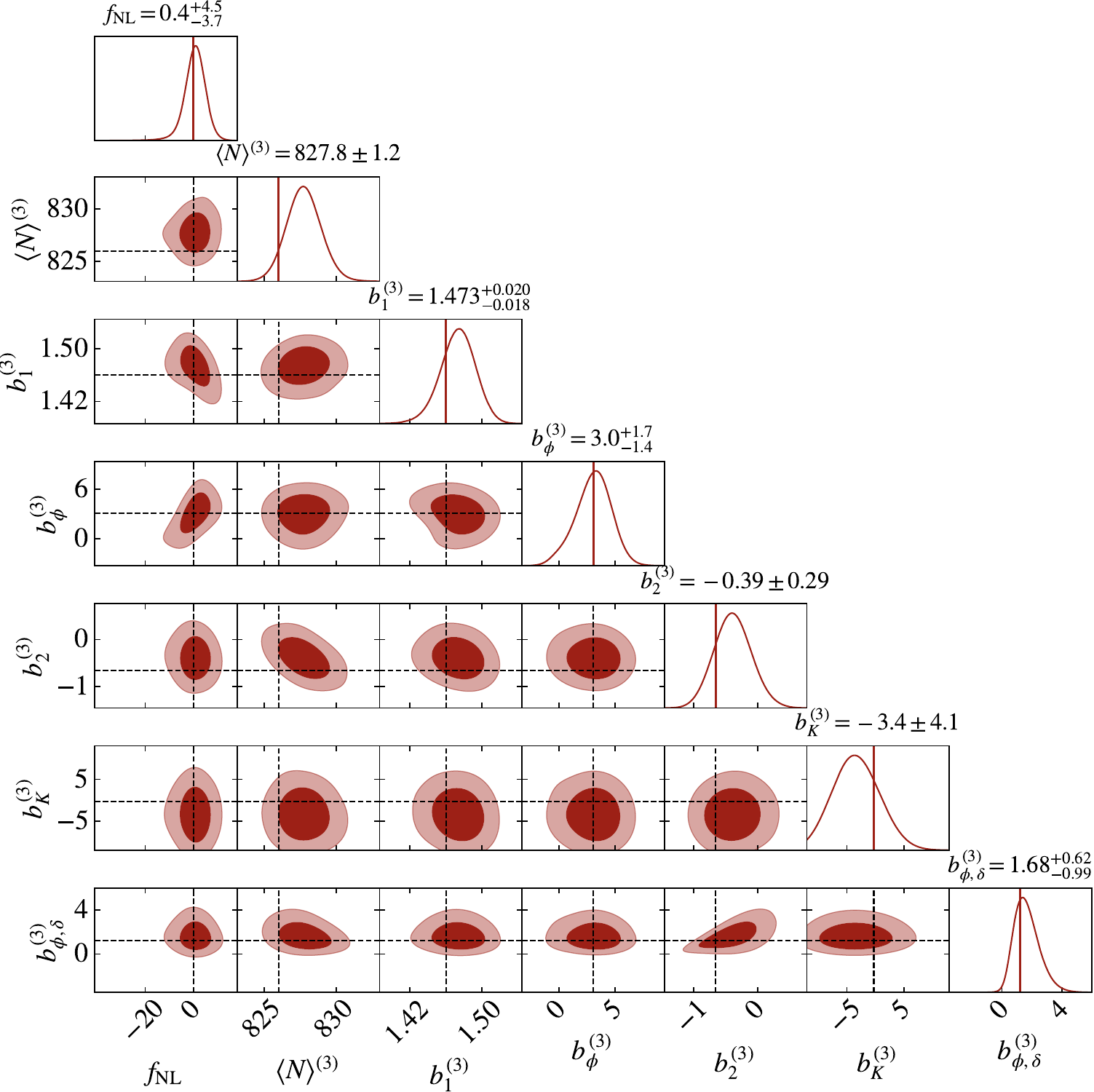}
\caption{Corner plot for $\fnl$ and bias parameters, for Run \#7, catalogue 3. Similarly to Fig. \ref{fig:pyramid_3_1}, there are few to no degeneracies between $\fnl$ and the bias parameters.}
\label{fig:pyramid_6_3}
\end{figure*}

On the same note, the corner plot depicts the interaction between the $\fnl$ and galaxy bias parameters and offers a visual representation of their joint distribution and correlations. This plot, visualised in Figs.~\ref{fig:pyramid_3_1}--\ref{fig:pyramid_3_4}, \ref{fig:pyramid_6_1}--\ref{fig:pyramid_6_3}, and \ref{fig:pyramid_6_4} displays the marginal distributions of each parameter on the diagonal and their joint distributions on the off-diagonal. It illustrates how changes in one parameter are associated with changes in the other, providing information on potential relationships and regions of interest between $\fnl$ and galaxy bias.



\section{Inferred density field and data projections}
\label{appendix_cosmic_fields}
This section showcases plots of the inferred density fields and a comprehensive comparison by including the ground truth mock data fields.

\begin{figure*}
	\centering
    \includegraphics[width=2.0\columnwidth]{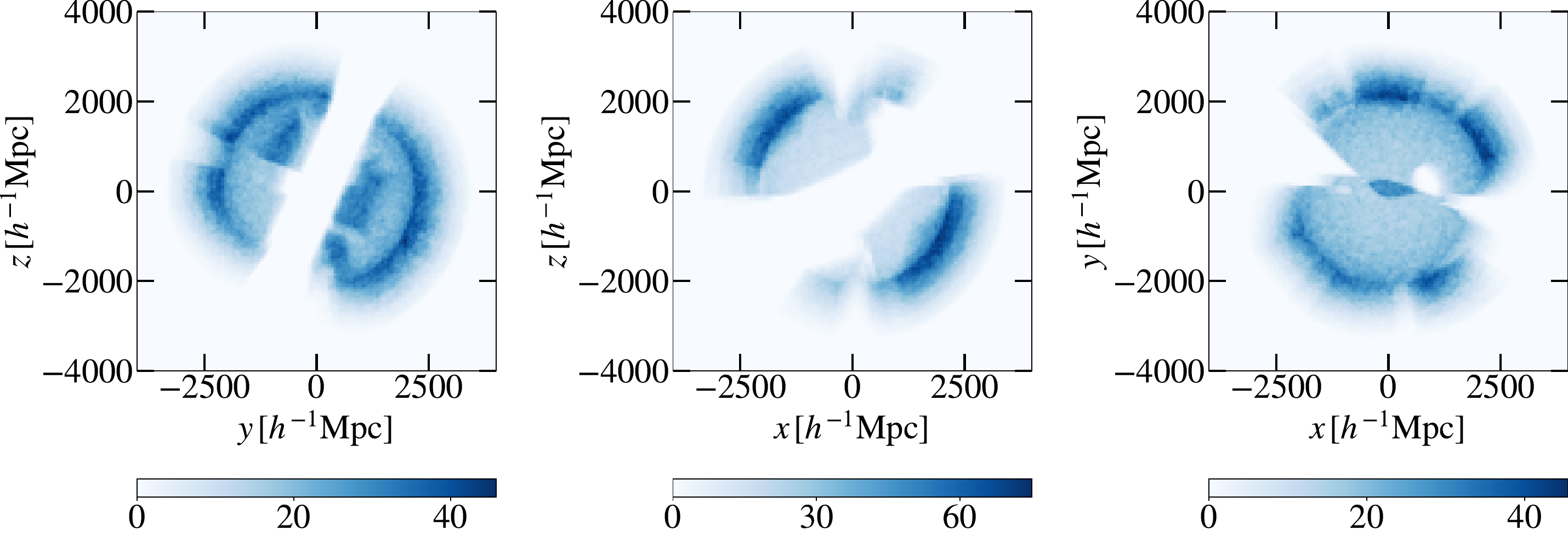}
	\caption{Averaged projections of the mock data fields. The colour bar displays the number of galaxies in each pixel, which contains the sum of all galaxies in the summed-over axis. The image is intended to demonstrate the effects of the window function on the observed data and how the method can account for it. The galaxy field projected here is used for Run \#3. Each pixel covers a width of $8000 \, \Mpch / 128 = 62.5 \, \Mpch$.}
	\label{fig:data_gallery}
\end{figure*}

Fig.~\ref{fig:data_gallery}: We present the averages of the mock data field in three different directions. Notice how the data are cut due to the combination of the radial selection function and the completeness mask.

\begin{figure*}
    \centering
    \includegraphics[width=2.0\columnwidth]{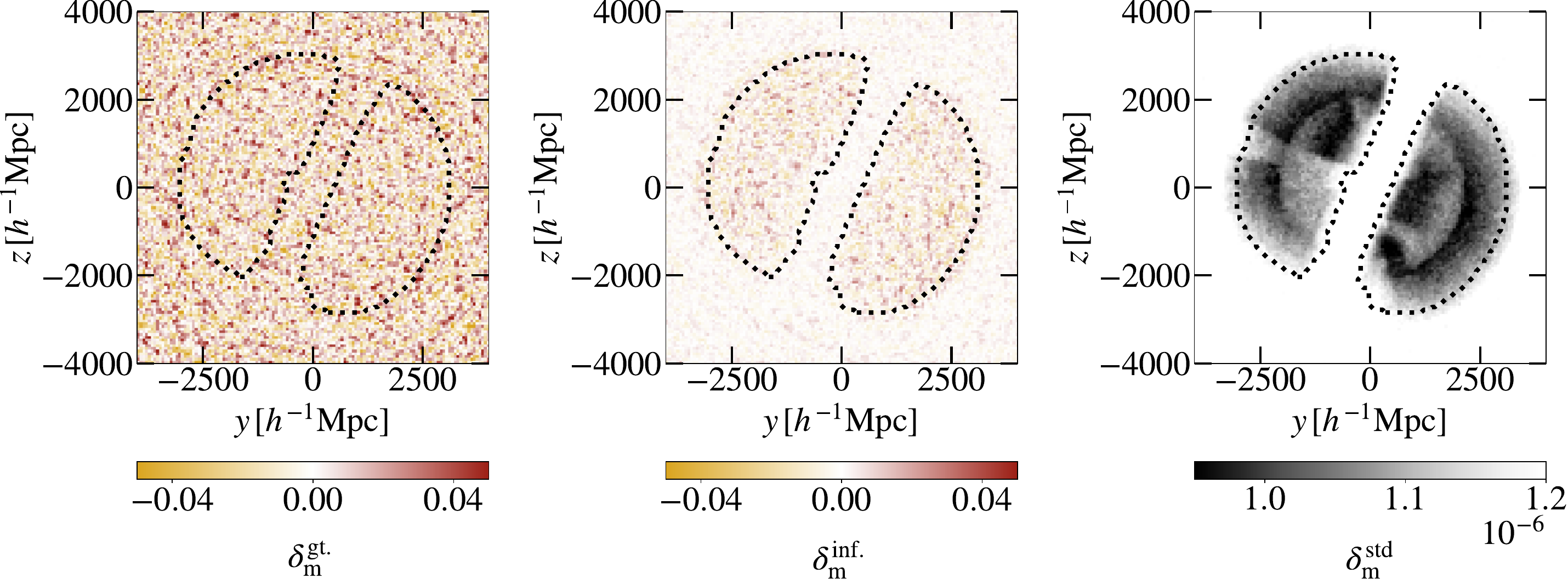}
    \caption{Averaged projections of the ground truth and statistical summaries of the inferred fields. The edge of the survey is highlighted with dotted lines, which means that voxels outside the edge are not observed. In the left panel, the ground truth density field is plotted. In the middle panel, the mean of the ensemble of the inferred fields is plotted. In the right panel, the standard deviation of the ensemble of the inferred fields is plotted. The image illustrates the method's capability to recover the ground truth density field within the regions of observed data. We note that the voxels within the window selection function have less uncertainty and larger inferred means. The inferred fields are the product of Run \#3. Each pixel covers a width and height of $8000 \, \Mpch / 128 = 62.5 \, \Mpch$.}
    \label{fig:deltam_gallery}
\end{figure*}

Fig.~\ref{fig:deltam_gallery}: We present the averages of the ground truth, the mean inferred, the variance of the inferred, and the residuals of the present-day dark matter field, $\delta_{\rm m}$, in a single direction.

To emphasise, the core outcome of our method is the inferred $\fnl$ distribution; these inferred cosmic fields, with uncertainty estimates, stem from the byproduct of the field-level inference approach.

\end{document}